\documentclass[aps,prl,twocolumn,twoside,superscriptaddress,floatfix,longbibliography]{revtex4-2}

\usepackage{amsmath,amssymb,amsthm,amsfonts}
\usepackage{mathtools}
\usepackage{bm}
\usepackage{dsfont,mathrsfs,array,etoolbox}
\usepackage[version=3]{mhchem} % \ce{}

\usepackage{graphicx}
\usepackage{subfigure}
\usepackage{psfrag}

\usepackage{longtable}
\usepackage{enumerate}
\usepackage{latexsym}
\usepackage{color}
\usepackage{setspace}
\usepackage{listings}
\usepackage{makecell}

\usepackage{physics}
\usepackage{braket}

\usepackage{changes}
\usepackage{comment}
\usepackage[normalem]{ulem}

\usepackage[colorlinks,bookmarks=false,citecolor=blue,linkcolor=red,urlcolor=blue]{hyperref}

\newcommand{\ii}{\mathrm{i}}

\newcommand{\bQ}{\mathbf{Q}}
\newcommand{\br}{\mathbf{r}}
\newcommand{\bR}{\mathbf{R}}
\newcommand{\bS}{\mathbf{S}}
\allowdisplaybreaks

\begin{document}
\title{Lee--Yang Zeros and Pseudocritical Drift in J-Q N\'eel Antiferromagnet--Valence-Bond Solid Transitions}
%\title{Lee--Yang Zeros and Pseudocritical Drift in J-Q N\'eel--VBS Transitions}
%\title{Lee--Yang Zeros and Pseudocritical Drift at the N\'eel--VBS Transition in J-Q Models}
\author{Chunhao Guo}
\affiliation{Department of Physics, School of Science and Research Center for Industries of the Future, Westlake University, Hangzhou 310030, China}
\affiliation{Institute of Natural Sciences, Westlake Institute for Advanced Study, Hangzhou 310024, China}

\author{Zhe Wang}
\affiliation{Department of Physics, School of Science and Research Center for Industries of the Future, Westlake University, Hangzhou 310030, China}
\affiliation{Institute of Natural Sciences, Westlake Institute for Advanced Study, Hangzhou 310024, China}

\author{Danhe Wang}
\affiliation{Key Laboratory of Polar Materials and Devices (MOE),
School of Physics, East China Normal University, Shanghai 200241, China}

\author{Zenan Liu}
\affiliation{Department of Physics, School of Science and Research Center for Industries of the Future, Westlake University, Hangzhou 310030, China}
\affiliation{Institute of Natural Sciences, Westlake Institute for Advanced Study, Hangzhou 310024, China}

\author{Haiyuan Zou}
\email{hyzou@phy.ecnu.edu.cn}
\affiliation{Key Laboratory of Polar Materials and Devices (MOE),
School of Physics, East China Normal University, Shanghai 200241, China}

\author{Zheng Yan}
\email{zhengyan@westlake.edu.cn}
\affiliation{Department of Physics, School of Science and Research Center for Industries of the Future, Westlake University, Hangzhou 310030, China}
\affiliation{Institute of Natural Sciences, Westlake Institute for Advanced Study, Hangzhou 310024, China}

\date{\today}

\begin{abstract} 
Square-lattice J-Q models provide a sign-problem-free setting for probing the quantum phase transition between N\'eel antiferromagnet and columnar valence-bond solid. We analyze this transition through the scaling of Lee--Yang zeros, computed within stochastic series expansion quantum Monte Carlo by reweighting configurations sampled near criticality in the presence of complex source fields. Benchmark studies of the dimerized Heisenberg model and the checkerboard J-Q model validate the method, yielding stable \(\mathrm{O}(3)\) critical scaling in the former and clear spacetime-volume scaling in the latter, as expected for a first-order transition. Applying the same analysis to the J-Q models, we find a pronounced and systematic drift of the leading-zero scaling with increasing system size, consistent with an extended pseudocritical regime. The Lee--Yang scaling implies an effective scaling dimension \(\Delta_{\phi}\) of the \(\mathrm{SO}(5)\) order-parameter field that decreases with size and is consistent with vanishing in the thermodynamic limit. Such behavior lies below the scalar unitarity bound of any unitary relativistic conformal field theory in \(2{+}1\) dimensions and enforces inverse spacetime-volume scaling of the zeros, the hallmark of a first-order transition. These results support a weakly first-order interpretation of the N\'eel--VBS transition and establish the finite-size Lee--Yang zeros as a sensitive, symmetry-resolved diagnostic of pseudocriticality and transition order in the J-Q family.
\end{abstract}

\maketitle

\textit{\color{blue} Introduction.}
Quantum magnets provide a controlled setting for studying quantum phase transitions beyond the standard Landau--Ginzburg--Wilson framework based on a single local order parameter \cite{sachdev2011quantum,chaikin1995principles,senthil2004deconfined,senthil2004quantum,senthil2024deconfinedreview}. A central example is the square-lattice transition between N\'eel antiferromagnet (AFM) and a columnar valence-bond solid (VBS), realized in sign-problem-free J-Q models and accessible to large-scale quantum Monte Carlo simulations \cite{sandvik2007evidence,melko2008scaling,lou2009antiferromagnetic,sandvik2010continuous,shao2016quantum,nahum2015emergent,sandvik2020consistent,takahashi2024so5}. Within conventional Landau theory, a direct continuous transition between these phases is not generically expected without additional fine tuning, since the N\'eel and VBS phases break distinct symmetries \cite{chaikin1995principles,sachdev2011quantum,senthil2004deconfined,senthil2004quantum}. The deconfined quantum critical point (DQCP) scenario instead posits a continuous transition governed by fractionalized spinons coupled to an emergent gauge field, with the N\'eel order parameter arising as a spinon bilinear and the VBS order associated with monopole operators \cite{senthil2004deconfined,senthil2004quantum,levin2004deconfined,kaul2013bridging,senthil2024deconfinedreview,qin2017duality,serna2019emergence,ma2019role,takahashi2020valencebond,wang2022scaling,demidio2024entanglement,zhou2024so5,chester2024bootstrapping,yang2025conformal,li2022bootstrapping,song2024extracting,tang2011method,tang2013confinement,chen2024phases}. Complementary to continuum field-theory approaches, Lee--Yang theory characterizes phase transitions through the analytic structure of the partition function $Z$ in the complex plane, where thermodynamic singularities are tied to the accumulation of zeros \cite{yang1952statistical,lee1952statistical,itzykson1983distribution,bena2005statistical,tang2025boundary,liu2025determination,he2026fermion,demidio2023lee,wada2025locating,li2023yanglee,li2025yanglee,kist2021leeyang,vecsei2022leeyang,vecsei2023leeyang,vecsei2025leeyang,gu2026fidelity,arguellocruz2026yanglee,abdelshafy2025yanglee,fan2025simulating,xu2025characterizing,liu2024imaginarytemperature,francis2021many,alsheikh2026carbm}. Importantly, Lee--Yang theory is naturally formulated at finite size and is therefore well suited to numerical simulations. For finite linear system size \(L\) and inverse temperature \(\beta\), Lee--Yang zeros are discrete and remain off the real axis, but their scaling with \(L\) provides a sharp, nonperturbative diagnostic of transition order and critical behavior \cite{fisher1972scaling,itzykson1983distribution,janke2001strength,kenna2013finite,biskup2000general,deger2019determination,wada2025locating,liu2023signatures,liu2024exact,liu2024from,meng2025detecting}. Lee--Yang zeros therefore provide a natural probe of the debated N\'eel--VBS criticality in J-Q models \cite{sandvik2010continuous,shao2016quantum,nahum2015emergent,zhao2019symmetry,ma2020theory,sen2010example,jiang2008antiferromagnet,kuklov2008generic,takahashi2024so5}. %Moreover, identifying the Lee-Yang edge is a key challenge in lattice QCD \cite{ejiri2006Lee,wakayama2019Lee}, with Lee-Yang zeros analyses providing essential insights and serving as a foundational tool.

Early quantum Monte Carlo studies of square-lattice \(\mathrm{SU}(2)\) J-Q models found nearly size-independent crossings of Binder cumulants and scaled correlation length, consistent with a putative continuous N\'eel--VBS transition \cite{sandvik2007evidence,melko2008scaling,lou2009antiferromagnetic,sandvik2010continuous}. At the same time, conventional finite-size diagnostics in these models remain difficult to interpret \cite{sandvik2010continuous,nahum2015deconfined,shao2016quantum,ma2020theory,sandvik2020consistent,takahashi2024so5}. Indeed, conventional indicators remain apparently critical to much larger sizes: Binder-cumulant scaling persists through \(L=512\) \cite{sandvik2020consistent}, while an apparent power-law regime in the spin-correlation derivative persists to \(L=1024\) in J-Q$_2$ \cite{takahashi2024so5}. The same large-scale study provides evidence for a very weak first-order transition, with correspondingly weak coexistence signatures and very small N\'eel and VBS order parameters \cite{takahashi2024so5}. Crossing points can look almost size independent on moderate lattices, yet continue to drift as larger sizes are included, while exponents from standard finite-size fits depend noticeably on the fitting window, indicating that the asymptotic regime is not cleanly reached. Joint AFM--VBS order-parameter distributions also exhibit approximate emergent symmetry, which is suggestive but does not by itself distinguish true criticality from a long-lived pseudocritical regime \cite{nahum2015deconfined,nahum2015emergent,wang2017deconfined,senthil2024deconfinedreview}. Conventional real-axis observables therefore provide important evidence, but do not directly isolate the finite-size singularity scale or decisively determine the character of the transition. Distinguishing a stable unitary fixed point from an anomalously weak first-order transition thus remains a central challenge for understanding N\'eel--VBS criticality in the J-Q family \cite{kaul2013bridging,nahum2015deconfined,shao2016quantum,ma2020theory,senthil2024deconfinedreview,wang2025extracting,zou2025unraveling}.

In this Letter we study the N\'eel--VBS transition through Lee--Yang zeros computed within stochastic series expansion (SSE) quantum Monte Carlo \cite{sandvik1999stochastic,syljuasen2002quantum,evertz2003loop,sandvik2010computational,yan2019sweeping,yan2022global,demidio2023lee,wada2025locating}. A statistically exact SSE estimator for partition-function ratios in complex fields was introduced in Ref.~\cite{demidio2023lee}, where it was applied to the leading N\'eel and VBS zeros of the J-Q$_2$ model. By evaluating complex generating functions in symmetry-resolved channels, we track the finite-size flow of the low-lying Lee--Yang zeros, which directly measures the singularity scale and distinguishes critical behavior from first-order coexistence \cite{itzykson1983distribution,privman1983finite,binder1987finite,borgs1990rigorous,biskup2000general,biskup2004partition}. Using amplitude-free two-size estimators constructed from low-lying zeros \cite{itzykson1983distribution,janke2001strength,kenna2013finite,deger2019determination,wada2025locating}, we analyze benchmark models together with the uniform J-Q$_3$ and J-Q$_2$ systems, resolving several low-lying zeros in both the N\'eel and VBS channels. While the benchmarks recover their expected behavior, J-Q$_2$ and J-Q$_3$ exhibit pronounced finite-size drift, indicating an extended pseudocritical regime. These results support a weakly first-order interpretation of the N\'eel--VBS transition and highlight the advantage of Lee--Yang zeros as a direct finite-size diagnostic of slow drift \cite{nahum2015deconfined,shao2016quantum,gorbenko2018walking,ma2020theory,senthil2024deconfinedreview,takahashi2024so5}.

%------------------------------------------------------------
% Figure: schematic diagrams
%------------------------------------------------------------
\begin{figure}[t]
  \centering
  \includegraphics[width=0.85\linewidth]{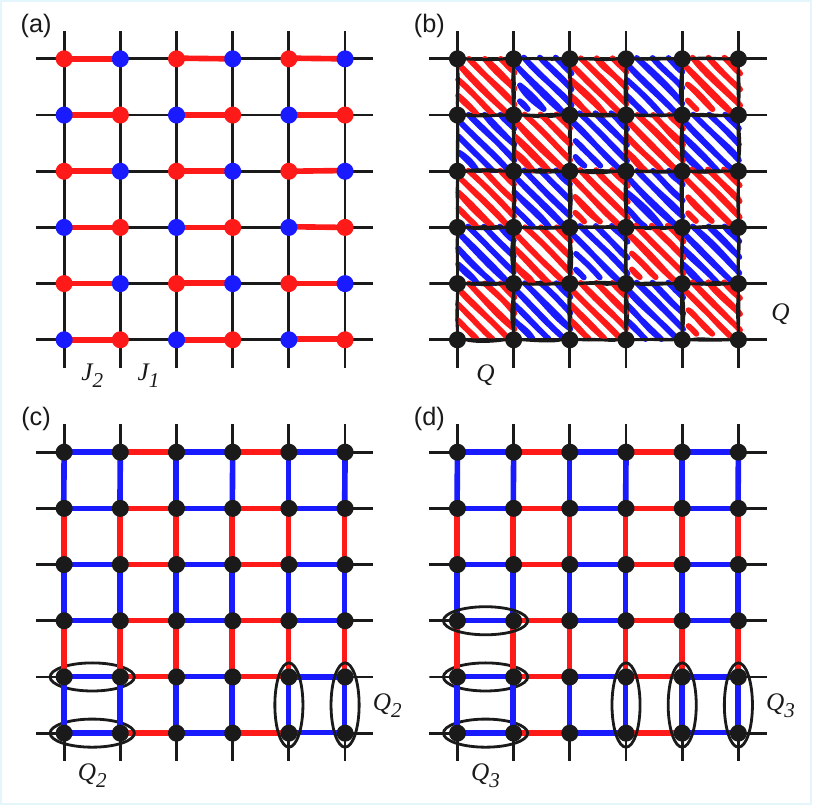}
  \caption{
    Schematic illustrations of the lattice Hamiltonians studied in this work. (a) Columnar dimerized Heisenberg model with alternating strong and weak nearest-neighbor bonds. (b) Checkerboard J-Q (CBJQ) model with \(Q\) interactions on a checkerboard plaquette pattern. (c) Uniform J-Q$_2$ model with \(Q_2\) interactions on every plaquette. (d) Uniform J-Q$_3$ model with \(Q_3\) interactions on \(2\times 3\) and \(3\times 2\) rectangles.
  }
  \label{fig:schematic}
\end{figure}

%==================== Computational methods ====================
\textit{\color{blue} Methods and Models.}
For finite \(L\) and inverse temperature \(\beta\), a symmetry-preserving anchor Hamiltonian \(H_0\) defines an ensemble that respects all microscopic
symmetries, so phases and criticality are inferred from the \(L\) dependence of symmetry-resolved fluctuations \cite{landau1980statistical,chaikin1995principles,sachdev2011quantum,kaul2013bridging}. We probe a channel specified by operator
support (site, bond, or plaquette) and translation sector \(\bQ\), defined from a local density \(O(x)\) at position \(\bR_x\) by the momentum projection
\(O_{\bQ}\equiv \sum_x e^{\ii\bQ\cdot\bR_x}O(x)\), which is a translation eigen-operator and does not mix with inequivalent \(\bQ\) sectors in correlators
\cite{tinkham1964group,landau1980statistical,chaikin1995principles,sachdev2011quantum}. We introduce an imaginary source \(\ii g_{\bQ}\) coupled to \(O_{\bQ}\), \(H(\ii g_{\bQ})=H_0-\ii g_{\bQ}O_{\bQ}\), and work with the normalized generator
\(G_{L,\beta}(\ii g_{\bQ})\equiv Z_{L,\beta}(\ii g_{\bQ})/Z_{L,\beta}(0)\), whose finite-\(L\) zeros encode the analytic structure of the complex $Z$ \cite{yang1952statistical,lee1952statistical,fisher1965nature,itzykson1983distribution,bena2005statistical} and whose thermodynamic pinching yields free-energy
nonanalyticities \cite{yang1952statistical,lee1952statistical,fisher1965nature,itzykson1983distribution}.\footnote{For the formal scaling discussion we adopt the convention \(H(\ii g_{\bQ})=H_0-\ii g_{\bQ}O_{\bQ}\). In the explicit SSE implementation the effective sign of the imaginary deformation is channel dependent. These differences amount only to channel-dependent relabelings of the scan coordinate (\(g\to -g\) where needed) and therefore do not affect the Lee--Yang zero set or its finite-size scaling.} We evaluate \(G\) using the statistically exact SSE estimator of Ref.~\cite{demidio2023lee} for partition-function ratios in complex fields, with configurations sampled
from the real anchor ensemble \cite{ding2024reweight,wang2025bipartite,ding2025tracking,ding2025evaluating,wang2026addressing}:
for each stored configuration \(\mathcal C\) we form the ratio \(R(\ii g_{\bQ};\mathcal C)\equiv W(\mathcal C;\ii g_{\bQ})/W(\mathcal C;0)\), where \(W\) is
the SSE weight including the source-dependent diagonal matrix elements \cite{sandvik1999stochastic,syljuasen2002quantum,sandvik2010computational}, and compute
\(G(\ii g_{\bQ})=\langle R(\ii g_{\bQ};\mathcal C)\rangle_{0}\). Lee--Yang zeros are then the values of \(\ii g_{\bQ}\) at which this estimator vanishes \cite{demidio2023lee,wada2025locating}. For each symmetry channel, we resolve several low-lying Lee--Yang zeros and construct paired-size estimators from them.

%\paragraph{Columnar dimerized Heisenberg benchmark and N\'eel-channel probe.}
As a continuous-transition example, we study the \(2{+}1\)D columnar dimerized Heisenberg model \cite{wenzel2008evidence,wenzel2009quantum}, shown in Fig.~\ref{fig:schematic}(a),
\begin{equation}
H_{\rm dim}
=
J_1\!\!\sum_{\langle ij\rangle\in{\rm weak}}\bS_i\!\cdot\!\bS_j
+
J_2\!\!\sum_{\langle ij\rangle\in{\rm strong}}\bS_i\!\cdot\!\bS_j ,
\label{eq:H_dim}
\end{equation}
which hosts an \(\mathrm{O}(3)\) quantum phase transition between a N\'eel AFM and a quantum-disordered dimer phase \cite{wenzel2008evidence,wenzel2009quantum,ma2018anomalous}. We probe the N\'eel channel at \(\bQ=(\pi,\pi)\) through \(M^z=\sum_{\br}(-1)^{x+y}S^z_{\br}\), by adding the imaginary source term \(\ii h M^z\), with \(h\in\mathbb R\). For each SSE configuration, we record the numbers of diagonal bond operators acting on the two antiparallel local spin configurations \((\uparrow_i\downarrow_j)\) and \((\downarrow_i\uparrow_j)\), denoted \(n^{(s)}_{\uparrow\downarrow}\) and \(n^{(s)}_{\downarrow\uparrow}\), respectively.\footnote{These counts correspond to diagonal SSE vertices in bond sector \(s\), with local bond states \((\uparrow_i\downarrow_j)\) or \((\downarrow_i\uparrow_j)\). The bond endpoints \(i \in A\) and \(j \in B\) on the bipartite lattice are labeled by \(s = \mu\epsilon\), where \(\mu \in \{x,y\}\) is the bond orientation and \(\epsilon = \pm\) denotes the translation-related sublattices, selected by the columnar staggering factor \((-1)^{r_\mu}\) \cite{demidio2023lee,wenzel2008evidence}.} With real anchor denominators \(J_s\) and \(\alpha_s=2/(N_cJ_s)\) (\(N_c=4\) on the square lattice) \cite{sandvik1991quantum,sandvik1999stochastic}, the per-configuration ratio factor is
\(
R_h(\ii h)=
\prod_{s}
\bigl(1-\ii \alpha_s h\bigr)^{n^{(s)}_{\uparrow\downarrow}}
\bigl(1+\ii \alpha_s h\bigr)^{n^{(s)}_{\downarrow\uparrow}} .
\)

%\paragraph{Checkerboard J-Q benchmark and PSS-channel probe.}
As a controlled first-order benchmark, we use the checkerboard J-Q (CBJQ) model \cite{zhao2019symmetry,sun2021emergent,li2024quantum}, shown in Fig.~\ref{fig:schematic}(b),
\begin{equation}
H_{\rm CBJQ}
=
-J\sum_{\langle ij\rangle} P_{ij}
-
Q\!\!\sum_{\langle ij,kl\rangle\in\square'} P_{ij}P_{kl},
\label{eq:H_CBJQ}
\end{equation}
where \(P_{ij}\equiv \tfrac14-\bS_i\!\cdot\!\bS_j\), and \(\square'\) denotes one of the two checkerboard plaquette subfamilies, for example the red shaded plaquettes in Fig.~\ref{fig:schematic}(b) \cite{zhao2019symmetry,sun2021emergent,kaul2013bridging}. This model exhibits a discontinuous transition between a N\'eel AFM and a plaquette-singlet-solid (PSS) phase \cite{zhao2019symmetry,sun2021emergent,li2024quantum}. Motivated by the emergent \(\mathrm{O}(4)\) symmetry reported at the transition \cite{zhao2019symmetry,sun2021emergent}, we probe Lee--Yang zeros for both the N\'eel and PSS order parameters. For the AFM channel, we couple an imaginary source \(\ii h\) to the staggered magnetization \(M^z=\sum_{\br}(-1)^{x+y}S^z_{\br}\). To detect PSS order, we resolve plaquettes by checkerboard and row parities, \(\epsilon_{\rm row}(\square')=(-1)^{y(\square')}\) \cite{zhao2019symmetry,sun2021emergent}, and define
\(
O^{\rm PSS}\equiv \sum_{\square'} \epsilon_{\rm row}(\square')\,\mathcal O(\square'),
\)
where \(\mathcal O(\square')\) is the local plaquette density associated with the microscopic interaction.\footnote{In the SSE implementation, \(\mathcal O(\square')\) corresponds to the four-spin operator \(P_{ij}P_{kl}\) on an active checkerboard plaquette (Eq.~\eqref{eq:H_CBJQ}). The PSS estimator is obtained by counting \(Q\)-vertex insertions with row parity \(\epsilon_{\rm row} = \pm 1\), denoted \(N_{Q+}\) and \(N_{Q-}\), which correspond to plaquette-operator insertions selected by the row-staggering factor.} In the PSS channel, we introduce an imaginary source term \(\ii\eta\,O^{\rm PSS}\). For each SSE configuration, the corresponding reweighting factor is determined by the real anchor coupling \(Q\) and by the row-parity-resolved numbers of \(Q\)-vertex insertions, \(N_{Q+}\) and \(N_{Q-}\), giving
\(
R_{\mathrm{pss}}(\ii \eta)=
\left(1+\ii\frac{\eta}{Q}\right)^{N_{Q+}}
\left(1-\ii\frac{\eta}{Q}\right)^{N_{Q-}} .
\)

%------------------------------------------------------------
% Figure: dimerized (strong--weak bond) Heisenberg benchmark
%------------------------------------------------------------
\begin{figure}[t]
  \centering
  \includegraphics[width=0.50\linewidth]{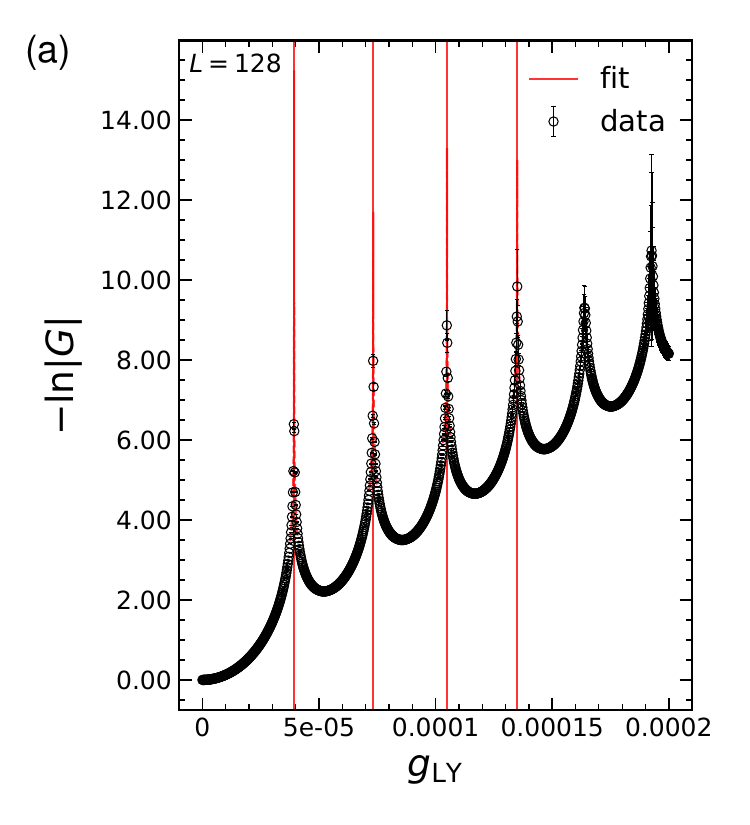}%
  \includegraphics[width=0.50\linewidth]{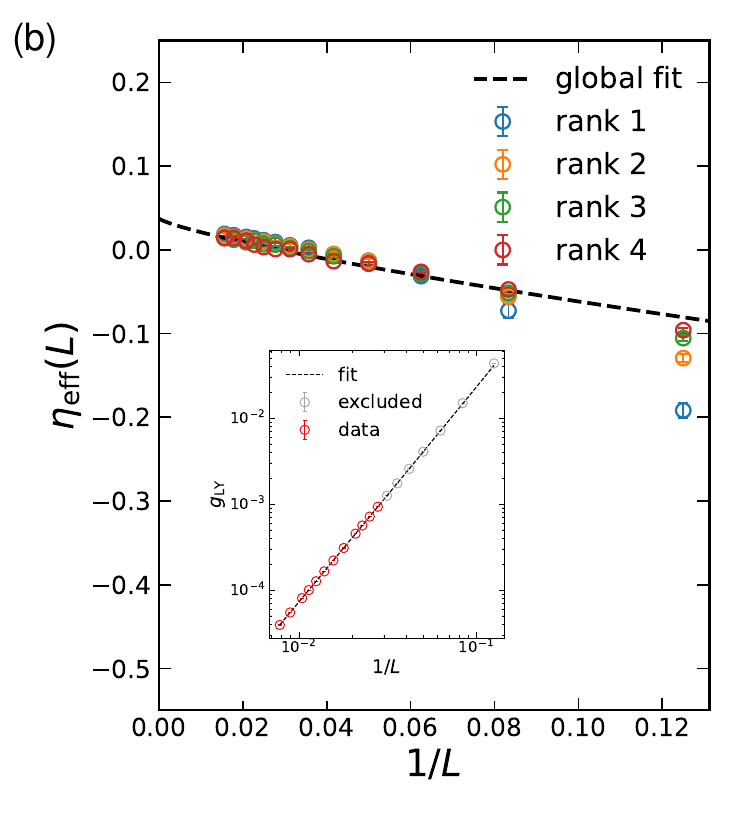}
  \caption{
    Leading Lee--Yang zeros \(g_{\mathrm{LY}}\) and two-size estimators \(\eta_{\mathrm{eff}}^{(k)}(L)\) of the \(2{+}1\)D columnar dimerized Heisenberg model at \(J_c=J_2/J_1=1.90951\) \cite{ma2018anomalous}. (a) \(-\ln |G|\) along the imaginary-source axis for \(L=128\); the red vertical lines mark the first four zeros, ordered from left to right as \(k=1,2,3,4\). Here \(G_{L,\beta}(\ii h)=Z_{L,\beta}(\ii h)/Z_{L,\beta}(0)\), so its zeros appear as divergences of \(-\ln |G|\). (b) \(\eta_{\mathrm{eff}}^{(k)}(L)\) from \((L,2L)\) pairs using the \(k\)th zero. This finite-size diagnostic approaches the anomalous dimension \(\eta\) in a critical regime. The dashed line shows the \(\mathrm{O}(3)\) reference form \(\eta+cL^{-\omega}\) with fixed \(\eta=0.0375\) and correction-to-scaling exponent \(\omega=0.78\). Inset: leading \(g_{\mathrm{LY}}\) versus \(1/L\), fitted to \(a\,L^{-y_h}\!\left(1+bL^{-\omega}\right)\), where \(y_h=(5-\eta)/2\) is the source-field scaling exponent; the fit gives \(\eta=0.0368\pm0.0037\). The fit uses fixed \(\omega=0.78\) and excludes \(L\le 32\).
  }
  \label{fig:heisenberg_dimer_maxima_eta_fit}
\end{figure}

%------------------------------------------------------------
% Figure: CBJQ
%------------------------------------------------------------
\begin{figure}[t]
  \centering
  \includegraphics[width=0.50\linewidth]{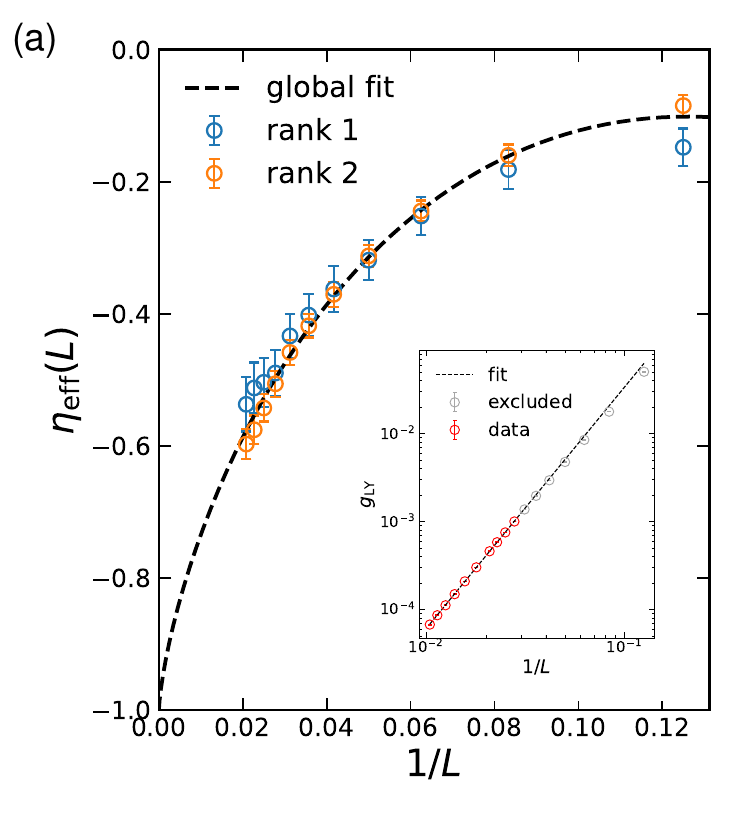}%
  \includegraphics[width=0.50\linewidth]{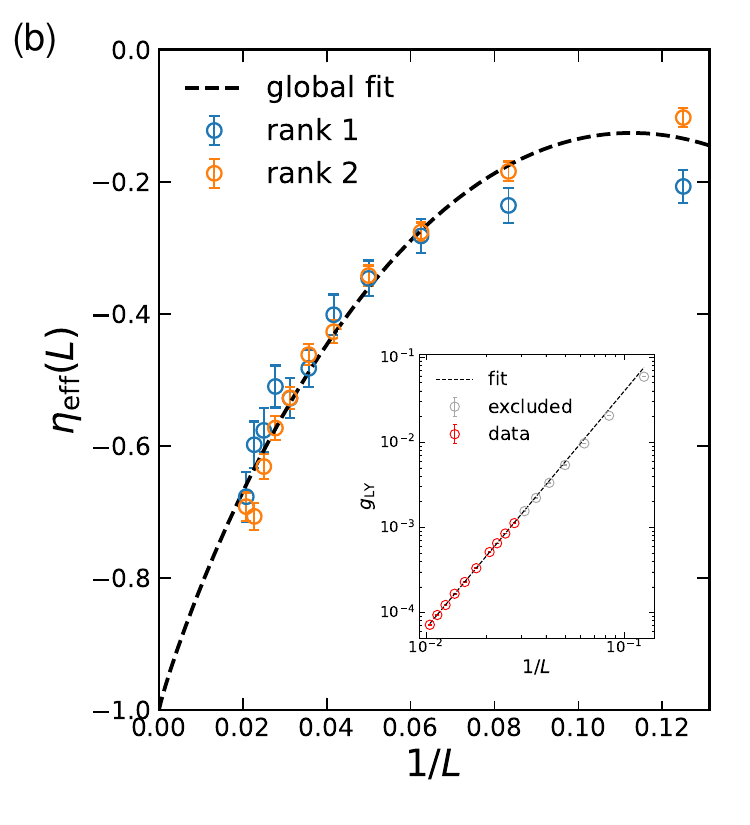}\\[0.5em]
  \includegraphics[width=0.50\linewidth]{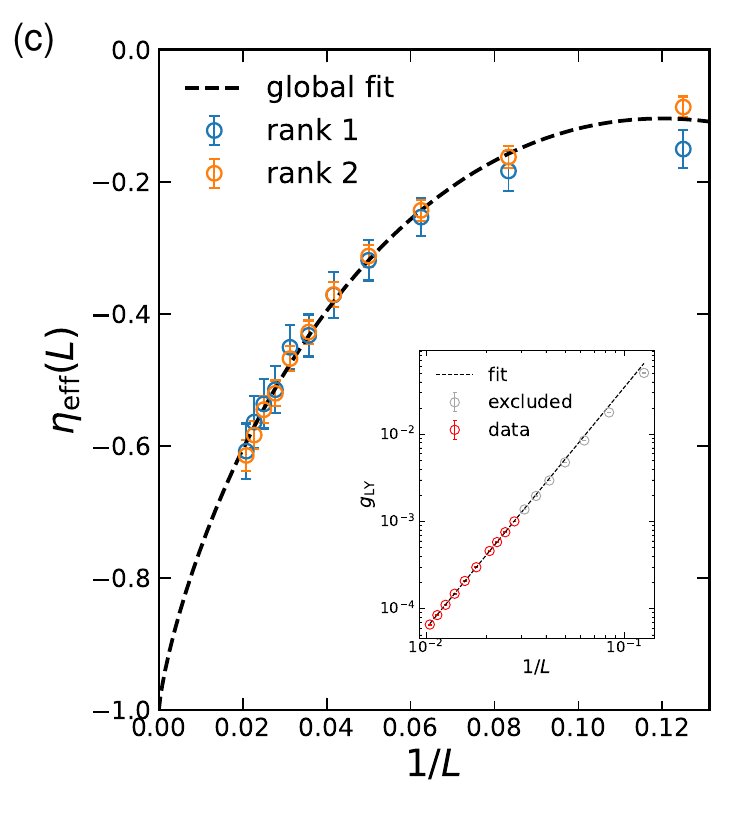}%
  \includegraphics[width=0.50\linewidth]{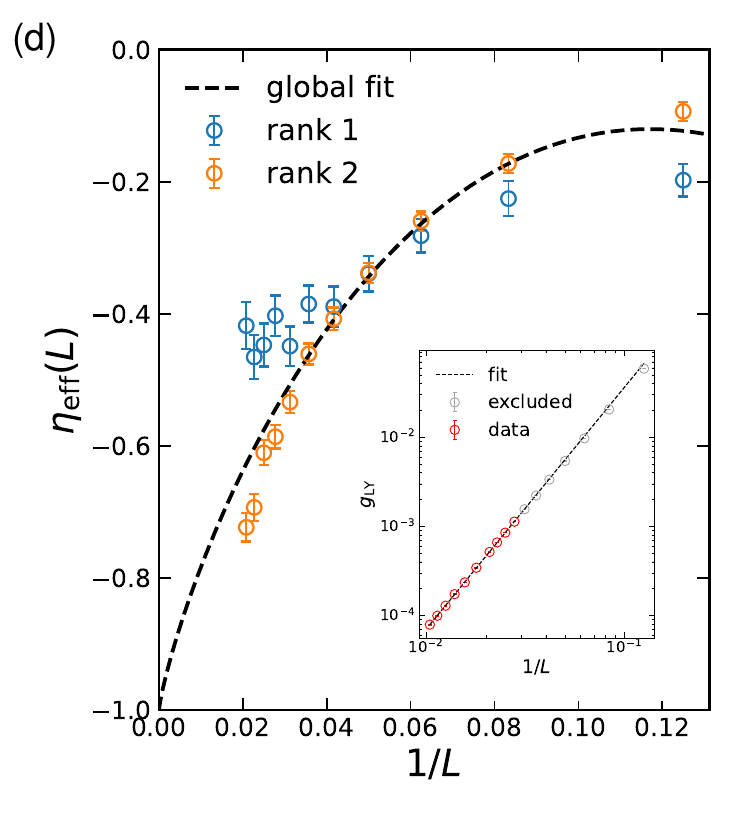}
  \caption{
    Leading Lee--Yang zeros \(g_{\mathrm{LY}}\) and two-size estimators \(\eta_{\mathrm{eff}}(L)\) of the \(2{+}1\)D CBJQ model. Panels (a,b) correspond to \(J/Q=0.2174\), and panels (c,d) to \(J/Q=0.2175\) \cite{zhao2019symmetry}. \(\eta_{\mathrm{eff}}(L)\) is obtained from \((L,2L)\) pairs using the first two zeros. The dashed lines show constrained fits to the first-order reference form \(\eta+c_1L^{-\omega}+c_2L^{-2\omega}\), with \(\eta=-1\) fixed. Panels (a,c) show zeros from analytic continuation of the N\'eel field \(h\) conjugate to \(M^z\); the fits give \(\omega=0.72\pm0.034\) at \(J/Q=0.2174\) and \(\omega=0.77\pm0.032\) at \(J/Q=0.2175\). Panels (b,d) show zeros from analytic continuation of the plaquette-singlet-solid (PSS) source conjugate to \(O^{\mathrm{PSS}}\); the corresponding values are \(\omega=0.90\pm0.028\) and \(\omega=0.82\pm0.029\), respectively. Insets: leading \(g_{\mathrm{LY}}\) versus \(1/L\), fitted to \(a\,L^{-y_h}\), with \(y_h=(5-\eta)/2\). The fits give \(\eta=-0.50\pm0.035\) (\(-0.58\pm0.045\)) in the N\'eel (PSS) channel at \(J/Q=0.2174\), and \(\eta=-0.54\pm0.045\) (\(-0.43\pm0.020\)) at \(J/Q=0.2175\). All fits exclude \(L\le 32\).
  }
  \label{fig:cbjq_maxima_eta_fit}
\end{figure}

%------------------------------------------------------------
% Figure: JQ3 and JQ2 combined
%------------------------------------------------------------
\begin{figure}[t]
  \centering
  \includegraphics[width=0.50\linewidth]{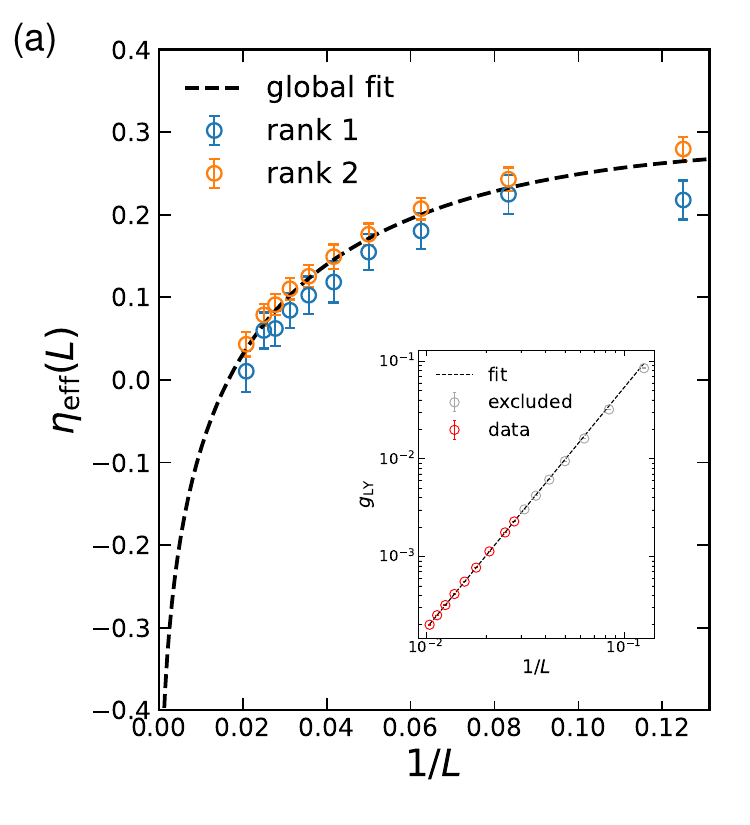}%
  \includegraphics[width=0.50\linewidth]{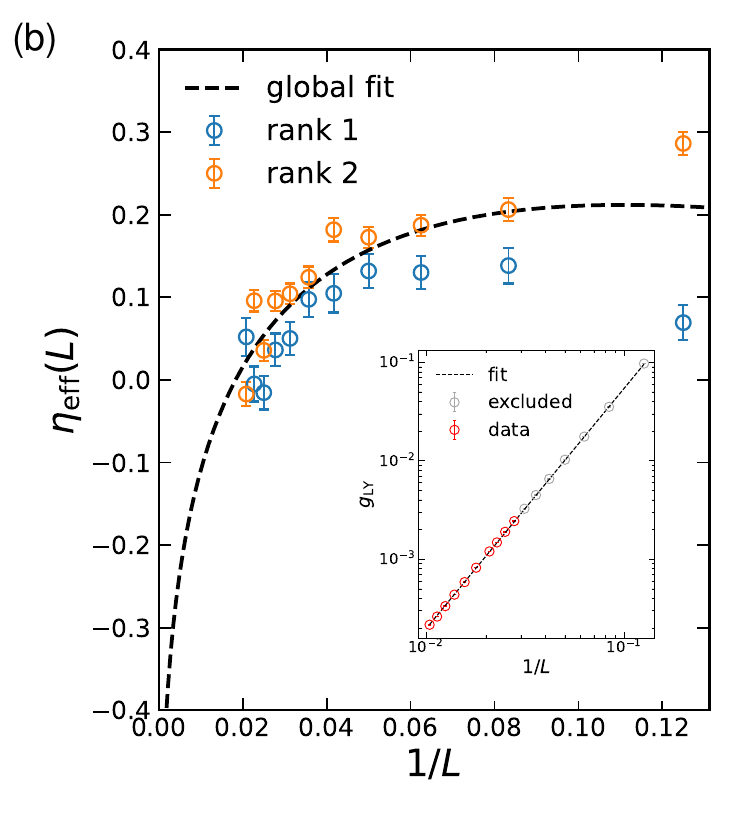}\\[0.5em]
  \includegraphics[width=0.50\linewidth]{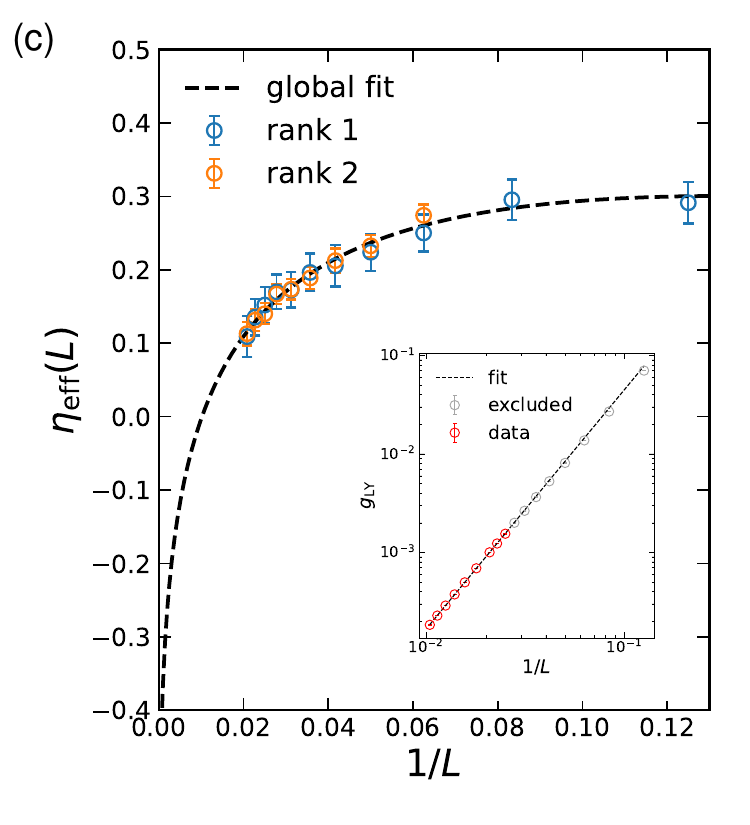}%
  \includegraphics[width=0.50\linewidth]{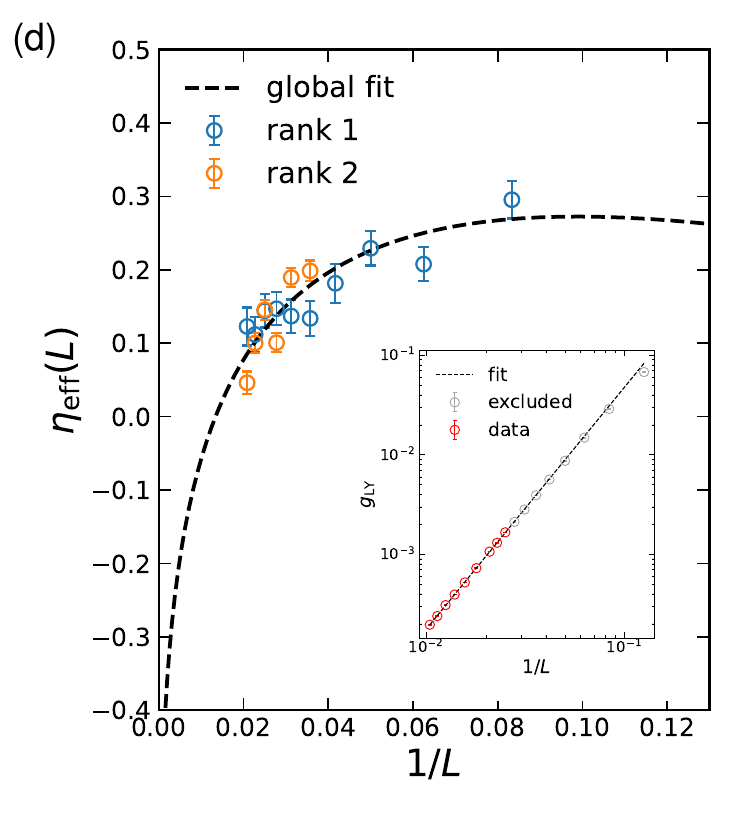}
  \caption{
    Leading Lee--Yang zeros \(g_{\mathrm{LY}}\) and two-size estimators \(\eta_{\mathrm{eff}}(L)\) of the \(2{+}1\)D J-Q$_3$ model at \(J_c=J/Q_3=0.67045\) \cite{lou2009antiferromagnetic,zhao2022scaling,takahashi2024so5} in panels (a,b), and of the \(2{+}1\)D J-Q$_2$ model at \(J_c=J/Q_2=0.045\) \cite{sandvik2010continuous,suwa2016level,shao2016quantum,sandvik2011thermodynamics,takahashi2024so5} in panels (c,d). \(\eta_{\mathrm{eff}}(L)\) is obtained from \((L,2L)\) pairs using the first two zeros. The dashed lines show constrained fits to the first-order reference form \(\eta+c_1L^{-\omega}+c_2L^{-2\omega}\), with \(\eta=-1\) fixed. Panels (a,c) show zeros from analytic continuation of the N\'eel field conjugate to \(M^z\); the fits give \(\omega=0.26\pm0.033\) for J-Q$_3$ and \(\omega=0.20\pm0.11\) for J-Q$_2$. Panels (b,d) show zeros from analytic continuation of the columnar VBS source \(\xi_x\) conjugate to \(D_x\); the corresponding values are \(\omega=0.37\pm0.032\) and \(\omega=0.43\pm0.058\), respectively. Insets: leading \(g_{\mathrm{LY}}\) versus \(1/L\), fitted to \(a\,L^{-y_h}\), with \(y_h=(5-\eta)/2\). The fits give \(\eta=0.050\pm0.022\) and \(0.027\pm0.024\) in the N\'eel and VBS channels, respectively, for J-Q$_3$, and \(\eta=0.138\pm0.021\) and \(0.134\pm0.018\), respectively, for J-Q$_2$. All fits exclude \(L\le 32\).
  }
  \label{fig:jq23_maxima_eta_fit}
\end{figure}

%\paragraph{Uniform J-Q$_2$ and J-Q$_3$ models and VBS-channel probes.}
Our primary targets are the uniform J-Q$_2$ and J-Q$_3$ models, which are canonical DQCP candidates \cite{sandvik2007evidence,melko2008scaling,lou2009antiferromagnetic,sandvik2010continuous,nahum2015deconfined,nahum2015emergent,shao2016quantum,sandvik2020consistent,takahashi2024so5,senthil2024deconfinedreview}, shown in Fig.~\ref{fig:schematic}(c) and (d). In these systems, conventional real-axis observables can appear stable over the accessible range of sizes and may yield seemingly converged critical exponents. As we show below, Lee--Yang zeros provide a more direct probe of the finite-size singularity scale and thereby expose slow drift more clearly. The Hamiltonians are
\begin{align}
H_{\text{JQ}_2}
&=
-J\sum_{\langle ij\rangle} P_{ij}
-
Q_2\!\!\sum_{\langle ij,kl\rangle\in\square} P_{ij}P_{kl},
\label{eq:H_JQ2}
\\
H_{\text{JQ}_3}
&=
-J\sum_{\langle ij\rangle} P_{ij}
-
Q_3\!\!\sum_{\langle ij,kl,mn\rangle\in\square} P_{ij}P_{kl}P_{mn},
\label{eq:H_JQ3}
\end{align}
where \(\langle ij,kl\rangle\in\square\) denotes the two parallel bonds of an elementary plaquette, in either orientation, and \(\langle ij,kl,mn\rangle\in\square\) denotes three adjacent parallel bonds on a \(2\times 3\) or \(3\times 2\) rectangle \cite{sandvik2007evidence,lou2009antiferromagnetic,shao2016quantum}. Tuning \(J/Q_p\), with \(p\in\{2,3\}\), drives the transition between the N\'eel AFM and the columnar VBS \cite{sandvik2007evidence,melko2008scaling,lou2009antiferromagnetic,sandvik2010continuous,nahum2015deconfined,shao2016quantum}. To detect VBS order, we probe the bond-supported columnar channels at \(\bQ=(\pi,0)\) and \((0,\pi)\), using
\(
D_{\mu}=\sum_{\br}(-1)^{r_{\mu}}\,\bS_{\br}\!\cdot\!\bS_{\br+\hat\mu}.
\)
Motivated by the emergent \(\mathrm{SO}(5)\) structure discussed for these transitions \cite{nahum2015deconfined,nahum2015emergent,wang2017deconfined,shao2016quantum,sandvik2020consistent,takahashi2024so5,senthil2024deconfinedreview}, one may equivalently probe the AFM sector or use plaquette-based definitions of VBS order; here we focus on the bond VBS channel and summarize alternative probes in the Supplemental Material~\cite{supplemental} (see also Refs.~\cite{lieb1961two,blythe2003lee,hastings2004lieb,ferrenberg1988new,ferrenberg1989optimized,oshikawa2000commensurability} therein). Implementing the imaginary source term \(\ii\xi_\mu D_\mu\), we separate bond insertions of orientation \(\mu\) into the two translation-related subclasses distinguished by the sign of \((-1)^{r_\mu}\), and denote the corresponding counts by \(N_{J+}\) and \(N_{J-}\)~\cite{demidio2023lee}.\footnote{In the SSE representation, \(N_{J+}\) and \(N_{J-}\) count diagonal \(J\)-bond operator insertions on bonds of orientation \(\mu\), corresponding to the two subclasses with \((-1)^{r_\mu} = \pm 1\).} The resulting factorized ratio contribution is \cite{demidio2023lee}
\(
R_{J\mu}(\ii \xi_\mu)=
\left(1+\ii\frac{\xi_\mu}{J}\right)^{N_{J+}}
\left(1-\ii\frac{\xi_\mu}{J}\right)^{N_{J-}}.
\)

At a continuous transition in a given symmetry sector, we denote by \(\ii g_{\bQ}\) the imaginary source coupled to the corresponding order parameter \(O_{\bQ}\) and  \(g_{\mathrm{LY}}^{(k)}(L)\) the \(k\)th Lee--Yang zero in the complex source plane for linear system size \(L\). A more detailed discussion of this scaling is given in the Supplemental Material. Since \(g_{\bQ}\) is the scaling field conjugate to \(O_{\bQ}\), with renormalization-group eigenvalue \(y_h\) \cite{fisher1972scaling,sachdev2011quantum,cardy1985conformal,itzykson1983distribution}, finite-size scaling gives
\(
g_{\mathrm{LY}}^{(k)}(L)\sim A_k\,L^{-y_h},
\)
up to nonuniversal metric factors and irrelevant-operator corrections \cite{itzykson1983distribution,kenna1994scaling,janke2001strength,kenna2013finite,deger2019determination}, with nonuniversal amplitude \(A_k\). Using
\(
y_h=(d+z+2-\eta)/2,
\)
where \(d\) is the spatial dimension, \(z\) is the dynamical exponent, and \(\eta\) is the anomalous dimension of the order parameter, the Lee--Yang zeros therefore provide direct access to \(\eta\) \cite{sachdev2011quantum,cardy1985conformal}.\footnote{Throughout this work, \(\eta\) extracted from \(y_h=(d+z+2-\eta)/2\) is used as a finite-size diagnostic. Only in a genuine critical regime does it coincide with the true anomalous dimension; in the coexistence regime, \(g_{\mathrm{LY}}(L,\beta)\sim(\beta L^d)^{-1}\) corresponds to the diagnostic limit \(\eta\to -1\).} At phase coexistence, by contrast, a small source biases competing phases by an amount proportional to the Euclidean spacetime volume \(V_\tau=\beta L^d\), where \(\beta\) is the inverse temperature, since the source term is summed over all sites and imaginary time. The zeros then scale as
\(
g_{\mathrm{LY}}^{(k)}(L,\beta)\sim B_k/(\beta L^d),
\)
with nonuniversal amplitude \(B_k\) \cite{privman1983finite,binder1987finite,borgs1990rigorous,biskup2000general,biskup2004partition}. This volume law follows because the imaginary source generates a relative phase proportional to \(g_{\bQ}\beta L^d\) between coexisting phases, and destructive interference occurs when this phase is of order unity. To eliminate \(B_k\), we use the two-size estimator\footnote{Starting from \(g_{\mathrm{LY}}^{(k)}(L)\sim A_kL^{-y_h}\), the ratio for a size pair \((L,2L)\) eliminates \(A_k\),
\(
g_{\mathrm{LY}}^{(k)}(2L)/g_{\mathrm{LY}}^{(k)}(L)\sim 2^{-y_h},
\)
so
\(
y_h=-\ln[g_{\mathrm{LY}}^{(k)}(2L)/g_{\mathrm{LY}}^{(k)}(L)]/\ln 2.
\)
Combining this with \(y_h=(d+z+2-\eta)/2\) gives the estimator below. At finite size, corrections to scaling make this a size-dependent effective estimate of \(\eta\), rather than its asymptotic value.} \cite{ma2018anomalous,deger2019determination,wada2025locating}
\(
\eta_{\mathrm{eff}}^{(k)}(L) = d + z + 2 + \frac{2}{\ln 2} \ln\left[\frac{g_{\mathrm{LY}}^{(k)}(2L)}{g_{\mathrm{LY}}^{(k)}(L)}\right].
\)
For \(d=2\) and \(z=1\), \(\eta_{\mathrm{eff}}^{(k)}(L)\) approaches \(\eta\) at a continuous transition, whereas for the first-order benchmarks studied here, with \(\beta\propto L\), the volume law \(g_{\mathrm{LY}}^{(k)}\sim L^{-(d+1)}\) gives the diagnostic value \(\eta=-1\) \cite{itzykson1983distribution,privman1983finite,binder1987finite,borgs1990rigorous,biskup2000general,biskup2004partition}. If the probed channel is not the ordering channel, the free energy remains analytic at \(g_{\bQ}=0\) in the thermodynamic limit, and the Lee--Yang zeros terminate at a finite edge \cite{fisher1978yang,cardy1985conformal,fonseca2003ising,bena2005statistical,rennecke2022universal}. In J-Q models, this gives a finite VBS-source (N\'eel-source) edge in the N\'eel (VBS) phase, closing only at the transition. In either neighboring phase, only the Lee--Yang zeros of the ordered channel exhibit asymptotic spacetime-volume scaling, while the competing channel retains a finite edge. The common downward flow found below in both the N\'eel and VBS channels therefore provides a direct internal check that the simulations probe the transition regime rather than either neighboring phase. We therefore focus on leading-zero scaling near criticality and leave edge behavior and its finite-size crossover to the Supplemental Material \cite{itzykson1983distribution,kenna1994scaling,janke2002density,bena2005statistical}.

This perspective also clarifies why the leading Lee--Yang zeros can diagnose slow pseudocritical drift more cleanly than conventional real-axis observables in a weakly first-order regime \cite{gorbenko2018walking,ma2020theory}. At first order, coexistence is governed by competition between phases rather than by a single scale-invariant state. Real-axis observables such as order-parameter moments or spin correlations therefore need not approach a unique critical power law and can be strongly affected by phase mixing and analytic backgrounds \cite{privman1983finite,binder1987finite,borgs1990rigorous}. By contrast, the leading Lee--Yang zeros directly encode the finite-size singularity scale through their distance from the real axis, and in the coexistence regime they cross over to the characteristic spacetime-volume scaling of a first-order transition.\footnote{Near coexistence, finite-volume observables are phase-weighted mixtures of competing states, while thermodynamic response functions, being derivatives of \(\ln Z\), are controlled by the nearby Lee--Yang zeros. In the ordering channel this gives \(g_{\mathrm{LY}}(L,\beta)\sim (\beta L^d)^{-1}\). See Supplemental Material for details.} The leading zeros can therefore reveal the slow crossover toward coexistence on size windows where conventional diagnostics still appear quasi-critical \cite{deger2019determination,wada2025locating}.

%==================== end Methods draft ====================

%==================== begin Results draft ====================

\textit{\color{blue} Results and Discussion.}
We first benchmark the Lee--Yang analysis in a system with well-established infrared behavior \cite{kaul2013bridging,sandvik2010computational}. The columnar dimerized Heisenberg model hosts a conventional \(2{+}1\)D \(\mathrm{O}(3)\) critical point \cite{wenzel2008evidence,wenzel2009quantum,kaul2013bridging}. As shown in Fig.~\ref{fig:heisenberg_dimer_maxima_eta_fit}, the leading zeros in the complex source conjugate to \(M^z\) are well described by
\(
g_{\mathrm{LY}}(L)=a\,L^{-y_h}\!\left(1+bL^{-\omega}\right),
\)
with \(y_h=(d+z+2-\eta)/2\) \cite{yang1952statistical,lee1952statistical,itzykson1983distribution,kenna1994scaling,deger2019determination}. Fitting with fixed \(\omega=0.78\) and excluding \(L\le 32\), we obtain \(\eta=0.0368\pm0.0037\), consistent with the established 3D \(\mathrm{O}(3)\) value \(\eta=0.0375(5)\) \cite{guida1998critical,campostrini2002critical,poland2019conformal}. We also construct \(\eta_{\mathrm{eff}}(L)\) from \((L,2L)\) pairs using the first four zeros, which removes the nonuniversal amplitude and exposes finite-size corrections more clearly \cite{ma2018anomalous,deger2019determination,wada2025locating}. The resulting sequence follows the \(\mathrm{O}(3)\) reference form \(\eta_{\mathrm{eff}}(L)=\eta+cL^{-\omega}\) with fixed \(\eta=0.0375\) and \(\omega=0.78\) \cite{kaul2013bridging,poland2019conformal}. Notably, \(\eta_{\mathrm{eff}}(L)\) drifts upward with increasing \(L\), in contrast to the first-order trend, where \(g_{\mathrm{LY}}(L)\sim L^{-(d+z)}\) implies \(\eta\to -1\) \cite{privman1983finite,binder1987finite,borgs1990rigorous,biskup2000general,biskup2004partition}.

The CBJQ transition provides a controlled first-order benchmark for our Lee--Yang diagnostics \cite{zhao2019symmetry,sun2021emergent,li2024quantum}. As shown in Fig.~\ref{fig:cbjq_maxima_eta_fit}, we examine two nearby couplings around the literature estimate \(J_c/Q=0.2175\pm0.0001\) \cite{zhao2019symmetry}. At \(J/Q=0.2174\), the leading zeros in both the N\'eel and PSS channels drift clearly toward the first-order spacetime-volume law
\(
g_{\mathrm{LY}}(L)\sim L^{-(d+z)}
\)
in the ground-state window \(\beta\propto L^z\) \cite{privman1983finite,binder1987finite,borgs1990rigorous,biskup2000general,biskup2004partition}, implying \(y_h=d+z\) and hence \(\eta\to -1\) for \(d=2\), \(z=1\) \cite{itzykson1983distribution,janke2001strength,kenna2013finite}. The paired-size flow is likewise well described by the first-order reference form
\(
\eta_{\mathrm{eff}}(L)=\eta+c_1L^{-\omega}+c_2L^{-2\omega}
\)
with fixed \(\eta=-1\) \cite{privman1983finite,binder1987finite,borgs1990rigorous,biskup2000general,biskup2004partition}. At \(J/Q=0.2175\), the N\'eel-channel zeros remain broadly consistent with this trend, while in the PSS channel the first and second zeros split visibly. We interpret this as indicating that \(J/Q=0.2175\) lies slightly on the N\'eel side of the transition, so that a small PSS-channel Lee--Yang edge remains finite on the accessible sizes: the first zero enters the edge-controlled regime earlier than the second, and the corresponding \(\eta_{\mathrm{eff}}(L)\) values should therefore not be viewed as following a single critical scaling form \cite{fisher1978yang,itzykson1983distribution,bena2005statistical,rennecke2022universal}. The \(J/Q=0.2174\) data thus provide the cleaner first-order benchmark in both channels. Their simultaneous flow in the N\'eel and PSS channels provides a direct check of this channel criterion, because detuning into either phase would asymptotically leave a finite Lee--Yang edge in the competing channel. Overall, the CBJQ results establish the expected first-order Lee--Yang phenomenology and highlight the sensitivity of Lee--Yang zeros to coexistence versus critical scaling even within a narrow coupling window \cite{zhao2019symmetry,sun2021emergent,privman1983finite,biskup2000general,biskup2004partition}. This provides a useful first-order baseline for the J-Q cases discussed below \cite{sandvik2010continuous,shao2016quantum,zhao2019symmetry,ma2020theory,takahashi2024so5}.

We now turn to the uniform J-Q$_3$ and J-Q$_2$ models near their N\'eel--VBS transitions. Figure~\ref{fig:jq23_maxima_eta_fit} shows the leading Lee--Yang zeros \(g_{\mathrm{LY}}(L)\) in the N\'eel and columnar VBS channels, obtained from the sources \(h\) and \(\xi_x\) conjugate to \(M^z\) and \(D_x\), respectively \cite{sandvik2007evidence,melko2008scaling,lou2009antiferromagnetic,sandvik2010continuous}. Although restricted size windows admit apparent power laws, the paired-size estimator \(\eta_{\mathrm{eff}}(L)\), constructed from the first two zeros, decreases systematically with \(L\) in both channels and both models toward the first-order benchmark \(\eta=-1\) \cite{ma2018anomalous,wada2025locating}. An independent VBS source constructed from the \(Q\)-interaction operators exhibits the same trend. For J-Q$_2$, the common downward flow in the N\'eel and VBS channels also persists over a narrow interval of couplings around the transition (see Supplemental Material). This robustness provides a stringent internal check on the choice of real coupling. A fixed detuning into either neighboring phase would asymptotically produce spacetime-volume pinching only in its ordered channel, while the competing channel would retain a finite Lee--Yang edge. The observed common flow in both channels is therefore characteristic of the finite-size transition regime. For J-Q$_3$, this behavior is consistent with the weakly first-order or pseudocritical interpretation discussed in the literature, namely a slow flow toward the spacetime-volume law \(g_{\mathrm{LY}}(L)\sim L^{-(d+z)}\), equivalently \(\eta\to -1\) and \(\Delta_{\phi}\to 0\) for \(d=2\), \(z=1\) \cite{biskup2000general,biskup2004partition,janke2001strength,nahum2015deconfined,shao2016quantum,ma2020theory,gorbenko2018walking,senthil2024deconfinedreview}. For J-Q$_2$, where the nature of the transition has remained less settled, we find closely analogous Lee--Yang flow in both the N\'eel and VBS probes, providing direct evidence that J-Q$_2$ exhibits the same slow drift toward the first-order benchmark \cite{sandvik2010continuous,suwa2016level,shao2016quantum,sandvik2011thermodynamics,takahashi2024so5}. The N\'eel-channel sequence is typically smoother than the VBS-channel sequence in our Lee--Yang analysis, suggesting stronger subleading finite-size effects in the VBS probe. Compared with CBJQ, however, the approach to \(\eta=-1\) is much slower in both J-Q$_2$ and J-Q$_3$, consistent with an extended pseudocritical regime in which accessible sizes remain quasi-critical over broad intermediate scales before eventually bending toward coexistence \cite{nahum2015deconfined,shao2016quantum,ma2020theory,gorbenko2018walking,senthil2024deconfinedreview,takahashi2024so5}. Calibrated by the continuous and first-order benchmarks, the direction and channel coherence of this flow provide strong evidence for a slow crossover toward spacetime-volume pinching. Overall, the Lee--Yang data support a weakly first-order interpretation of both transitions and show that Lee--Yang zeros provide a particularly sensitive probe of this slow drift, especially in the less settled J-Q$_2$ case.

\textit{\color{blue} Conclusion.}
We introduced a symmetry-resolved Lee--Yang framework for quantum magnets that combines SSE sampling at a symmetry-preserving real anchor Hamiltonian with analytic continuation of conjugate sources to evaluate complex generating functions and extract leading zeros. 
Benchmark calculations recover conventional \(\mathrm{O}(3)\) scaling in the columnar dimerized Heisenberg model and the expected first-order scaling behavior in the CBJQ model.
For the uniform J-Q$_3$ and J-Q$_2$ models, we find systematic downward drift of the leading-zero scaling in both N\'eel and VBS channels. In J-Q$_3$, this behavior is consistent with the weakly first-order or pseudocritical scenario discussed in the literature; in J-Q$_2$, we find closely analogous Lee--Yang flow providing direct evidence for the same slow drift toward the first-order benchmark, with the two-size estimator \(\eta_{\mathrm{eff}}\) remaining compatible with a slow approach toward \(\eta\to -1\), equivalently \(\Delta_{\phi}\to 0\), consistent with an extended pseudocritical regime and a weakly first-order interpretation of the N\'eel--VBS transition.
Our work therefore clarifies a long-standing question from the perspective of complexified partition functions and demonstrates the power of the Lee--Yang zero method.
%More broadly, Lee--Yang zeros provide an amplitude-free and sensitive diagnostic for distinguishing continuous criticality from slow drift toward coexistence in finite size simulations.

\textit{\color{blue} Acknowledgments.}
We thank Q. Ye, X. Li, J. Rong and W. Zhu for helpful discussions. 
 Z.W. acknowledges support from the China Postdoctoral Science Foundation under Grant No. 2024M752898. Z.L. is supported by the China Postdoctoral Science Foundation under Grants No. 2024M762935. This project is supported by the Scientific Research Project (No. WU2025B011) and the Start-up Funding of Westlake University. H.Z. is supported by the National Natural Science Foundation of China (Grant No. 12274126).
The authors thank the IT service office and the high-performance computing center of Westlake University.

\textit{\color{blue} Data availability.}
The data supporting this study are available in the repository cited in Ref.~\cite{guo2026data}.

\bibliography{ref}

@article{sandvik2007evidence,
  title = {Evidence for Deconfined Quantum Criticality in a Two-Dimensional {Heisenberg} Model with Four-Spin Interactions},
  author = {Sandvik, Anders W.},
  journal = {Phys. Rev. Lett.},
  volume = {98},
  issue = {22},
  pages = {227202},
  numpages = {4},
  year = {2007},
  month = {Jun},
  publisher = {American Physical Society},
  doi = {10.1103/PhysRevLett.98.227202},
  url = {https://link.aps.org/doi/10.1103/PhysRevLett.98.227202}
}

@article{wang2025extracting,
  title = {Finite-size scaling theory of {R{\'e}nyi} thermal entropy for second-order, first-order, and weakly first-order quantum phase transitions},
  author = {Wang, Zhe and Zhu, Yanzhang and Guo, Chunhao and Chen, Bin-Bin and Ding, Yi-Ming and Liu, Zenan and Yan, Zheng},
  journal = {Phys. Rev. X},
  year = {2026},
  note = {Accepted 6 October 2026},
  doi = {10.1103/42kr-81xw},
  url = {https://journals.aps.org/prx/accepted/10.1103/42kr-81xw},
  eprint = {2506.16111},
  archivePrefix = {arXiv},
  primaryClass = {cond-mat.str-el}
}

@article{yan2019sweeping,
  title = {Sweeping cluster algorithm for quantum spin systems with strong geometric restrictions},
  author = {Yan, Zheng and Wu, Yongzheng and Liu, Chenrong and Sylju{\aa}sen, Olav F. and Lou, Jie and Chen, Yan},
  journal = {Phys. Rev. B},
  volume = {99},
  issue = {16},
  pages = {165135},
  numpages = {14},
  year = {2019},
  month = {Apr},
  publisher = {American Physical Society},
  doi = {10.1103/PhysRevB.99.165135},
  url = {https://link.aps.org/doi/10.1103/PhysRevB.99.165135}
}

@article{ding2025tracking,
  title = {Tracking the variation of entanglement {R{\'e}nyi} negativity: A quantum {Monte Carlo} study},
  author = {Ding, Yi-Ming and Tang, Yin and Wang, Zhe and Wang, Zhiyan and Mao, Bin-Bin and Yan, Zheng},
  journal = {Phys. Rev. B},
  volume = {111},
  issue = {24},
  pages = {L241108},
  year = {2025},
  publisher = {American Physical Society},
  doi = {10.1103/PhysRevB.111.L241108},
  url = {https://link.aps.org/doi/10.1103/PhysRevB.111.L241108}
}

@article{ding2025evaluating,
  title = {Evaluating many-body stabilizer {R{\'e}nyi} entropy by sampling reduced {Pauli} strings: Singularities, volume law, and nonlocal magic},
  author = {Ding, Yi-Ming and Wang, Zhe and Yan, Zheng},
  journal = {PRX Quantum},
  volume = {6},
  issue = {3},
  pages = {030328},
  year = {2025},
  publisher = {American Physical Society},
  doi = {10.1103/pyzr-jmvw},
  url = {https://link.aps.org/doi/10.1103/pyzr-jmvw}
}

@article{wang2026addressing,
  title = {Addressing general measurements in quantum {Monte Carlo}},
  author = {Wang, Zhiyan and Liu, Zenan and Mao, Bin-Bin and Wang, Zhe and Yan, Zheng},
  journal = {Nat. Commun.},
  volume = {17},
  issue = {1},
  pages = {679},
  year = {2026},
  publisher = {Springer Nature},
  doi = {10.1038/s41467-025-67324-0},
  url = {https://www.nature.com/articles/s41467-025-67324-0}
}

@article{wang2025bipartite,
  title = {Bipartite reweight-annealing algorithm of quantum {Monte Carlo} to extract large-scale data of entanglement entropy and its derivative},
  author = {Wang, Zhe and Wang, Zhiyan and Ding, Yi-Ming and Mao, Bin-Bin and Yan, Zheng},
  journal = {Nat. Commun.},
  volume = {16},
  issue = {1},
  pages = {5880},
  year = {2025},
  publisher = {Springer Nature},
  doi = {10.1038/s41467-025-61084-7},
  url = {https://www.nature.com/articles/s41467-025-61084-7}
}

@article{ding2024reweight,
  title = {Reweight-annealing method for evaluating the partition function via quantum {Monte Carlo} calculations},
  author = {Ding, Yi-Ming and Sun, Jun-Song and Ma, Nvsen and Pan, Gaopei and Cheng, Chen and Yan, Zheng},
  journal = {Phys. Rev. B},
  volume = {110},
  issue = {16},
  pages = {165152},
  numpages = {10},
  year = {2024},
  month = {Oct},
  publisher = {American Physical Society},
  doi = {10.1103/PhysRevB.110.165152},
  url = {https://link.aps.org/doi/10.1103/PhysRevB.110.165152}
}

@article{yan2022global,
  title = {Global scheme of sweeping cluster algorithm to sample among topological sectors},
  author = {Yan, Zheng},
  journal = {Phys. Rev. B},
  volume = {105},
  issue = {18},
  pages = {184432},
  numpages = {11},
  year = {2022},
  month = {May},
  publisher = {American Physical Society},
  doi = {10.1103/PhysRevB.105.184432},
  url = {https://link.aps.org/doi/10.1103/PhysRevB.105.184432}
}

@article{liu2025determination,
  title = {Determination of the melting temperature of hexagonal ice using {Lee-Yang} phase transition theory},
  author = {Liu, Ling and Dong, Yihua and Ye, Qi-Jun and Li, Xin-Zheng},
  journal = {Phys. Rev. B},
  volume = {112},
  issue = {10},
  pages = {104102},
  numpages = {10},
  year = {2025},
  month = {Sep},
  publisher = {American Physical Society},
  doi = {10.1103/fr9q-df52},
  url = {https://link.aps.org/doi/10.1103/fr9q-df52}
}

@article{tang2025boundary,
  title = {Boundary criticality of complex conformal field theory: A case study in the non-{Hermitian} 5-state {Potts} model},
  author = {Tang, Yin and Liu, Qianyu and Tang, Qicheng and Zhu, Wei},
  journal = {SciPost Phys.},
  volume = {19},
  issue = {6},
  pages = {164},
  year = {2025},
  publisher = {SciPost Foundation},
  doi = {10.21468/SciPostPhys.19.6.164},
  url = {https://scipost.org/10.21468/SciPostPhys.19.6.164}
}

@article{he2026fermion,
  title = {Fermion sign problem and the structure of {Lee-Yang} zeros: The form of the partition function for indistinguishable particles and its zeros at 0 {K}},
  author = {He, Ran-Chen and Zeng, Jia-Xi and Yang, Shu and Wang, Cong and Ye, Qi-Jun and Li, Xin-Zheng},
  journal = {Phys. Rev. E},
  volume = {113},
  issue = {2},
  pages = {024115},
  numpages = {11},
  year = {2026},
  month = {Feb},
  publisher = {American Physical Society},
  doi = {10.1103/m1py-qtt5},
  url = {https://link.aps.org/doi/10.1103/m1py-qtt5}
}

@article{nahum2015emergent,
  title = {Emergent {SO}(5) Symmetry at the {N{\'e}el} to Valence-Bond-Solid Transition},
  author = {Nahum, Adam and Serna, Pablo and Chalker, J. T. and Ortu{\~n}o, M. and Somoza, A. M.},
  journal = {Phys. Rev. Lett.},
  volume = {115},
  issue = {26},
  pages = {267203},
  numpages = {5},
  year = {2015},
  month = {Dec},
  publisher = {American Physical Society},
  doi = {10.1103/PhysRevLett.115.267203},
  url = {https://link.aps.org/doi/10.1103/PhysRevLett.115.267203}
}

@article{nahum2015deconfined,
  title = {Deconfined Quantum Criticality, Scaling Violations, and Classical Loop Models},
  author = {Nahum, Adam and Chalker, J. T. and Serna, Pablo and Ortu{\~n}o, M. and Somoza, A. M.},
  journal = {Phys. Rev. X},
  volume = {5},
  issue = {4},
  pages = {041048},
  numpages = {21},
  year = {2015},
  month = {Dec},
  publisher = {American Physical Society},
  doi = {10.1103/PhysRevX.5.041048},
  url = {https://link.aps.org/doi/10.1103/PhysRevX.5.041048}
}

@article{shao2016quantum,
  title = {Quantum criticality with two length scales},
  author = {Shao, Hui and Guo, Wenan and Sandvik, Anders W.},
  journal = {Science},
  volume = {352},
  number = {6282},
  pages = {213--216},
  year = {2016},
  doi = {10.1126/science.aad5007},
  url = {https://www.science.org/doi/10.1126/science.aad5007}
}

@article{zhao2019symmetry,
  title = {Symmetry-enhanced discontinuous phase transition in a two-dimensional quantum magnet},
  author = {Zhao, Bowen and Weinberg, Phillip and Sandvik, Anders W.},
  journal = {Nat. Phys.},
  volume = {15},
  issue = {7},
  pages = {678--682},
  year = {2019},
  doi = {10.1038/s41567-019-0484-x},
  url = {https://www.nature.com/articles/s41567-019-0484-x}
}

@article{ma2020theory,
  title = {Theory of deconfined pseudocriticality},
  author = {Ma, Ruochen and Wang, Chong},
  journal = {Phys. Rev. B},
  volume = {102},
  issue = {2},
  pages = {020407},
  numpages = {6},
  year = {2020},
  month = {Jul},
  publisher = {American Physical Society},
  doi = {10.1103/PhysRevB.102.020407},
  url = {https://link.aps.org/doi/10.1103/PhysRevB.102.020407}
}

@article{senthil2004deconfined,
  title = {Deconfined quantum critical points},
  author = {Senthil, T. and Vishwanath, A. and Balents, L. and Sachdev, S. and Fisher, M. P. A.},
  journal = {Science},
  volume = {303},
  number = {5663},
  pages = {1490--1494},
  year = {2004},
  doi = {10.1126/science.1091806},
  url = {https://www.science.org/doi/10.1126/science.1091806}
}

@article{senthil2004quantum,
  title = {Quantum criticality beyond the {Landau-Ginzburg-Wilson} paradigm},
  author = {Senthil, T. and Balents, L. and Sachdev, S. and Vishwanath, A. and Fisher, M. P. A.},
  journal = {Phys. Rev. B},
  volume = {70},
  issue = {14},
  pages = {144407},
  numpages = {22},
  year = {2004},
  month = {Oct},
  publisher = {American Physical Society},
  doi = {10.1103/PhysRevB.70.144407},
  url = {https://link.aps.org/doi/10.1103/PhysRevB.70.144407}
}

@article{levin2004deconfined,
  title = {Deconfined quantum criticality and {N{\'e}el} order via dimer disorder},
  author = {Levin, Michael and Senthil, T.},
  journal = {Phys. Rev. B},
  volume = {70},
  issue = {22},
  pages = {220403},
  numpages = {4},
  year = {2004},
  month = {Dec},
  publisher = {American Physical Society},
  doi = {10.1103/PhysRevB.70.220403},
  url = {https://link.aps.org/doi/10.1103/PhysRevB.70.220403}
}

@article{gorbenko2018walking,
  title = {Walking, weak first-order transitions, and complex {CFT}s},
  author = {Gorbenko, Victor and Rychkov, Slava and Zan, Bernardo},
  journal = {J. High Energy Phys.},
  volume = {2018},
  number = {10},
  pages = {108},
  year = {2018},
  doi = {10.1007/JHEP10(2018)108},
  url = {https://doi.org/10.1007/JHEP10(2018)108}
}

@misc{demidio2023lee,
  title = {{Lee--Yang} zeros at {$O(3)$} and deconfined quantum critical points},
  author = {D'Emidio, Jonathan},
  year = {2023},
  eprint = {2308.00575},
  archivePrefix = {arXiv},
  primaryClass = {cond-mat.str-el},
  url = {https://arxiv.org/abs/2308.00575}
}

@article{wang2017deconfined,
  title = {Deconfined Quantum Critical Points: Symmetries and Dualities},
  author = {Wang, Chong and Nahum, Adam and Metlitski, Max A. and Xu, Cenke and Senthil, T.},
  journal = {Phys. Rev. X},
  volume = {7},
  issue = {3},
  pages = {031051},
  numpages = {47},
  year = {2017},
  month = {Sep},
  publisher = {American Physical Society},
  doi = {10.1103/PhysRevX.7.031051},
  url = {https://link.aps.org/doi/10.1103/PhysRevX.7.031051}
}

@article{wada2025locating,
  title = {Locating Critical Points Using Ratios of {Lee-Yang} Zeros},
  author = {Wada, Tatsuya and Kitazawa, Masakiyo and Kanaya, Kazuyuki},
  journal = {Phys. Rev. Lett.},
  volume = {134},
  issue = {16},
  pages = {162302},
  numpages = {6},
  year = {2025},
  month = {Apr},
  publisher = {American Physical Society},
  doi = {10.1103/PhysRevLett.134.162302},
  url = {https://link.aps.org/doi/10.1103/PhysRevLett.134.162302}
}

@article{yang1952statistical,
  title = {Statistical Theory of Equations of State and Phase Transitions. {I}. Theory of Condensation},
  author = {Yang, C. N. and Lee, T. D.},
  journal = {Phys. Rev.},
  volume = {87},
  issue = {3},
  pages = {404--409},
  year = {1952},
  month = {Aug},
  publisher = {American Physical Society},
  doi = {10.1103/PhysRev.87.404},
  url = {https://link.aps.org/doi/10.1103/PhysRev.87.404}
}

@article{lee1952statistical,
  title = {Statistical Theory of Equations of State and Phase Transitions. {II}. Lattice Gas and {Ising} Model},
  author = {Lee, T. D. and Yang, C. N.},
  journal = {Phys. Rev.},
  volume = {87},
  issue = {3},
  pages = {410--419},
  year = {1952},
  month = {Aug},
  publisher = {American Physical Society},
  doi = {10.1103/PhysRev.87.410},
  url = {https://link.aps.org/doi/10.1103/PhysRev.87.410}
}

@article{itzykson1983distribution,
  title = {Distribution of zeros in {Ising} and gauge models},
  author = {Itzykson, C. and Pearson, R. B. and Zuber, J.-B.},
  journal = {Nucl. Phys. B},
  volume = {220},
  issue = {4},
  pages = {415--433},
  year = {1983},
  doi = {10.1016/0550-3213(83)90499-6},
  url = {https://doi.org/10.1016/0550-3213(83)90499-6}
}

@article{sun2021emergent,
  title = {Emergent {O(4)} symmetry at the phase transition from plaquette-singlet to antiferromagnetic order in quasi-two-dimensional quantum magnets},
  author = {Sun, Guangyu and Ma, Nvsen and Zhao, Bowen and Sandvik, Anders W. and Meng, Zi Yang},
  journal = {Chin. Phys. B},
  volume = {30},
  issue = {6},
  pages = {067505},
  year = {2021},
  doi = {10.1088/1674-1056/abf3b8},
  url = {https://doi.org/10.1088/1674-1056/abf3b8}
}

@book{landau1980statistical,
  title = {Statistical Physics, Part 1},
  author = {Landau, L. D. and Lifshitz, E. M.},
  series = {Course of Theoretical Physics},
  volume = {5},
  edition = {3},
  publisher = {Pergamon Press},
  address = {Oxford},
  year = {1980}
}

@book{chaikin1995principles,
  title = {Principles of Condensed Matter Physics},
  author = {Chaikin, P. M. and Lubensky, T. C.},
  publisher = {Cambridge University Press},
  address = {Cambridge},
  year = {1995}
}

@book{tinkham1964group,
  title = {Group Theory and Quantum Mechanics},
  author = {Tinkham, Michael},
  publisher = {McGraw-Hill},
  address = {New York},
  year = {1964}
}

@book{sachdev2011quantum,
  title = {Quantum Phase Transitions},
  author = {Sachdev, Subir},
  edition = {2},
  publisher = {Cambridge University Press},
  address = {Cambridge},
  year = {2011}
}

@inbook{senthil2024deconfinedreview,
  title = {Deconfined Quantum Critical Points: A Review},
  author = {Senthil, T.},
  booktitle = {50 Years of the Renormalization Group},
  chapter = {14},
  pages = {169--195},
  publisher = {World Scientific},
  year = {2024},
  doi = {10.1142/9789811282386_0014},
  url = {https://doi.org/10.1142/9789811282386_0014}
}

@incollection{fisher1965nature,
  title = {The Nature of Critical Points},
  author = {Fisher, Michael E.},
  booktitle = {Lectures in Theoretical Physics},
  editor = {Brittin, W. E.},
  volume = {7C},
  pages = {1--159},
  publisher = {University of Colorado Press},
  address = {Boulder},
  year = {1965}
}

@article{bena2005statistical,
  title = {Statistical Mechanics of Equilibrium and Nonequilibrium Phase Transitions: The {Y}ang--{L}ee Formalism},
  author = {Bena, I. and Droz, M. and Lipowski, A.},
  journal = {Int. J. Mod. Phys. B},
  volume = {19},
  issue = {29},
  pages = {4269--4329},
  year = {2005},
  doi = {10.1142/S0217979205032759},
  url = {https://doi.org/10.1142/S0217979205032759}
}

@article{cardy1985conformal,
  title = {Conformal Invariance and the {Y}ang--{L}ee Edge Singularity in Two Dimensions},
  author = {Cardy, John L.},
  journal = {Phys. Rev. Lett.},
  volume = {54},
  issue = {13},
  pages = {1354--1356},
  year = {1985},
  month = {Apr},
  publisher = {American Physical Society},
  doi = {10.1103/PhysRevLett.54.1354},
  url = {https://link.aps.org/doi/10.1103/PhysRevLett.54.1354}
}

@article{janke2001strength,
  title = {The Strength of First and Second Order Phase Transitions from Partition Function Zeroes},
  author = {Janke, Wolfhard and Kenna, Ralph},
  journal = {J. Stat. Phys.},
  volume = {102},
  issue = {5--6},
  pages = {1211--1227},
  year = {2001},
  doi = {10.1023/A:1004836227767},
  url = {https://doi.org/10.1023/A:1004836227767}
}

@article{kenna2013finite,
  title = {A new critical exponent `coppa' and its logarithmic counterpart `hat coppa'},
  author = {Kenna, Ralph and Berche, Bertrand},
  journal = {Condens. Matter Phys.},
  volume = {16},
  issue = {2},
  pages = {23601},
  year = {2013},
  doi = {10.5488/CMP.16.23601},
  url = {https://doi.org/10.5488/CMP.16.23601}
}

@article{privman1983finite,
  title = {Finite-Size Effects at First-Order Transitions},
  author = {Privman, V. and Fisher, Michael E.},
  journal = {J. Stat. Phys.},
  volume = {33},
  issue = {2},
  pages = {385--417},
  year = {1983},
  doi = {10.1007/BF01009803},
  url = {https://doi.org/10.1007/BF01009803}
}

@article{binder1987finite,
  title = {Theory of first-order phase transitions},
  author = {Binder, Kurt},
  journal = {Rep. Prog. Phys.},
  volume = {50},
  issue = {7},
  pages = {783--859},
  year = {1987},
  doi = {10.1088/0034-4885/50/7/001},
  url = {https://doi.org/10.1088/0034-4885/50/7/001}
}

@article{borgs1990rigorous,
  title = {A Rigorous Theory of Finite-Size Scaling at First-Order Phase Transitions},
  author = {Borgs, C. and Koteck{\'y}, R.},
  journal = {J. Stat. Phys.},
  volume = {61},
  issue = {1--2},
  pages = {79--119},
  year = {1990},
  doi = {10.1007/BF01013955},
  url = {https://doi.org/10.1007/BF01013955}
}

@article{blythe2003lee,
  title = {The {Lee--Yang} theory of equilibrium and nonequilibrium phase transitions},
  author = {Blythe, Richard A. and Evans, Martin R.},
  journal = {Braz. J. Phys.},
  volume = {33},
  issue = {3},
  pages = {464--475},
  year = {2003},
  doi = {10.1590/S0103-97332003000300008},
  url = {https://doi.org/10.1590/S0103-97332003000300008}
}

@article{biskup2000general,
  title = {General Theory of {Lee--Yang} Zeros in Models with First-Order Phase Transitions},
  author = {Biskup, Marek and Borgs, Christian and Chayes, Jennifer T. and Kleinwaks, Logan J. and Koteck{\'y}, Roman},
  journal = {Phys. Rev. Lett.},
  volume = {84},
  issue = {21},
  pages = {4794--4797},
  year = {2000},
  month = {May},
  publisher = {American Physical Society},
  doi = {10.1103/PhysRevLett.84.4794},
  url = {https://link.aps.org/doi/10.1103/PhysRevLett.84.4794}
}

@article{fisher1972scaling,
  title = {Scaling Theory for Finite-Size Effects in the Critical Region},
  author = {Fisher, Michael E. and Barber, Michael N.},
  journal = {Phys. Rev. Lett.},
  volume = {28},
  issue = {23},
  pages = {1516--1519},
  year = {1972},
  month = {Jun},
  publisher = {American Physical Society},
  doi = {10.1103/PhysRevLett.28.1516},
  url = {https://link.aps.org/doi/10.1103/PhysRevLett.28.1516}
}

@article{fisher1978yang,
  title = {{Yang--Lee} Edge Singularity and $\phi^3$ Field Theory},
  author = {Fisher, Michael E.},
  journal = {Phys. Rev. Lett.},
  volume = {40},
  issue = {25},
  pages = {1610--1613},
  year = {1978},
  month = {Jun},
  publisher = {American Physical Society},
  doi = {10.1103/PhysRevLett.40.1610},
  url = {https://link.aps.org/doi/10.1103/PhysRevLett.40.1610}
}

@article{kenna1994scaling,
  title = {Scaling and density of {Lee--Yang} zeros in the four-dimensional {Ising} model},
  author = {Kenna, R. and Lang, C. B.},
  journal = {Phys. Rev. E},
  volume = {49},
  issue = {6},
  pages = {5012--5017},
  year = {1994},
  month = {Jun},
  publisher = {American Physical Society},
  doi = {10.1103/PhysRevE.49.5012},
  url = {https://link.aps.org/doi/10.1103/PhysRevE.49.5012}
}

@article{janke2002density,
  title = {Density of partition function zeroes and phase transition strength},
  author = {Janke, W. and Kenna, R.},
  journal = {Comput. Phys. Commun.},
  volume = {147},
  issue = {1--2},
  pages = {443--446},
  year = {2002},
  doi = {10.1016/S0010-4655(02)00323-5},
  url = {https://doi.org/10.1016/S0010-4655(02)00323-5}
}

@article{lieb1961two,
  title = {Two soluble models of an antiferromagnetic chain},
  author = {Lieb, Elliott H. and Schultz, Theodore D. and Mattis, Daniel C.},
  journal = {Ann. Phys. (N. Y.)},
  volume = {16},
  issue = {3},
  pages = {407--466},
  year = {1961},
  doi = {10.1016/0003-4916(61)90115-4},
  url = {https://doi.org/10.1016/0003-4916(61)90115-4}
}

@article{oshikawa2000commensurability,
  title = {Commensurability, Excitation Gap, and Topology in Quantum Many-Particle Systems on a Periodic Lattice},
  author = {Oshikawa, Masaki},
  journal = {Phys. Rev. Lett.},
  volume = {84},
  issue = {7},
  pages = {1535--1538},
  year = {2000},
  month = {Feb},
  publisher = {American Physical Society},
  doi = {10.1103/PhysRevLett.84.1535},
  url = {https://link.aps.org/doi/10.1103/PhysRevLett.84.1535}
}

@article{hastings2004lieb,
  title = {{Lieb--Schultz--Mattis} in higher dimensions},
  author = {Hastings, Matthew B.},
  journal = {Phys. Rev. B},
  volume = {69},
  issue = {10},
  pages = {104431},
  numpages = {6},
  year = {2004},
  month = {Mar},
  publisher = {American Physical Society},
  doi = {10.1103/PhysRevB.69.104431},
  url = {https://link.aps.org/doi/10.1103/PhysRevB.69.104431}
}

@article{fonseca2003ising,
  title = {{Ising} Field Theory in a Magnetic Field: Analytic Properties of the Free Energy},
  author = {Fonseca, P. and Zamolodchikov, A. B.},
  journal = {J. Stat. Phys.},
  volume = {110},
  pages = {527--590},
  year = {2003},
  doi = {10.1023/A:1022147532606},
  url = {https://doi.org/10.1023/A:1022147532606}
}

@article{rennecke2022universal,
  title = {Universal location of {Yang--Lee} edge singularity for a one-component field theory in 1 \(\le\) d \(\le\) 4},
  author = {Rennecke, Fabian and Skokov, Vladimir V.},
  journal = {Ann. Phys. (N. Y.)},
  volume = {444},
  pages = {169010},
  year = {2022},
  doi = {10.1016/j.aop.2022.169010},
  url = {https://doi.org/10.1016/j.aop.2022.169010}
}

@article{sandvik1991quantum,
  title = {Quantum {Monte Carlo} simulation method for spin systems},
  author = {Sandvik, Anders W. and Kurkij{\"a}rvi, Juhani},
  journal = {Phys. Rev. B},
  volume = {43},
  issue = {7},
  pages = {5950--5961},
  year = {1991},
  month = {Mar},
  publisher = {American Physical Society},
  doi = {10.1103/PhysRevB.43.5950},
  url = {https://link.aps.org/doi/10.1103/PhysRevB.43.5950}
}

@article{sandvik1999stochastic,
  title = {Stochastic series expansion method with operator-loop update},
  author = {Sandvik, Anders W.},
  journal = {Phys. Rev. B},
  volume = {59},
  issue = {22},
  pages = {R14157--R14160},
  year = {1999},
  month = {Jun},
  publisher = {American Physical Society},
  doi = {10.1103/PhysRevB.59.R14157},
  url = {https://link.aps.org/doi/10.1103/PhysRevB.59.R14157}
}

@article{syljuasen2002quantum,
  title = {Quantum {Monte Carlo} with directed loops},
  author = {Sylju{\aa}sen, Olav F. and Sandvik, Anders W.},
  journal = {Phys. Rev. E},
  volume = {66},
  issue = {4},
  pages = {046701},
  numpages = {10},
  year = {2002},
  month = {Oct},
  publisher = {American Physical Society},
  doi = {10.1103/PhysRevE.66.046701},
  url = {https://link.aps.org/doi/10.1103/PhysRevE.66.046701}
}

@inproceedings{sandvik2010computational,
  title = {Computational Studies of Quantum Spin Systems},
  author = {Sandvik, Anders W.},
  booktitle = {Lectures on the Physics of Strongly Correlated Systems XIV},
  series = {AIP Conference Proceedings},
  volume = {1297},
  pages = {135--338},
  year = {2010},
  publisher = {AIP},
  doi = {10.1063/1.3518900},
  url = {https://pubs.aip.org/aip/acp/article/1297/1/135/854814/Computational-Studies-of-Quantum-Spin-Systems}
}

@article{evertz2003loop,
  title = {The loop algorithm},
  author = {Evertz, Hans Gerd},
  journal = {Adv. Phys.},
  volume = {52},
  issue = {1},
  pages = {1--66},
  year = {2003},
  doi = {10.1080/0001873021000049195},
  url = {https://doi.org/10.1080/0001873021000049195}
}

@article{melko2008scaling,
  title = {Scaling in the Fan of an Unconventional Quantum Critical Point},
  author = {Melko, Roger G. and Kaul, Ribhu K.},
  journal = {Phys. Rev. Lett.},
  volume = {100},
  issue = {1},
  pages = {017203},
  numpages = {4},
  year = {2008},
  month = {Jan},
  publisher = {American Physical Society},
  doi = {10.1103/PhysRevLett.100.017203},
  url = {https://link.aps.org/doi/10.1103/PhysRevLett.100.017203}
}

@article{lou2009antiferromagnetic,
  title = {Antiferromagnetic to valence-bond-solid transitions in two-dimensional {SU}({$N$}) {Heisenberg} models with multispin interactions},
  author = {Lou, Jie and Sandvik, Anders W. and Kawashima, Naoki},
  journal = {Phys. Rev. B},
  volume = {80},
  issue = {18},
  pages = {180414},
  numpages = {4},
  year = {2009},
  month = {Nov},
  publisher = {American Physical Society},
  doi = {10.1103/PhysRevB.80.180414},
  url = {https://link.aps.org/doi/10.1103/PhysRevB.80.180414}
}

@article{sen2010example,
  title = {Example of a first-order {N{\'e}el} to valence-bond-solid transition in two dimensions},
  author = {Sen, Arnab and Sandvik, Anders W.},
  journal = {Phys. Rev. B},
  volume = {82},
  issue = {17},
  pages = {174428},
  numpages = {8},
  year = {2010},
  month = {Nov},
  publisher = {American Physical Society},
  doi = {10.1103/PhysRevB.82.174428},
  url = {https://link.aps.org/doi/10.1103/PhysRevB.82.174428}
}

@article{jiang2008antiferromagnet,
  title = {From an antiferromagnet to a valence bond solid: Evidence for a first order phase transition},
  author = {Jiang, F.-J. and Nyfeler, Matthias and Chandrasekharan, Shailesh and Wiese, U.-J.},
  journal = {J. Stat. Mech.: Theory Exp.},
  volume = {2008},
  number = {02},
  pages = {P02009},
  year = {2008},
  doi = {10.1088/1742-5468/2008/02/P02009},
  url = {https://doi.org/10.1088/1742-5468/2008/02/P02009}
}

@article{kuklov2008generic,
  title = {Deconfined Criticality: Generic First-Order Transition in the {SU(2)} Symmetry Case},
  author = {Kuklov, Anatoly B. and Matsumoto, Masaki and Prokof'ev, Nikolay V. and Svistunov, Boris V. and Troyer, Matthias},
  journal = {Phys. Rev. Lett.},
  volume = {101},
  issue = {5},
  pages = {050405},
  numpages = {4},
  year = {2008},
  month = {Aug},
  publisher = {American Physical Society},
  doi = {10.1103/PhysRevLett.101.050405},
  url = {https://link.aps.org/doi/10.1103/PhysRevLett.101.050405}
}

@article{kaul2013bridging,
  title = {Bridging lattice-scale physics and continuum field theory with quantum {Monte Carlo} simulations},
  author = {Kaul, Ribhu K. and Melko, Roger G. and Sandvik, Anders W.},
  journal = {Annu. Rev. Condens. Matter Phys.},
  volume = {4},
  pages = {179--215},
  year = {2013},
  doi = {10.1146/annurev-conmatphys-030212-184215},
  url = {https://doi.org/10.1146/annurev-conmatphys-030212-184215}
}

@article{deger2019determination,
  title = {Determination of universal critical exponents using {Lee--Yang} theory},
  author = {Deger, Aydin and Flindt, Christian},
  journal = {Phys. Rev. Res.},
  volume = {1},
  issue = {2},
  pages = {023004},
  numpages = {7},
  year = {2019},
  month = {Sep},
  publisher = {American Physical Society},
  doi = {10.1103/PhysRevResearch.1.023004},
  url = {https://link.aps.org/doi/10.1103/PhysRevResearch.1.023004}
}

@article{poland2019conformal,
  title = {The Conformal Bootstrap: Theory, Numerical Techniques, and Applications},
  author = {Poland, David and Rychkov, Slava and Vichi, Alessandro},
  journal = {Rev. Mod. Phys.},
  volume = {91},
  issue = {1},
  pages = {015002},
  numpages = {74},
  year = {2019},
  month = {Jan},
  publisher = {American Physical Society},
  doi = {10.1103/RevModPhys.91.015002},
  url = {https://link.aps.org/doi/10.1103/RevModPhys.91.015002}
}

@article{ferrenberg1988new,
  title = {New {Monte Carlo} technique for studying phase transitions},
  author = {Ferrenberg, Alan M. and Swendsen, Robert H.},
  journal = {Phys. Rev. Lett.},
  volume = {61},
  issue = {23},
  pages = {2635--2638},
  year = {1988},
  month = {Dec},
  publisher = {American Physical Society},
  doi = {10.1103/PhysRevLett.61.2635},
  url = {https://link.aps.org/doi/10.1103/PhysRevLett.61.2635}
}

@article{ferrenberg1989optimized,
  title = {Optimized {Monte Carlo} data analysis},
  author = {Ferrenberg, Alan M. and Swendsen, Robert H.},
  journal = {Phys. Rev. Lett.},
  volume = {63},
  issue = {12},
  pages = {1195--1198},
  year = {1989},
  month = {Sep},
  publisher = {American Physical Society},
  doi = {10.1103/PhysRevLett.63.1195},
  url = {https://link.aps.org/doi/10.1103/PhysRevLett.63.1195}
}

@article{wenzel2008evidence,
  title = {Evidence for an Unconventional Universality Class from a Two-Dimensional Dimerized Quantum {Heisenberg} Model},
  author = {Wenzel, Sandro and Bogacz, Leszek and Janke, Wolfhard},
  journal = {Phys. Rev. Lett.},
  volume = {101},
  issue = {12},
  pages = {127202},
  numpages = {4},
  year = {2008},
  month = {Sep},
  publisher = {American Physical Society},
  doi = {10.1103/PhysRevLett.101.127202},
  url = {https://link.aps.org/doi/10.1103/PhysRevLett.101.127202}
}

@article{wenzel2009quantum,
  title = {Comprehensive quantum {Monte Carlo} study of the quantum critical points in planar dimerized/quadrumerized {Heisenberg} models},
  author = {Wenzel, Sandro and Janke, Wolfhard},
  journal = {Phys. Rev. B},
  volume = {79},
  issue = {1},
  pages = {014410},
  numpages = {10},
  year = {2009},
  month = {Jan},
  publisher = {American Physical Society},
  doi = {10.1103/PhysRevB.79.014410},
  url = {https://link.aps.org/doi/10.1103/PhysRevB.79.014410}
}

@article{ma2018anomalous,
  title = {Anomalous Quantum-Critical Scaling Corrections in Two-Dimensional Antiferromagnets},
  author = {Ma, Nvsen and Weinberg, Phillip and Shao, Hui and Guo, Wenan and Yao, Dao-Xin and Sandvik, Anders W.},
  journal = {Phys. Rev. Lett.},
  volume = {121},
  issue = {11},
  pages = {117202},
  numpages = {6},
  year = {2018},
  month = {Sep},
  publisher = {American Physical Society},
  doi = {10.1103/PhysRevLett.121.117202},
  url = {https://link.aps.org/doi/10.1103/PhysRevLett.121.117202}
}

@article{guida1998critical,
  title = {Critical exponents of the {N}-vector model},
  author = {Guida, R. and Zinn-Justin, J.},
  journal = {J. Phys. A},
  volume = {31},
  pages = {8103--8121},
  year = {1998},
  doi = {10.1088/0305-4470/31/40/006},
  url = {https://doi.org/10.1088/0305-4470/31/40/006}
}

@article{campostrini2002critical,
  title = {Critical exponents and equation of state of the three-dimensional {Heisenberg} universality class},
  author = {Campostrini, M. and Hasenbusch, M. and Pelissetto, A. and Rossi, P. and Vicari, E.},
  journal = {Phys. Rev. B},
  volume = {65},
  issue = {14},
  pages = {144520},
  numpages = {11},
  year = {2002},
  month = {Apr},
  publisher = {American Physical Society},
  doi = {10.1103/PhysRevB.65.144520},
  url = {https://link.aps.org/doi/10.1103/PhysRevB.65.144520}
}

@article{li2024quantum,
  author  = {Chengchen Li and Huihang Lin and Rong Yu},
  title   = {Quantum scaling of the spin lattice relaxation rate in the checkerboard {$J$-$Q$} model},
  journal = {J. Phys.: Condens. Matter},
  volume  = {36},
  number  = {35},
  pages   = {355805},
  year    = {2024},
  doi     = {10.1088/1361-648X/ad4ccd},
  url     = {https://doi.org/10.1088/1361-648X/ad4ccd}
}

@article{biskup2004partition,
  title = {Partition Function Zeros at First-Order Phase Transitions: A General Analysis},
  author = {Biskup, Marek and Borgs, Christian and Chayes, Jennifer T. and Kleinwaks, Logan J. and Koteck{\'y}, Roman},
  journal = {Commun. Math. Phys.},
  volume = {251},
  pages = {79--131},
  year = {2004},
  doi = {10.1007/s00220-004-1169-5},
  url = {https://doi.org/10.1007/s00220-004-1169-5}
}

@article{zhao2022scaling,
  title = {Scaling of Entanglement Entropy at Deconfined Quantum Criticality},
  author = {Zhao, Jiarui and Wang, Yan-Cheng and Yan, Zheng and Cheng, Meng and Meng, Zi Yang},
  journal = {Phys. Rev. Lett.},
  volume = {128},
  issue = {1},
  pages = {010601},
  numpages = {6},
  year = {2022},
  month = {Jan},
  publisher = {American Physical Society},
  doi = {10.1103/PhysRevLett.128.010601},
  url = {https://link.aps.org/doi/10.1103/PhysRevLett.128.010601}
}

@article{sandvik2011thermodynamics,
  title = {Thermodynamics of a Gas of Deconfined Bosonic Spinons in Two Dimensions},
  author = {Sandvik, Anders W. and Kotov, Valeri N. and Sushkov, Oleg P.},
  journal = {Phys. Rev. Lett.},
  volume = {106},
  issue = {20},
  pages = {207203},
  numpages = {4},
  year = {2011},
  month = {May},
  publisher = {American Physical Society},
  doi = {10.1103/PhysRevLett.106.207203},
  url = {https://link.aps.org/doi/10.1103/PhysRevLett.106.207203}
}

@article{sandvik2020consistent,
  title = {Consistent Scaling Exponents at the Deconfined Quantum-Critical Point},
  author = {Sandvik, Anders W. and Zhao, Bowen},
  journal = {Chin. Phys. Lett.},
  volume = {37},
  issue = {5},
  pages = {057502},
  year = {2020},
  doi = {10.1088/0256-307X/37/5/057502},
  url = {https://doi.org/10.1088/0256-307X/37/5/057502}
}

@misc{takahashi2024so5,
  title = {{SO(5)} multicriticality in two-dimensional quantum magnets},
  author = {Takahashi, Jun and Shao, Hui and Zhao, Bowen and Guo, Wenan and Sandvik, Anders W.},
  year = {2024},
  eprint = {2405.06607},
  archivePrefix = {arXiv},
  primaryClass = {cond-mat.str-el},
  url = {https://arxiv.org/abs/2405.06607}
}

@article{sandvik2010continuous,
  title = {Continuous Quantum Phase Transition between an Antiferromagnet and a Valence-Bond Solid in Two Dimensions: Evidence for Logarithmic Corrections to Scaling},
  author = {Sandvik, Anders W.},
  journal = {Phys. Rev. Lett.},
  volume = {104},
  issue = {17},
  pages = {177201},
  numpages = {4},
  year = {2010},
  month = {Apr},
  publisher = {American Physical Society},
  doi = {10.1103/PhysRevLett.104.177201},
  url = {https://link.aps.org/doi/10.1103/PhysRevLett.104.177201}
}

@article{suwa2016level,
  title = {Level spectroscopy in a two-dimensional quantum magnet: Linearly dispersing spinons at the deconfined quantum critical point},
  author = {Suwa, Hidemaro and Sen, Arnab and Sandvik, Anders W.},
  journal = {Phys. Rev. B},
  volume = {94},
  issue = {14},
  pages = {144416},
  numpages = {9},
  year = {2016},
  month = {Oct},
  publisher = {American Physical Society},
  doi = {10.1103/PhysRevB.94.144416},
  url = {https://link.aps.org/doi/10.1103/PhysRevB.94.144416}
}

@misc{zou2025unraveling,
  title = {Unraveling deconfined quantum criticality in non-{Hermitian} easy-plane {$J$--$Q$} model},
  author = {Zou, Xuan and Yin, Shuai and Li, Zi-Xiang and Yao, Hong},
  year = {2025},
  eprint = {2511.03456},
  archivePrefix = {arXiv},
  primaryClass = {cond-mat.str-el},
  url = {https://arxiv.org/abs/2511.03456}
}

@article{li2023yanglee,
  title = {{Yang-Lee} Zeros, Semicircle Theorem, and Nonunitary Criticality in {Bardeen-Cooper-Schrieffer} Superconductivity},
  author = {Li, Hongchao and Yu, Xie-Hang and Nakagawa, Masaya and Ueda, Masahito},
  journal = {Phys. Rev. Lett.},
  volume = {131},
  issue = {21},
  pages = {216001},
  numpages = {6},
  year = {2023},
  month = {Nov},
  publisher = {American Physical Society},
  doi = {10.1103/PhysRevLett.131.216001},
  url = {https://link.aps.org/doi/10.1103/PhysRevLett.131.216001}
}

@article{li2025yanglee,
  title = {{Yang-Lee} zeros in quantum phase transitions: An entanglement perspective},
  author = {Li, Hongchao},
  journal = {Phys. Rev. B},
  volume = {111},
  issue = {4},
  pages = {045139},
  numpages = {12},
  year = {2025},
  month = {Jan},
  publisher = {American Physical Society},
  doi = {10.1103/PhysRevB.111.045139},
  url = {https://link.aps.org/doi/10.1103/PhysRevB.111.045139}
}

@article{kist2021leeyang,
  title = {{Lee-Yang} theory of criticality in interacting quantum many-body systems},
  author = {Kist, Timo and Lado, Jose L. and Flindt, Christian},
  journal = {Phys. Rev. Res.},
  volume = {3},
  issue = {3},
  pages = {033206},
  numpages = {6},
  year = {2021},
  month = {Sep},
  publisher = {American Physical Society},
  doi = {10.1103/PhysRevResearch.3.033206},
  url = {https://link.aps.org/doi/10.1103/PhysRevResearch.3.033206}
}

@article{vecsei2022leeyang,
  title = {{Lee-Yang} theory of the two-dimensional quantum {Ising} model},
  author = {Vecsei, Pascal M. and Lado, Jose L. and Flindt, Christian},
  journal = {Phys. Rev. B},
  volume = {106},
  issue = {5},
  pages = {054402},
  numpages = {10},
  year = {2022},
  month = {Aug},
  publisher = {American Physical Society},
  doi = {10.1103/PhysRevB.106.054402},
  url = {https://link.aps.org/doi/10.1103/PhysRevB.106.054402}
}

@article{vecsei2023leeyang,
  title = {{Lee-Yang} theory of quantum phase transitions with neural network quantum states},
  author = {Vecsei, Pascal M. and Flindt, Christian and Lado, Jose L.},
  journal = {Phys. Rev. Res.},
  volume = {5},
  issue = {3},
  pages = {033116},
  numpages = {9},
  year = {2023},
  month = {Aug},
  publisher = {American Physical Society},
  doi = {10.1103/PhysRevResearch.5.033116},
  url = {https://link.aps.org/doi/10.1103/PhysRevResearch.5.033116}
}

@article{vecsei2025leeyang,
  title = {{Lee-Yang} formalism for phase transitions of interacting fermions using tensor networks},
  author = {Vecsei, Pascal M. and Lado, Jose L. and Flindt, Christian},
  journal = {Phys. Rev. B},
  volume = {111},
  issue = {7},
  pages = {075134},
  numpages = {9},
  year = {2025},
  month = {Feb},
  publisher = {American Physical Society},
  doi = {10.1103/PhysRevB.111.075134},
  url = {https://link.aps.org/doi/10.1103/PhysRevB.111.075134}
}

@article{gu2026fidelity,
  title = {Fidelity zeros and {Lee-Yang} theory of quantum phase transitions},
  author = {Gu, Tian-Yi and Sun, Gaoyong},
  journal = {Phys. Rev. B},
  volume = {113},
  issue = {1},
  pages = {014417},
  numpages = {8},
  year = {2026},
  month = {Jan},
  publisher = {American Physical Society},
  doi = {10.1103/3x34-f53v},
  url = {https://link.aps.org/doi/10.1103/3x34-f53v}
}

@article{arguellocruz2026yanglee,
  title = {{Yang-Lee} Quantum Criticality in Various Dimensions},
  author = {Arguello Cruz, Erick and Klebanov, Igor R. and Tarnopolsky, Grigory and Xin, Yuan},
  journal = {Phys. Rev. X},
  volume = {16},
  issue = {1},
  pages = {011022},
  numpages = {27},
  year = {2026},
  month = {Feb},
  publisher = {American Physical Society},
  doi = {10.1103/w4qg-2xwn},
  url = {https://link.aps.org/doi/10.1103/w4qg-2xwn}
}

@article{abdelshafy2025yanglee,
  title = {{Yang-Lee} zeros of two-dimensional nearest-neighbor antiferromagnetic {Ising} models: A numerical linked cluster expansion study},
  author = {Abdelshafy, Mahmoud and Sedik, Muhammad},
  journal = {Phys. Rev. B},
  volume = {112},
  issue = {5},
  pages = {054445},
  numpages = {12},
  year = {2025},
  month = {Aug},
  publisher = {American Physical Society},
  doi = {10.1103/x38c-w4z2},
  url = {https://link.aps.org/doi/10.1103/x38c-w4z2}
}

@article{fan2025simulating,
  title = {Simulating the non-unitary {Yang--Lee} conformal field theory on the fuzzy sphere},
  author = {Fan, Ruihua and Dong, Junkai and Vishwanath, Ashvin},
  journal = {J. High Energy Phys.},
  volume = {2026},
  number = {09},
  pages = {041},
  year = {2026},
  month = {Sep},
  doi = {10.1007/JHEP09(2026)041},
  url = {https://link.springer.com/article/10.1007/JHEP09(2026)041},
  eprint = {2505.06342},
  archivePrefix = {arXiv},
  primaryClass = {cond-mat.str-el}
}

@article{xu2025characterizing,
  title = {Characterizing the {Yang-Lee} zeros of the classical {Ising} model through dynamic quantum phase transitions},
  author = {Xu, Mingtao and Yi, Wei and Cai, De-Huan},
  journal = {Phys. Rev. A},
  volume = {111},
  issue = {4},
  pages = {042204},
  numpages = {9},
  year = {2025},
  month = {Apr},
  publisher = {American Physical Society},
  doi = {10.1103/PhysRevA.111.042204},
  url = {https://link.aps.org/doi/10.1103/PhysRevA.111.042204}
}

@article{liu2024imaginarytemperature,
  title = {Imaginary-temperature zeros for quantum phase transitions},
  author = {Liu, Jinghu and Yin, Shuai and Chen, Li},
  journal = {Phys. Rev. B},
  volume = {110},
  issue = {13},
  pages = {134313},
  numpages = {10},
  year = {2024},
  month = {Oct},
  publisher = {American Physical Society},
  doi = {10.1103/PhysRevB.110.134313},
  url = {https://link.aps.org/doi/10.1103/PhysRevB.110.134313}
}

@article{liu2023signatures,
  author  = {Yang Liu and Songtai Lv and Yang Yang and Haiyuan Zou},
  title   = {Signatures of Quantum Criticality in the Complex Inverse Temperature Plane},
  journal = {Chin. Phys. Lett.},
  volume  = {40},
  number  = {5},
  pages   = {050502},
  year    = {2023},
  doi     = {10.1088/0256-307X/40/5/050502},
  url     = {https://doi.org/10.1088/0256-307X/40/5/050502}
}

@article{liu2024exact,
  title = {Exact {Fisher} zeros and thermofield dynamics across a quantum critical point},
  author = {Liu, Yang and Lv, Songtai and Meng, Yuchen and Tan, Zefan and Zhao, Erhai and Zou, Haiyuan},
  journal = {Phys. Rev. Res.},
  volume = {6},
  issue = {4},
  pages = {043139},
  numpages = {6},
  year = {2024},
  month = {Nov},
  publisher = {American Physical Society},
  doi = {10.1103/PhysRevResearch.6.043139},
  url = {https://link.aps.org/doi/10.1103/PhysRevResearch.6.043139}
}

@article{liu2024from,
  author  = {Yang Liu and Erhai Zhao and Haiyuan Zou},
  title   = {From Complexification to Self-Similarity: New Aspects of Quantum Criticality},
  journal = {Chin. Phys. Lett.},
  volume  = {41},
  number  = {10},
  pages   = {100501},
  year    = {2024},
  doi     = {10.1088/0256-307X/41/10/100501},
  url     = {https://doi.org/10.1088/0256-307X/41/10/100501}
}

@article{meng2025detecting,
  title = {Detecting Many-Body Scars from {Fisher} Zeros},
  author = {Meng, Yuchen and Lv, Songtai and Liu, Yang and Tan, Zefan and Zhao, Erhai and Zou, Haiyuan},
  journal = {Phys. Rev. Lett.},
  volume = {135},
  issue = {7},
  pages = {070402},
  numpages = {7},
  year = {2025},
  month = {Aug},
  publisher = {American Physical Society},
  doi = {10.1103/glc5-hv2m},
  url = {https://link.aps.org/doi/10.1103/glc5-hv2m}
}

@article{qin2017duality,
  title = {Duality between the Deconfined Quantum-Critical Point and the Bosonic Topological Transition},
  author = {Qin, Yan Qi and He, Yuan-Yao and You, Yi-Zhuang and Lu, Zhong-Yi and Sen, Arnab and Sandvik, Anders W. and Xu, Cenke and Meng, Zi Yang},
  journal = {Phys. Rev. X},
  volume = {7},
  issue = {3},
  pages = {031052},
  numpages = {18},
  year = {2017},
  month = {Sep},
  publisher = {American Physical Society},
  doi = {10.1103/PhysRevX.7.031052},
  url = {https://link.aps.org/doi/10.1103/PhysRevX.7.031052}
}

@article{serna2019emergence,
  title = {Emergence and spontaneous breaking of approximate $\mathrm{O}(4)$ symmetry at a weakly first-order deconfined phase transition},
  author = {Serna, Pablo and Nahum, Adam},
  journal = {Phys. Rev. B},
  volume = {99},
  issue = {19},
  pages = {195110},
  numpages = {14},
  year = {2019},
  month = {May},
  publisher = {American Physical Society},
  doi = {10.1103/PhysRevB.99.195110},
  url = {https://link.aps.org/doi/10.1103/PhysRevB.99.195110}
}

@article{ma2019role,
  title = {Role of {Noether}'s Theorem at the Deconfined Quantum Critical Point},
  author = {Ma, Nvsen and You, Yi-Zhuang and Meng, Zi Yang},
  journal = {Phys. Rev. Lett.},
  volume = {122},
  issue = {17},
  pages = {175701},
  numpages = {6},
  year = {2019},
  month = {May},
  publisher = {American Physical Society},
  doi = {10.1103/PhysRevLett.122.175701},
  url = {https://link.aps.org/doi/10.1103/PhysRevLett.122.175701}
}

@article{takahashi2020valencebond,
  title = {Valence-bond solids, vestigial order, and emergent {SO(5)} symmetry in a two-dimensional quantum magnet},
  author = {Takahashi, Jun and Sandvik, Anders W.},
  journal = {Phys. Rev. Res.},
  volume = {2},
  issue = {3},
  pages = {033459},
  numpages = {29},
  year = {2020},
  month = {Sep},
  publisher = {American Physical Society},
  doi = {10.1103/PhysRevResearch.2.033459},
  url = {https://link.aps.org/doi/10.1103/PhysRevResearch.2.033459}
}

@Article{wang2022scaling,
	title={{Scaling of the disorder operator at deconfined quantum criticality}},
	author={Yan-Cheng Wang and Nvsen Ma and Meng Cheng and Zi Yang Meng},
	journal={SciPost Phys.},
	volume={13},
	pages={123},
	year={2022},
	publisher={SciPost},
	doi={10.21468/SciPostPhys.13.6.123},
	url={https://scipost.org/10.21468/SciPostPhys.13.6.123},
}

@article{demidio2024entanglement,
  title = {Entanglement Entropy and Deconfined Criticality: Emergent {SO(5)} Symmetry and Proper Lattice Bipartition},
  author = {D'Emidio, Jonathan and Sandvik, Anders W.},
  journal = {Phys. Rev. Lett.},
  volume = {133},
  issue = {16},
  pages = {166702},
  numpages = {7},
  year = {2024},
  month = {Oct},
  publisher = {American Physical Society},
  doi = {10.1103/PhysRevLett.133.166702},
  url = {https://link.aps.org/doi/10.1103/PhysRevLett.133.166702}
}

@article{zhou2024so5,
  title = {{SO(5)} Deconfined Phase Transition under the Fuzzy-Sphere Microscope: Approximate Conformal Symmetry, Pseudo-Criticality, and Operator Spectrum},
  author = {Zhou, Zheng and Hu, Liangdong and Zhu, W. and He, Yin-Chen},
  journal = {Phys. Rev. X},
  volume = {14},
  issue = {2},
  pages = {021044},
  numpages = {22},
  year = {2024},
  month = {Jun},
  publisher = {American Physical Society},
  doi = {10.1103/PhysRevX.14.021044},
  url = {https://link.aps.org/doi/10.1103/PhysRevX.14.021044}
}

@article{chester2024bootstrapping,
  title = {Bootstrapping Deconfined Quantum Tricriticality},
  author = {Chester, Shai M. and Su, Ning},
  journal = {Phys. Rev. Lett.},
  volume = {132},
  issue = {11},
  pages = {111601},
  numpages = {7},
  year = {2024},
  month = {Mar},
  publisher = {American Physical Society},
  doi = {10.1103/PhysRevLett.132.111601},
  url = {https://link.aps.org/doi/10.1103/PhysRevLett.132.111601}
}

@article{yang2025conformal,
  title = {Conformal Operator Flows of the Deconfined Quantum Criticality from $\mathrm{SO}(5)$ to $\mathrm{O}(4)$},
  author = {Yang, Shuai and Hu, Liang-dong and Han, Chao and Zhu, W. and Chen, Yan},
  journal = {Phys. Rev. Lett.},
  volume = {136},
  issue = {7},
  pages = {076505},
  numpages = {7},
  year = {2026},
  month = {Feb},
  publisher = {American Physical Society},
  doi = {10.1103/l6vw-6z79},
  url = {https://link.aps.org/doi/10.1103/l6vw-6z79}
}

@article{li2022bootstrapping,
  author  = {Zhijin Li},
  title   = {Bootstrapping conformal {QED}$_3$ and deconfined quantum critical point},
  journal = {J. High Energy Phys.},
  volume  = {2022},
  number  = {11},
  pages   = {005},
  year    = {2022},
  doi     = {10.1007/JHEP11(2022)005},
  url     = {https://doi.org/10.1007/JHEP11(2022)005}
}

@Article{song2024extracting,
	title={{Extracting subleading corrections in entanglement entropy at quantum phase transitions}},
	author={Menghan Song and Jiarui Zhao and Zi Yang Meng and Cenke Xu and Meng Cheng},
	journal={SciPost Phys.},
	volume={17},
	pages={010},
	year={2024},
	publisher={SciPost},
	doi={10.21468/SciPostPhys.17.1.010},
	url={https://scipost.org/10.21468/SciPostPhys.17.1.010},
}

@article{tang2011method,
  title = {Method to Characterize Spinons as Emergent Elementary Particles},
  author = {Tang, Ying and Sandvik, Anders W.},
  journal = {Phys. Rev. Lett.},
  volume = {107},
  issue = {15},
  pages = {157201},
  numpages = {5},
  year = {2011},
  month = {Oct},
  publisher = {American Physical Society},
  doi = {10.1103/PhysRevLett.107.157201},
  url = {https://link.aps.org/doi/10.1103/PhysRevLett.107.157201}
}

@article{tang2013confinement,
  title = {Confinement and Deconfinement of Spinons in Two Dimensions},
  author = {Tang, Ying and Sandvik, Anders W.},
  journal = {Phys. Rev. Lett.},
  volume = {110},
  issue = {21},
  pages = {217213},
  numpages = {5},
  year = {2013},
  month = {May},
  publisher = {American Physical Society},
  doi = {10.1103/PhysRevLett.110.217213},
  url = {https://link.aps.org/doi/10.1103/PhysRevLett.110.217213}
}

@article{chen2024phases,
  title = {Phases of $(2+1)\mathrm{D}$ {SO(5)} Nonlinear Sigma Model with a Topological Term on a Sphere: Multicritical Point and Disorder Phase},
  author = {Chen, Bin-Bin and Zhang, Xu and Wang, Yuxuan and Sun, Kai and Meng, Zi Yang},
  journal = {Phys. Rev. Lett.},
  volume = {132},
  issue = {24},
  pages = {246503},
  numpages = {7},
  year = {2024},
  month = {Jun},
  publisher = {American Physical Society},
  doi = {10.1103/PhysRevLett.132.246503},
  url = {https://link.aps.org/doi/10.1103/PhysRevLett.132.246503}
}

@article{francis2021many,
author = {Akhil Francis and Daiwei Zhu and Huerta Alderete, Cinthia and Sonika Johri and Xiao Xiao and James K. Freericks and Christopher Monroe and Norbert M. Linke and Alexander F. Kemper},
title = {Many-body thermodynamics on quantum computers via partition function zeros},
journal = {Sci. Adv.},
volume = {7},
number = {34},
pages = {eabf2447},
year = {2021},
doi = {10.1126/sciadv.abf2447},
URL = {https://www.science.org/doi/abs/10.1126/sciadv.abf2447}}

@misc{alsheikh2026carbm,
      title={{CaRBM}: A fixed-depth quantum algorithm with partial correction for thermal state preparation}, 
      author={Omar Alsheikh and A. F. Kemper and Ermal Rrapaj and Evan J. Rule and Goksu C. Toga},
      year={2026},
      eprint={2603.17971},
      archivePrefix={arXiv},
      primaryClass={quant-ph},
      url={https://arxiv.org/abs/2603.17971}, 
}

@misc{supplemental,
  note = {See Supplemental Material for the symmetry-resolved formulation, reweighting estimators, numerical details, and additional scaling results}
}

@misc{guo2026data,
  author = {Guo, Chunhao and Wang, Zhe and Wang, Danhe and Liu, Zenan and Zou, Haiyuan and Yan, Zheng},
  title = {Data for ``{Lee--Yang} Zeros and Pseudocritical Drift in {J-Q} {N\'eel} Antiferromagnet--Valence-Bond Solid Transitions''},
  year = {2026},
  url = {https://github.com/cguo313/Manuscript_Data_Oct_2026}
}

\end{document}

\title{Supplemental Material: Lee--Yang Zeros and Pseudocritical Drift in J-Q N\'eel Antiferromagnet--Valence-Bond Solid Transitions}
%\title{Supplemental Material: Lee--Yang Zeros and Pseudocritical Drift in J-Q N\'eel--VBS Transitions}

\author{Chunhao Guo}
\affiliation{Department of Physics, School of Science and Research Center for Industries of the Future, Westlake University, Hangzhou 310030, China}
\affiliation{Institute of Natural Sciences, Westlake Institute for Advanced Study, Hangzhou 310024, China}

\author{Zhe Wang}
\affiliation{Department of Physics, School of Science and Research Center for Industries of the Future, Westlake University, Hangzhou 310030, China}
\affiliation{Institute of Natural Sciences, Westlake Institute for Advanced Study, Hangzhou 310024, China}

\author{Danhe Wang}
\affiliation{Key Laboratory of Polar Materials and Devices (MOE),
School of Physics, East China Normal University, Shanghai 200241, China}

\author{Zenan Liu}
\affiliation{Department of Physics, School of Science and Research Center for Industries of the Future, Westlake University, Hangzhou 310030, China}
\affiliation{Institute of Natural Sciences, Westlake Institute for Advanced Study, Hangzhou 310024, China}

\author{Haiyuan Zou}
\email{hyzou@phy.ecnu.edu.cn}
\affiliation{Key Laboratory of Polar Materials and Devices (MOE),
School of Physics, East China Normal University, Shanghai 200241, China}

\author{Zheng Yan}
\email{zhengyan@westlake.edu.cn}
\affiliation{Department of Physics, School of Science and Research Center for Industries of the Future, Westlake University, Hangzhou 310030, China}
\affiliation{Institute of Natural Sciences, Westlake Institute for Advanced Study, Hangzhou 310024, China}

\date{\today}
\maketitle

%============================================================
\section{Lee--Yang zeros and thermodynamic singularities}
\label{sec:LY_intro_pinch_density}
%============================================================

Throughout this Supplemental Material, \(L\) denotes the linear system size, \(d\) the spatial dimension, \(N=L^d\) the total number of sites or unit cells, and \(\beta\) the inverse temperature. A generally complex source \(\lambda_{\bQ}\in\mathbb C\) couples to a symmetry-resolved operator \(O_{\bQ}\) in momentum (translation) sector \(\bQ\). Unless a subscript \(0\) is written explicitly, expectation values \(\langle\cdots\rangle\) are taken in the ensemble of the deformed Hamiltonian currently under discussion; \(\langle\cdots\rangle_0\) denotes the anchor ensemble of \(H_0\). When a low-lying Lee--Yang zero is pinned to the imaginary axis, we parameterize it as
\begin{equation}
\lambda_{\bQ}^{(k)}=\ii g_{\mathrm{LY}}^{(k)},
\qquad
g_{\mathrm{LY}}^{(k)}\in\mathbb R.
\label{eq:imag_zero_parameterization_global}
\end{equation}

%------------------------------------------------------------
\subsection{Finite-size analyticity and zero pinching}
\label{subsec:zeros_analyticity_finite_size}
%------------------------------------------------------------

For any finite system size \(L\) and inverse temperature \(\beta\), the finite-size free energy is analytic for real source \(\lambda_{\bQ}\), so no genuine singularity can occur on the real axis \cite{yang1952statistical,lee1952statistical,fisher1965nature}. A phase transition therefore requires a nonuniform thermodynamic limit: the convergence \(f_{L,\beta}\to f\) fails to be uniform in any neighborhood of the transition point because the limiting free energy becomes nonanalytic there. In the Lee--Yang framework, this nonuniformity arises when complex zeros of the generating function condense into curves that approach and pinch the real axis as \(L\to\infty\) \cite{itzykson1983distribution,bena2005statistical,biskup2004partition}. Spontaneous symmetry breaking provides the canonical example in the source direction. In an ordered phase, the limits \(L\to\infty\) and \(\lambda_{\bQ}\to0\) do not commute, producing a cusp in \(f(\lambda_{\bQ})\) and a discontinuity in the intensive conjugate expectation value
\(m_{\bQ}(\lambda_{\bQ})\equiv-\partial f/\partial \lambda_{\bQ}\) at \(\lambda_{\bQ}=0\). Microscopically, this is enforced by Lee--Yang zeros pinching the origin on the inverse spacetime-volume scale \cite{yang1952statistical,lee1952statistical,privman1983finite,biskup2000general}.

We consider a generally complex source \(\lambda_{\bQ}\in\mathbb C\) coupled to a symmetry-channel operator \(O_{\bQ}\) \cite{yang1952statistical,lee1952statistical,fisher1965nature,bena2005statistical},
\begin{equation}
H(\lambda_{\bQ})=H_0-\lambda_{\bQ}O_{\bQ},
\qquad
Z_{L,\beta}(\lambda_{\bQ})=\Tr\,e^{-\beta H(\lambda_{\bQ})}.
\label{eq:Z_source_def_LYintro}
\end{equation}
We work with the normalized generating function
\begin{equation}
G_{L,\beta}(\lambda_{\bQ})\equiv \frac{Z_{L,\beta}(\lambda_{\bQ})}{Z_{L,\beta}(0)},
\qquad
G_{L,\beta}(0)=1,
\label{eq:G_def_LYintro}
\end{equation}
whose zeros coincide with those of \(Z_{L,\beta}\) \cite{yang1952statistical,lee1952statistical,bena2005statistical}. We also define the free-energy density and normalized free-energy shift,
\begin{equation}
f_{L,\beta}(\lambda_{\bQ})\equiv -\frac{1}{\beta N}\ln Z_{L,\beta}(\lambda_{\bQ}),
\qquad
\phi_{L,\beta}(\lambda_{\bQ})\equiv -\frac{1}{\beta N}\ln G_{L,\beta}(\lambda_{\bQ})
= f_{L,\beta}(\lambda_{\bQ})-f_{L,\beta}(0),
\label{eq:f_phi_def_LYintro}
\end{equation}
with \(N=L^d\) \cite{fisher1965nature,itzykson1983distribution,bena2005statistical}.

For finite \((L,\beta)\), if \(\lambda_{\bQ}\in\mathbb R\) and \(H(\lambda_{\bQ})\) is Hermitian, then \(e^{-\beta H(\lambda_{\bQ})}\) is positive and
\begin{equation}
Z_{L,\beta}(\lambda_{\bQ})=\Tr\,e^{-\beta H(\lambda_{\bQ})}>0
\qquad
(\beta>0,\ \lambda_{\bQ}\in\mathbb R),
\label{eq:Z_positive_real}
\end{equation}
so neither \(Z_{L,\beta}\) nor \(G_{L,\beta}\) has real zeros at finite size \cite{yang1952statistical,lee1952statistical,fisher1965nature,bena2005statistical}. Consequently, \(\phi_{L,\beta}(\lambda_{\bQ})\) is smooth on the real axis. If \(H(\lambda_{\bQ})\) depends analytically on \(\lambda_{\bQ}\) in a neighborhood of the real axis, as in Eq.~\eqref{eq:Z_source_def_LYintro}, then \(Z_{L,\beta}\) and \(G_{L,\beta}\) are analytic there as traces of analytic matrix-valued functions \cite{fisher1965nature,itzykson1983distribution,blythe2003lee}. Thermodynamic nonanalyticities must therefore arise from a nonuniform limit as \(L\to\infty\) \cite{fisher1965nature,bena2005statistical}.

At finite size, \(Z_{L,\beta}(\lambda_{\bQ})\) is an entire function of \(\lambda_{\bQ}\) and has isolated zeros \(\{\lambda_n\}\) with finite multiplicity \cite{itzykson1983distribution,bena2005statistical}. Because \(Z_{L,\beta}\) is real and strictly positive on the real axis, the zero set is invariant under complex conjugation, so zeros occur in pairs \(\lambda_n\) and \(\lambda_n^\ast\) \cite{yang1952statistical,lee1952statistical,itzykson1983distribution}. In some variables, \(Z\) is a polynomial and admits a global factorization in terms of its zeros \cite{yang1952statistical,lee1952statistical}. More generally, one may isolate the zeros in a chosen neighborhood and absorb the remaining nonvanishing part into an analytic prefactor. Concretely, suppose the chosen neighborhood contains \(n_{\mathrm z}\) zeros, denoted \(\lambda_1,\ldots,\lambda_{n_{\mathrm z}}\). Then
\begin{equation}
G_{L,\beta}(\lambda_{\bQ})
=
\mathcal A_{L,\beta}^{\mathrm{loc}}(\lambda_{\bQ})
\prod_{j=1}^{n_{\mathrm z}}(\lambda_{\bQ}-\lambda_j),
\qquad
\mathcal A_{L,\beta}^{\mathrm{loc}}(\lambda_{\bQ})\neq 0
\ \text{and analytic},
\label{eq:local_product_G_LYintro}
\end{equation}
which gives
\begin{equation}
\phi_{L,\beta}(\lambda_{\bQ})
=
-\frac{1}{\beta N}\sum_{j=1}^{n_{\mathrm z}} \ln(\lambda_{\bQ}-\lambda_j)
+\phi_{L,\beta}^{\rm an}(\lambda_{\bQ}),
\label{eq:phi_logzeros_LYintro}
\end{equation}
with analytic background
\begin{equation}
\phi_{L,\beta}^{\rm an}(\lambda_{\bQ})
=
-\frac{1}{\beta N}\ln \mathcal A_{L,\beta}^{\mathrm{loc}}(\lambda_{\bQ}).
\end{equation}
Here \(n_{\mathrm z}\) is simply the number of zeros retained in that local factorization. A global factorization, when needed, is provided by the appropriate Weierstrass/Hadamard canonical product for the entire function \(G_{L,\beta}\) \cite{itzykson1983distribution,bena2005statistical}.
Thus all singularities of \(\ln G_{L,\beta}\) lie at complex zeros and do not affect the real axis at finite size. A thermodynamic singularity can occur only if, as \(L\to\infty\), the zeros condense into a continuum whose support approaches and pinches the real axis at a real impact point \(\lambda_{\bQ}=\lambda_c\) \cite{yang1952statistical,lee1952statistical,fisher1965nature,itzykson1983distribution,bena2005statistical}.

The scaling variable controlling the approach to \(\lambda_c\) depends on the origin of the pinch. In a coexistence regime, finite-size rounding is controlled by the competition between macroscopically distinct sectors whose free-energy difference is extensive in the spacetime volume \(\beta N\) \cite{privman1983finite,binder1987finite,borgs1990rigorous,biskup2004partition}. A small shift \(\delta\lambda_{\bQ}=\lambda_{\bQ}-\lambda_c\) biases their relative weights by an amount of order \(\beta N\,\delta\lambda_{\bQ}\), so the natural scaled coordinate is
\begin{equation}
u\equiv \beta N(\lambda_{\bQ}-\lambda_c).
\label{eq:u_def_g_LYintro}
\end{equation}
In this regime the leading zeros satisfy \(u_n=O(1)\), equivalently
\begin{equation}
\lambda_n-\lambda_c=O\!\left(\frac{1}{\beta N}\right),
\label{eq:coexistence_scale_g_LYintro}
\end{equation}
which is the characteristic inverse spacetime-volume approach to the real axis \cite{borgs1990rigorous,biskup2004partition}.

By contrast, at a continuous critical point the leading zero scales as
\begin{equation}
\lambda_1(L)\sim L^{-y_h},
\label{eq:g1_critical_scaling}
\end{equation}
where \(y_h\) is the RG eigenvalue of the source \(\lambda_{\bQ}\). The corresponding finite-size scaling variable is \(\lambda_{\bQ}L^{y_h}\) \cite{fisher1965nature,itzykson1983distribution,kenna2013finite,deger2019determination}. For ground-state quantum criticality one typically scales imaginary time as \(\beta\sim L^z\) \cite{sachdev2011quantum,sandvik2007evidence,sandvik2010continuous}. In that case the coexistence scale \(\Im(\lambda_1)\sim(\beta N)^{-1}\sim L^{-(d+z)}\) is parametrically distinct from the critical law \(L^{-y_h}\). These two behaviors therefore provide a sharp finite-size discriminator between coexistence and critical pinching \cite{privman1983finite,binder1987finite,kenna2013finite,biskup2004partition}.

%------------------------------------------------------------
\subsection{Zero density, first-order pinching, and symmetry breaking}
\label{subsec:cauchy_transform_LYintro}
%------------------------------------------------------------

Differentiating Eq.~\eqref{eq:phi_logzeros_LYintro} in this local representation gives
\begin{equation}
\frac{\partial \phi_{L,\beta}}{\partial \lambda_{\bQ}}
=
-\frac{1}{\beta N}\sum_{j=1}^{n_{\mathrm z}} \frac{1}{\lambda_{\bQ}-\lambda_j}
+
\frac{\partial \phi^{\rm an}_{L,\beta}}{\partial \lambda_{\bQ}},
\label{eq:dphi_sum_LYintro}
\end{equation}
so the nonanalytic structure of \(\phi_{L,\beta}\) is carried by simple poles at the Lee--Yang zeros \cite{fisher1965nature,itzykson1983distribution,bena2005statistical}. On the other hand, differentiating Eq.~\eqref{eq:f_phi_def_LYintro} and using
\(\partial_{\lambda_{\bQ}}\ln Z_{L,\beta}=\beta\avg{O_{\bQ}}\) yields
\begin{equation}
\frac{\partial \phi_{L,\beta}}{\partial \lambda_{\bQ}}
=
-\frac{1}{N}\avg{O_{\bQ}},
\label{eq:dphi_orderparam_LYintro}
\end{equation}
so any thermodynamic nonanalyticity in \(\partial\phi/\partial\lambda_{\bQ}\) is equivalently a nonanalyticity in the intensive channel expectation value \cite{fisher1965nature,itzykson1983distribution}.

In a coexistence regime it is natural to work in the scaled coordinate \(u=\beta N(\lambda_{\bQ}-\lambda_c)\). Writing \(u_n=\beta N(\lambda_n-\lambda_c)\), we focus on the zeros controlling the local pinch and assume that near \(u=0\) they lie close to an axis-pinned locus,
\begin{equation}
u_n\approx \ii s_n,
\qquad
s_n\in\mathbb R,
\label{eq:un_is_g_LYintro}
\end{equation}
as occurs when the limiting set is invariant under \(u\mapsto -u^\ast\) \cite{yang1952statistical,lee1952statistical,itzykson1983distribution,biskup2000general}. We define the cumulative counting function for the positive branch,
\begin{equation}
\mathcal N_L(s)\equiv \#\{n:\ 0<s_n<s\},
\qquad
s>0,
\label{eq:counting_def_g_LYintro}
\end{equation}
that is, \(\mathcal N_L(s)\) counts how many positive-branch zeros satisfy \(0<s_n<s\). We assume that as \(L\to\infty\) it converges in the distributional sense to a limiting density
\begin{equation}
\rho(s)\equiv \lim_{L\to\infty}\frac{d\mathcal N_L(s)}{ds}.
\label{eq:rho_def_g_LYintro}
\end{equation}
When the pointwise limit exists, we also write
\begin{equation}
\mathcal N(s)\equiv \lim_{L\to\infty}\mathcal N_L(s),
\end{equation}
so that \(\rho(s)=d\mathcal N(s)/ds\) in the regular case. Using \(\lambda_{\bQ}-\lambda_n=(u-u_n)/(\beta N)\) in Eq.~\eqref{eq:dphi_sum_LYintro} and replacing the discrete sum over pinching zeros by the corresponding continuum description gives
\begin{equation}
\left.\frac{\partial \phi}{\partial \lambda_{\bQ}}\right|_{\rm sing}
=
-\int_{-\infty}^{\infty}\!ds\,\frac{\rho(s)}{u-\ii s},
\qquad
u=\beta N(\lambda_{\bQ}-\lambda_c),
\label{eq:dphi_integral_LYintro}
\end{equation}
which is the Cauchy transform of the limiting zero density \cite{fisher1965nature,itzykson1983distribution,janke2001strength,biskup2004partition}.

To expose the nonanalyticity, we compare the two limits of the Cauchy transform as \(u\) approaches the accumulation line from opposite sides. Write
\begin{equation}
u=\Delta+\ii\eps,
\qquad
\Delta=\Re u,
\quad
\eps=\Im u.
\label{eq:u_decomp_LYintro}
\end{equation}
The Sokhotski--Plemelj formula then gives, with \(\mathcal P\) denoting the Cauchy principal value,
\begin{equation}
\lim_{\Delta\to0^\pm}\frac{1}{u-\ii s}
=
\ii\,\mathcal P\frac{1}{s-\eps}\ \pm\ \pi\,\delta(s-\eps),
\label{eq:SP_specialized_eps_LYintro}
\end{equation}
and hence
\begin{equation}
\lim_{\Delta\to0^\pm}
\left.\frac{\partial \phi}{\partial \lambda_{\bQ}}\right|_{\rm sing}
=
-\ii\,\mathcal P\!\int_{-\infty}^{\infty}\!ds\,\frac{\rho(s)}{s-\eps}
\mp
\pi\,\rho(\eps).
\label{eq:dphi_boundary_values_LYintro}
\end{equation}
Therefore the discontinuity across the accumulation line is
\begin{equation}
\left[\left.\frac{\partial \phi}{\partial \lambda_{\bQ}}\right|_{\rm sing}\right]_{\Delta\to0^+}^{\Delta\to0^-}
=
-2\pi\,\rho(\eps),
\label{eq:jump_general_eps_LYintro}
\end{equation}
which is the standard relation between the jump across the cut and the local zero density \cite{fisher1965nature,itzykson1983distribution}.

If the density approaches a finite nonzero limit at the impact point,
\begin{equation}
\rho(s)=\rho(0)+O(s),
\qquad
\rho(0)\neq0,
\label{eq:rho0_nonzero_g_LYintro}
\end{equation}
then Eq.~\eqref{eq:dphi_boundary_values_LYintro} implies a finite jump at \(\eps=0\). Here \(\lambda_c^\pm\) denotes approaching the real impact point \(\lambda_c\) from the two sides transverse to the pinching line, equivalently \(\Delta\to0^\pm\) at fixed \(\eps=0\):
\begin{equation}
\left[\frac{\partial \phi}{\partial \lambda_{\bQ}}\right]_{\lambda_c^+}^{\lambda_c^-}
=
-2\pi\,\rho(0).
\label{eq:jump_rho0_g_LYintro}
\end{equation}
Using Eq.~\eqref{eq:dphi_orderparam_LYintro}, the intensive channel expectation value is discontinuous,
\begin{equation}
\left[\frac{1}{N}\avg{O_{\bQ}}\right]_{\lambda_c^-}^{\lambda_c^+}
=
2\pi\,\rho(0),
\label{eq:jump_orderparam_rho0}
\end{equation}
so a finite \(\rho(0)\) produces a first-order singularity in the \(\lambda_{\bQ}\) direction \cite{fisher1965nature,privman1983finite,binder1987finite,janke2001strength,biskup2004partition}. This is also the Lee--Yang form of spontaneous symmetry breaking. In a broken-symmetry phase,
\begin{equation}
m_\pm
=
\lim_{\lambda_{\bQ}\to0^\pm}\lim_{L\to\infty}\frac{1}{N}\avg{O_{\bQ}},
\end{equation}
and if the channel is odd under \(\lambda_{\bQ}\mapsto-\lambda_{\bQ}\), then \(m_\pm=\pm m_0\) and
\begin{equation}
m_0=\pi\,\rho(0),
\label{eq:m0_rho0_subsec}
\end{equation}
up to the overall normalization of \(u\) \cite{yang1952statistical,lee1952statistical,itzykson1983distribution,biskup2000general}. In this language, spontaneous symmetry breaking means that the thermodynamic limit produces a finite zero density at the impact point in the coexistence-scaled variable.

The same criterion is equivalent to inverse spacetime-volume scaling of the leading zero. If \(\rho(0)\in(0,\infty)\), then \(\mathcal N(s)=\rho(0)s+o(s)\) as \(s\to0^+\), so the first positive zero \(s_1\) satisfies \(s_1=O(1)\). Mapping back gives
\begin{equation}
\Im(\lambda_1-\lambda_c)
=
\frac{s_1}{\beta N}
+o\!\left(\frac{1}{\beta N}\right)
\sim
\frac{1}{\beta N},
\label{eq:volumelaw_from_rho0}
\end{equation}
which is the characteristic inverse spacetime-volume scaling of the leading Lee--Yang zero \cite{privman1983finite,binder1987finite,borgs1990rigorous,biskup2004partition}. Conversely, if the leading zeros satisfy
\begin{equation}
\Im(\lambda_n-\lambda_c)\sim \frac{1}{\beta N}
\qquad\Longleftrightarrow\qquad
s_n=\beta N\,\Im(\lambda_n-\lambda_c)=O(1),
\label{eq:volumelaw_assumption}
\end{equation}
and if \(\mathcal N(s)\) exists for fixed \(s=O(1)\) and is differentiable at \(s=0^+\), then
\begin{equation}
\rho(0)\equiv \left.\frac{d\mathcal N}{ds}\right|_{s=0^+}
=
\lim_{s\to0^+}\frac{\mathcal N(s)}{s}
\label{eq:rho0_from_counting}
\end{equation}
exists and is finite. Under these mild regularity assumptions, volume-law pinching is therefore equivalent to a finite nonzero zero density at the impact point \cite{privman1983finite,binder1987finite,biskup2004partition}.

A compact derivation of the volume law follows from the standard two-sector coexistence approximation \cite{privman1983finite,binder1987finite,borgs1990rigorous,biskup2004partition}. For real \(\lambda_{\bQ}\) near \(\lambda_c\), assume two competing macrostates \(\alpha=\pm\) with analytic free-energy densities \(f_\pm(\lambda_{\bQ})\) that cross at \(\lambda_c\),
\begin{equation}
f_+(\lambda_c)=f_-(\lambda_c)\equiv f_c,
\qquad
f_\pm(\lambda_{\bQ})\ \text{analytic near }\lambda_c,
\label{eq:coexistence_free_energy_crossing}
\end{equation}
and that the partition function is dominated by these contributions,
\begin{equation}
Z_{L,\beta}(\lambda_{\bQ})
\simeq
\mathcal A_+(\lambda_{\bQ})\,e^{-\beta N f_+(\lambda_{\bQ})}
+
\mathcal A_-(\lambda_{\bQ})\,e^{-\beta N f_-(\lambda_{\bQ})},
\label{eq:two_phase_partition_function}
\end{equation}
with nonvanishing, slowly varying prefactors \(\mathcal A_\pm\). Define
\begin{equation}
m_\pm
\equiv
\frac{1}{N}\avg{O_{\bQ}}_{\pm}
=
-\left.\frac{\partial f_\pm}{\partial \lambda_{\bQ}}\right|_{\lambda_c},
\qquad
\Delta m\equiv m_+-m_-.
\label{eq:sector_m_def}
\end{equation}
Expanding about \(\lambda_c\), one finds
\begin{equation}
\lambda_n-\lambda_c
=
\frac{1}{\beta N\,\Delta m}
\Bigl[
\ln|R|+\ii\bigl(\pi(2n+1)-\theta\bigr)
\Bigr]
+
o\!\left(\frac{1}{\beta N}\right),
\qquad
n\in\mathbb Z,
\label{eq:zeros_from_coexistence}
\end{equation}
where
\begin{equation}
R=\frac{\mathcal A_-(\lambda_c)}{\mathcal A_+(\lambda_c)}=|R|e^{\ii\theta}.
\end{equation}
Here the integer \(n\in\mathbb Z\) labels successive zeros along the pinching line. Thus
\begin{equation}
\Im(\lambda_n-\lambda_c)=O\!\left(\frac{1}{\beta N}\right),
\label{eq:volumelaw_from_coexistence}
\end{equation}
independent of microscopic details \cite{privman1983finite,binder1987finite,borgs1990rigorous,biskup2004partition}. In the axis-pinned case \(|R|=1\) (equivalently \(\ln|R|=0\)), the zeros lie asymptotically along the imaginary \(u\) axis,
\begin{equation}
u_n\simeq \ii\,\frac{\pi(2n+1)-\theta}{\Delta m},
\label{eq:un_coexistence_simple}
\end{equation}
and are equally spaced with separation \(2\pi/|\Delta m|\). The corresponding limiting density at the impact point is therefore
\begin{equation}
\rho(0)=\frac{|\Delta m|}{2\pi}
\qquad
\text{for }u=\beta N(\lambda_{\bQ}-\lambda_c).
\label{eq:rho0_from_coexistence}
\end{equation}
This agrees with the jump relation above and makes explicit that a first-order singularity in the \(\lambda_{\bQ}\) direction is equivalent to a finite density of Lee--Yang zeros at the impact point \cite{privman1983finite,binder1987finite,borgs1990rigorous,biskup2004partition}.

If instead the zero density vanishes at the impact point,
\begin{equation}
\rho(s)\sim A_{\rho}|s|^a,
\qquad
A_{\rho}>0,\quad a>0,
\qquad
(s\to0),
\label{eq:rho_power_g_LYintro}
\end{equation}
then \(\rho(0)=0\), and Eq.~\eqref{eq:jump_general_eps_LYintro} immediately implies
\begin{equation}
\left[\frac{\partial \phi}{\partial \lambda_{\bQ}}\right]_{\lambda_c^+}^{\lambda_c^-}=0.
\label{eq:continuous_first_derivative_g_LYintro}
\end{equation}
Hence \(\avg{O_{\bQ}}/N\) is continuous at \(\lambda_c\), and any singularity in the source direction can appear only in higher derivatives, equivalently higher integrated cumulants \cite{fisher1965nature,itzykson1983distribution,janke2001strength}. Determining the detailed scaling of \(\partial\phi/\partial\lambda_{\bQ}\) then requires more information than the local form \(\rho(s)\sim A_{\rho}|s|^a\) alone, because the full Cauchy transform depends on the density away from the impact point as well. In such a case coexistence scaling need not apply; for a conventional RG-controlled continuous critical point one instead expects the leading zero to follow the critical power law \(\lambda_1(L)\sim L^{-y_h}\).

%============================================================
\section{Scaling of leading zeros and Lee--Yang edges}
\label{sec:LY_scaling_and_edge}
%============================================================

%------------------------------------------------------------
\subsection{Critical scaling and coexistence scaling of leading zeros}
\label{subsec:RG_FSS_leading_zeros}
%------------------------------------------------------------

Lee--Yang zeros provide a compact, symmetry-resolved probe of how fluctuations reorganize with system size \cite{itzykson1983distribution,janke2001strength,kenna2013finite,deger2019determination}. When the low-lying zeros are pinned to the imaginary axis, we write the \(k\)th zero as
\begin{equation}
\lambda_{\bQ}^{(k)}(L)=\ii g_{\mathrm{LY}}^{(k)}(L),
\qquad
g_{\mathrm{LY}}^{(k)}(L)\in\mathbb R,
\end{equation}
which is the convention used in the numerical analysis below. Their finite-size flow distinguishes critical pinching from coexistence scaling \cite{itzykson1983distribution,privman1983finite,biskup2000general,biskup2004partition}.

Near a continuous transition, \(g_{\bQ}\) is the RG scaling field conjugate to the channel operator \(O_{\bQ}\) \cite{sachdev2011quantum,cardy1985conformal,itzykson1983distribution}. Under coarse graining by a factor \(b>1\),
\[
g_{\bQ}' = b^{y_h} g_{\bQ},
\]
so the source has length dimension \(-y_h\), equivalently \(g_{\bQ}L^{y_h}\) is the natural dimensionless scaling variable. If \(O_{\bQ}\) has scaling dimension \(\Delta_O\) in effective spacetime dimension \(d+z\), then
\begin{equation}
y_h=d+z-\Delta_O.
\label{eq:yh_def}
\end{equation}
For an order parameter with anomalous dimension \(\eta\),
\begin{equation}
\Delta_O=\frac{d+z-2+\eta}{2},
\qquad
y_h=\frac{d+z+2-\eta}{2}.
\label{eq:yh_eta}
\end{equation}
Throughout this Supplemental Material, whenever leading-zero scaling is re-expressed through Eq.~\eqref{eq:yh_eta} outside a genuine critical regime, \(\eta\) should be understood as a finite-size diagnostic rather than a true anomalous dimension. In particular, the coexistence law \(g_{\mathrm{LY}}(L,\beta)\sim (\beta L^d)^{-1}\) corresponds to the diagnostic limit \(\eta\to -1\).

Define
\begin{equation}
\phi_{L,\beta}(g_{\bQ})\equiv-(\beta N)^{-1}\ln G_{L,\beta}(\ii g_{\bQ}),
\qquad N=L^d .
\label{eq:phi_def_sm}
\end{equation}
At a quantum critical point with dynamical exponent \(z\), RG implies the homogeneity relation \cite{fisher1972scaling,sachdev2011quantum,itzykson1983distribution,cardy1985conformal}
\begin{equation}
\phi_{\rm sing}\!\bigl(g_{\bQ},L,\beta;w_i\bigr)
=
b^{-(d+z)}
\phi_{\rm sing}\!\bigl(g_{\bQ} b^{y_h},\,L/b,\,\beta/b^{z};\,w_i b^{-\omega_i}\bigr),
\label{eq:phi_homogeneity}
\end{equation}
where \(w_i\) are irrelevant scaling fields with correction exponents \(\omega_i>0\). Choosing \(b=L\) and fixing the ground-state scaling window \(\beta/L^z=\mathrm{const}\) gives \cite{sachdev2011quantum,fisher1972scaling}
\begin{equation}
\phi_{\rm sing}(g_{\bQ},L)
=
L^{-(d+z)}
\Phi\!\left(g_{\bQ}L^{y_h},\,w_i L^{-\omega_i}\right).
\label{eq:phi_FSS}
\end{equation}
Thus \(g_{\bQ}L^{y_h}\) is the relevant finite-size scaling variable. Since
\begin{equation}
G_{L,\beta}(\ii g_{\bQ})
=
\exp[-\beta N\,\phi_{L,\beta}(g_{\bQ})],
\end{equation}
and \(\beta N\sim L^{d+z}\) in the ground-state window,
\begin{equation}
G_{L,\beta}(\ii g_{\bQ})
=
\mathcal F\!\left(g_{\bQ}L^{y_h},\,w_i L^{-\omega_i}\right),
\label{eq:G_scaling_form}
\end{equation}
for some scaling function \(\mathcal F\). The \(k\)th axis-pinned Lee--Yang zero is therefore determined by a root of \(\mathcal F\), implying
\begin{equation}
g_{\mathrm{LY}}^{(k)}(L)\sim A_k\,L^{-y_h}
\qquad (L\to\infty),
\label{eq:LY_scaling_general}
\end{equation}
up to nonuniversal metric factors and scaling corrections
\cite{cardy1985conformal,itzykson1983distribution,janke2001strength,kenna2013finite}, where \(A_k\) is a nonuniversal amplitude. Systematic deviations from a pure power law over accessible sizes then provide a sensitive diagnostic of pseudocritical drift \cite{deger2019determination,wada2025locating}.

In a coexistence regime, the source biases the relative weights of competing macrostates through an extensive order parameter, so the natural finite-size scale is the Euclidean spacetime volume \cite{privman1983finite,binder1987finite,borgs1990rigorous,biskup2004partition}
\begin{equation}
V_\tau \equiv \beta L^d .
\end{equation}
Equivalently, in the configuration representation the phase factor involves the time-integrated observable
\begin{equation}
X[\mathcal C]=\int_0^\beta d\tau\,O_{\bQ}[\mathcal C,\tau]\sim \beta L^d,
\end{equation}
so zeros occur when the dominant macroscopic contributions acquire an \(\mathcal O(1)\) relative phase. This gives the standard spacetime-volume law \cite{privman1983finite,binder1987finite,borgs1990rigorous,biskup2004partition}
\begin{equation}
g_{\mathrm{LY}}^{(k)}(L,\beta)\sim \frac{B_k}{\beta L^d},
\label{eq:LY_ordered_scaling}
\end{equation}
where \(B_k\) is a nonuniversal amplitude for the \(k\)th zero in the coexistence regime. If one further sets \(\beta\propto L^z\), Eq.~\eqref{eq:LY_ordered_scaling} reduces to \(g_{\mathrm{LY}}^{(k)}\sim L^{-(d+z)}\), but the more general statement is the \(1/(\beta L^d)\) scaling set by \(V_\tau\).

Leading-zero scaling therefore separates two parametrically distinct regimes \cite{itzykson1983distribution,privman1983finite,biskup2000general,biskup2004partition}. At a continuous transition,
\begin{equation}
g_{\mathrm{LY}}^{(k)}(L)\sim A_k\,L^{-y_h},
\end{equation}
up to nonuniversal metrics and irrelevant-operator corrections. In a coexistence regime, the natural scaling variable is \(g_{\bQ}\beta L^d\), and the leading zeros obey
\begin{equation}
g_{\mathrm{LY}}^{(k)}(L,\beta)\sim \frac{B_k}{\beta L^d}.
\end{equation}
This distinction is the basic finite-size criterion we use to separate critical scaling from coexistence scaling in a given symmetry channel \cite{janke2001strength,kenna2013finite,deger2019determination}.

\subsection{Lee--Yang edges in gapped symmetric regimes}
\label{sec:SI_LY_edge_disordered}

For a local operator \(O(\br)\), the connected equal-time correlator is
\begin{equation}
C(\br)\equiv \langle O(\br)O(\mathbf 0)\rangle - \langle O \rangle^2.
\end{equation}
Writing \(r\equiv |\br|\), a convenient scaling form in a gapped symmetric phase near a continuous transition is
\begin{equation}
C(r)=r^{-2\Delta_O}\,F(r/\xi),
\label{eq:edge_corr_scaling}
\end{equation}
where \(\Delta_O\) is the scaling dimension of \(O\) at the nearby critical theory, \(\xi\) is the correlation length, and \(F(x)\) decays exponentially for \(x\gg 1\) \cite{sachdev2011quantum,chaikin1995principles}. Thus the large-distance correlator is exponentially small, possibly multiplied by algebraic prefactors. As one approaches a continuous quantum critical point by tuning a real control parameter \(r_{\mathrm{ctl}}\to r_{\mathrm{ctl},c}\),
\begin{equation}
\xi \sim |\delta r_{\mathrm{ctl}}|^{-\nu},
\qquad
\delta r_{\mathrm{ctl}} \equiv r_{\mathrm{ctl}}-r_{\mathrm{ctl},c},
\end{equation}
while the characteristic gap scales as
\begin{equation}
\Delta_{\rm gap}\sim \xi^{-z}.
\end{equation}
For Lorentz-invariant (\(z=1\)) systems this reduces to \(\xi\sim v/\Delta_{\rm gap}\) up to a nonuniversal velocity \(v\) \cite{sachdev2011quantum}. The correlation time likewise obeys
\begin{equation}
\xi_\tau \sim \xi^z.
\end{equation}

A source \(\lambda_{\bQ}\) coupled to a Hermitian operator \(O_{\bQ}\) deforms the Hamiltonian as
\begin{equation}
H(\lambda_{\bQ}) = H_0 - \lambda_{\bQ} O_{\bQ},
\quad
Z(\lambda_{\bQ}) = \Tr e^{-\beta H(\lambda_{\bQ})}.
\end{equation}
By Hermiticity,
\begin{equation}
Z(\lambda_{\bQ}^*) = Z(\lambda_{\bQ})^*,
\end{equation}
so zeros occur in complex-conjugate pairs \cite{itzykson1983distribution,bena2005statistical}. In a gapped symmetric regime, the thermodynamic free energy is analytic for real \(\lambda_{\bQ}\) in a finite interval around the origin, so no zeros accumulate on the real axis there. Instead, the zeros approach a limiting locus in the complex plane that remains a finite distance from \(\lambda_{\bQ}=0\). The point of minimal distance from the origin defines the Lee--Yang edge, and that distance is the Lee--Yang gap \cite{fisher1978yang,bena2005statistical}. When the limiting locus is axis-pinned, the edge may be parameterized as
\begin{equation}
\lambda_{\bQ,\mathrm{edge}} = \ii g_{\mathrm{edge}},
\end{equation}
with \(g_{\mathrm{edge}}\neq 0\), so the leading zeros approach \(\pm \ii g_{\mathrm{edge}}\) in the thermodynamic limit at fixed \(\delta r_{\mathrm{ctl}}\neq 0\) \cite{fisher1978yang,bena2005statistical,rennecke2022universal}.

Near a continuous quantum critical point, the singular part of the free-energy density obeys \cite{fisher1972scaling,sachdev2011quantum,itzykson1983distribution,kenna1994scaling}
\begin{equation}
f_{\rm sing}(\delta r_{\mathrm{ctl}}, \lambda_{\bQ}; L, \beta)
=
L^{-(d+z)}
\Phi\!\left(\delta r_{\mathrm{ctl}} L^{1/\nu}, \lambda_{\bQ} L^{y_h}, \frac{\beta}{L^z}\right),
\end{equation}
where \(y_h\) is the RG eigenvalue of the source \(\lambda_{\bQ}\) \cite{itzykson1983distribution,kenna1994scaling}. Since \(Z=\exp[-\beta L^d f]\), the leading zeros have the finite-size form
\begin{equation}
\lambda_{\bQ}^{(1)}(L,\delta r_{\mathrm{ctl}})
\sim
L^{-y_h}\,
\mathcal F\!\left(x,\frac{\beta}{L^z}\right),
\qquad
x\equiv \delta r_{\mathrm{ctl}} L^{1/\nu},
\end{equation}
which is the quantum analogue of standard Lee--Yang scaling \cite{itzykson1983distribution,kenna1994scaling,janke2002density}. The crossover is controlled by \(L/\xi\), equivalently by \(x\sim (L/\xi)^{1/\nu}\). For \(L\ll \xi\), the system appears quasi-critical and \(\lambda_{\bQ}^{(1)}\sim L^{-y_h}\). For \(L\gg \xi\), the leading zeros saturate to an \(L\)-independent edge scale \cite{kenna1994scaling,janke2002density}. Requiring a finite thermodynamic limit,
\begin{equation}
\lambda_{\bQ,\mathrm{edge}}(\delta r_{\mathrm{ctl}})
=
\lim_{L\to\infty}\lambda_{\bQ}^{(1)}(L,\delta r_{\mathrm{ctl}})\neq 0,
\end{equation}
fixes the large-\(x\) asymptotic behavior \(\mathcal F(x,\infty)\sim x^{\nu y_h}\mathcal C_{\mathrm{edge}}\), with \(\mathcal C_{\mathrm{edge}}\) a nonuniversal edge amplitude, and gives
\begin{equation}
\lambda_{\bQ,\mathrm{edge}}(\delta r_{\mathrm{ctl}})
\sim
|\delta r_{\mathrm{ctl}}|^{\nu y_h}\,\mathcal C_{\mathrm{edge}}
\sim
\xi^{-y_h}.
\end{equation}

If a probe source couples to a nonordering channel, analyticity near real \(\lambda_{\bQ}=0\) implies a finite Lee--Yang gap and hence a nonzero edge \cite{fisher1978yang,bena2005statistical,rennecke2022universal}. For the source conjugate to the ordering field itself, the symmetric/disordered phase still generally has a finite Yang--Lee edge at nonzero imaginary field; the edge closes only as criticality is approached. In the ordered or coexistence regime, by contrast, zeros can pinch the real axis at the origin and produce the source-direction nonanalyticity \cite{yang1952statistical,lee1952statistical,fisher1965nature,itzykson1983distribution,blythe2003lee,bena2005statistical}. In the J-Q setting, this means that VBS-source zeros are expected to remain gapped throughout the N\'eel phase, while N\'eel-source zeros are gapped throughout the VBS phase; for the respective ordering field, the edge closes when the transition is approached from the opposite symmetric side. Separately, spin-\(\tfrac12\) square-lattice J-Q models cannot realize a trivial, fully symmetric, short-range-entangled gapped phase with one spin-\(\tfrac12\) per unit cell; symmetry breaking or topological order must intervene \cite{lieb1961two,oshikawa2000commensurability,hastings2004lieb}. In the specific square-lattice J-Q models studied numerically, the nonmagnetic side is observed to be a VBS phase \cite{sandvik2007evidence,sandvik2010continuous,shao2016quantum}, consistent with those constraints. We therefore focus on leading-zero scaling near the N\'eel--VBS transition, where drift and crossover most directly constrain the character of the transition, and leave a systematic study of deep-phase edges for future work.

\section{Symmetry channels and Lee--Yang generators}
%============================================================

\subsection{Translation-resolved operator channels}
\label{subsec:order_parameter_symmetry_channels_clean}

Our Lee--Yang probes are specified by the microscopic support of the operator, site, bond, or plaquette, together with a translation sector of a symmetry-preserving anchor Hamiltonian \(H_0\) \cite{tinkham1964group,landau1980statistical,chaikin1995principles,sachdev2011quantum}. Since \(H_0\) preserves the microscopic symmetries at any finite \(L\), phases and criticality are inferred from the finite-size behavior of symmetry-resolved fluctuations \cite{landau1980statistical,chaikin1995principles,sachdev2011quantum}.

Let \(a\) denote a microscopic object carrying a local operator, with position \(\bR_a\): a site \(a=\br\), an oriented bond \(a=(\br,\mu)\) with \(\mu\in\{x,y\}\), or a plaquette \(a=\square\). A translation sector is isolated by
\begin{equation}
O_{\bQ}\equiv \sum_a e^{\ii \bQ\cdot \bR_a}\,O(a).
\label{eq:OQ_def_channel_clean}
\end{equation}
Under translation by lattice vector \(\ba\),
\begin{equation}
T_{\ba} O_{\bQ} T_{\ba}^{-1}=e^{\ii\bQ\cdot\ba}O_{\bQ},
\end{equation}
so \(O_{\bQ}\) is a translation eigen-operator and does not mix with inequivalent \(\bQ\) sectors in correlators \cite{tinkham1964group,landau1980statistical,sachdev2011quantum}. Further point-group resolution may be imposed through the choice of the local density \(O(a)\) \cite{tinkham1964group,sachdev2011quantum}. The corresponding equal-time structure factor is
\begin{equation}
S_O(\bQ)\equiv \frac{1}{N}\,\avg{O_{\bQ}O_{-\bQ}}.
\label{eq:structure_factor_channel_clean}
\end{equation}

The representative channels used in this work are standard \cite{sandvik2007evidence,sandvik2010continuous,zhao2019symmetry,sachdev2011quantum}. For N\'eel order on the square lattice, \(\bQ=(\pi,\pi)\), and a convenient site-supported channel is
\begin{equation}
M^z=\sum_{\br}(-1)^{x+y}S^z_{\br}.
\end{equation}
For columnar VBS order we first define the nearest-neighbor bond energies
\begin{equation}
B_x(\br)\equiv \bS_{\br}\!\cdot\!\bS_{\br+\hat x},
\qquad
B_y(\br)\equiv \bS_{\br}\!\cdot\!\bS_{\br+\hat y},
\end{equation}
where \(\hat x\) and \(\hat y\) are the unit lattice vectors of the square lattice, and the bond-supported channel operators are
\begin{equation}
D_x=\sum_{\br}(-1)^x\,B_x(\br),
\qquad
D_y=\sum_{\br}(-1)^y\,B_y(\br),
\label{eq:Dxy_clean}
\end{equation}
which lie in the \((\pi,0)\) and \((0,\pi)\) sectors \cite{sandvik2007evidence,sandvik2010continuous,shao2016quantum,nahum2015emergent}. Plaquette-supported probes are defined analogously from a local plaquette density \(\mathcal O(\square)\). Writing \(\bR_\square=(x(\square),y(\square))\) for a fixed reference coordinate of plaquette \(\square\) (for example its lower-left corner), we define
\begin{equation}
O^{\rm plaq}_{(\pi,0)}=\sum_{\square}(-1)^{x(\square)}\mathcal O(\square),
\qquad
O^{\rm plaq}_{(0,\pi)}=\sum_{\square}(-1)^{y(\square)}\mathcal O(\square).
\label{eq:plaq_clean}
\end{equation}
These operators furnish alternative microscopic realizations of the same columnar VBS symmetry sectors \cite{zhao2019symmetry,sun2021emergent}.

Representative symmetry-resolved channels are summarized in Table~\ref{tab:clean_channels}.

\begin{table}[t]
\caption{Representative symmetry-resolved channels.}
\label{tab:clean_channels}
\begin{ruledtabular}
\begin{tabular}{lll}
Support & \(\bQ\) sector & Operator \\
\hline
Site & \((\pi,\pi)\) &
\(M^z=\sum_{\br}(-1)^{x+y}S^z_{\br}\) \\
Bond & \((\pi,0)\), \((0,\pi)\) &
\(D_x=\sum_{\br}(-1)^x B_x(\br)\),\;
\(D_y=\sum_{\br}(-1)^y B_y(\br)\) \\
Plaquette & \((\pi,0)\), \((0,\pi)\) &
\(O^{\rm plaq}_{(\pi,0)}\),\;
\(O^{\rm plaq}_{(0,\pi)}\) \\
\end{tabular}
\end{ruledtabular}
\end{table}

\subsection{Checkerboard-resolved plaquette channels}
\label{subsec:order_parameter_symmetry_channels_pss}

For the checkerboard J-Q (CBJQ) model it is convenient to resolve plaquettes into the two checkerboard families, since the multi-spin interaction acts only on one family \cite{zhao2019symmetry,sun2021emergent}. When both families are present, this basis is related by an exact change of basis to the usual clean columnar plaquette sectors \cite{zhao2019symmetry}. Here PSS stands for plaquette singlet solid.

For a plaquette at reference coordinate \(\bR_\square=(x(\square),y(\square))\), define
\begin{equation}
\epsilon_{\rm cb}(\square)\equiv (-1)^{x(\square)+y(\square)},
\qquad
\epsilon_{\rm row}(\square)\equiv (-1)^{y(\square)}.
\label{eq:eps_cb_row_def}
\end{equation}
Here \(\epsilon_{\rm cb}(\square)=\pm1\) labels the two checkerboard families, while \(\epsilon_{\rm row}(\square)=\pm1\) labels the row parity. They satisfy
\begin{equation}
(-1)^{x(\square)}=\epsilon_{\rm cb}(\square)\,\epsilon_{\rm row}(\square).
\label{eq:eps_relation}
\end{equation}
With \(\mathcal O(\square)\) a local plaquette density, define the projected operators
\begin{equation}
O^{\rm PSS}_{s}\equiv \sum_{\square}\frac{1+s\,\epsilon_{\rm cb}(\square)}{2}\,
\epsilon_{\rm row}(\square)\,\mathcal O(\square),
\qquad s=\pm 1.
\label{eq:OPSS_s_def_SM}
\end{equation}
The label \(s=+1\) selects the \(\epsilon_{\rm cb}=+1\) checkerboard family and \(s=-1\) selects the \(\epsilon_{\rm cb}=-1\) family.
The corresponding two-source deformation is
\begin{equation}
H(\eta_{+},\eta_{-})=H_0-\eta_{+}\,O^{\rm PSS}_{+}-\eta_{-}\,O^{\rm PSS}_{-}.
\label{eq:PSS_sources_SM}
\end{equation}

When both checkerboard families are present, the family basis is related exactly to the clean columnar basis,
\begin{equation}
O^{\rm PSS}_{+}+O^{\rm PSS}_{-}=O^{\rm plaq}_{(0,\pi)},
\qquad
O^{\rm PSS}_{+}-O^{\rm PSS}_{-}=O^{\rm plaq}_{(\pi,0)},
\label{eq:PSS_to_clean_SM}
\end{equation}
so one may equivalently write
\begin{equation}
H(\eta_{+},\eta_{-})=
H_0-\eta_{(0,\pi)}\,O^{\rm plaq}_{(0,\pi)}-\eta_{(\pi,0)}\,O^{\rm plaq}_{(\pi,0)},
\label{eq:eta_clean_reparam_SM}
\end{equation}
with \(\eta_{(0,\pi)}=(\eta_{+}+\eta_{-})/2\) and \(\eta_{(\pi,0)}=(\eta_{+}-\eta_{-})/2\) \cite{zhao2019symmetry}.

In CBJQ only one checkerboard family carries the \(Q\) interaction \cite{zhao2019symmetry}. The active-family operator, for example \(O^{\rm PSS}_{+}\), is therefore the natural PSS probe, and in practice we set the inactive source to zero \cite{zhao2019symmetry,sun2021emergent}.

\begin{table}[t]
\caption{Plaquette \(\mathbb Z_2\) labels in the PSS decomposition.}
\label{tab:pss_labels}
\begin{ruledtabular}
\begin{tabular}{ll}
Label & Meaning \\
\hline
\(\epsilon_{\rm cb}(\square)=(-1)^{x(\square)+y(\square)}\) & checkerboard family \\
\(\epsilon_{\rm row}(\square)=(-1)^{y(\square)}\) & row parity \\
\end{tabular}
\end{ruledtabular}
\end{table}

\subsection{Conjugate sources and normalized generating functions}
\label{subsec:conjugate_sources_general}

A source \(\lambda_{\bQ}\) conjugate to \(O_{\bQ}\) deforms the anchor Hamiltonian as
\begin{equation}
H(\lambda_{\bQ}) = H_0 - \lambda_{\bQ}\,O_{\bQ}.
\label{eq:source_coupling}
\end{equation}
Derivatives of \(\ln Z(\lambda_{\bQ})\) generate the corresponding responses \cite{chaikin1995principles,sachdev2011quantum},
\begin{equation}
\frac{\partial \ln Z}{\partial \lambda_{\bQ}}=\beta\,\langle O_{\bQ}\rangle,
\qquad
\frac{\partial^2 \ln Z}{\partial \lambda_{\bQ}^2}
=
\beta\int_{0}^{\beta}\! d\tau\,
\langle \delta O_{\bQ}(\tau)\,\delta O_{\bQ}(0)\rangle,
\label{eq:susceptibility_relation}
\end{equation}
where in this paragraph \(O_{\bQ}(\tau)=e^{\tau H(\lambda_{\bQ})}O_{\bQ}e^{-\tau H(\lambda_{\bQ})}\) is the imaginary-time Heisenberg operator of the deformed Hamiltonian and \(\delta O_{\bQ}(\tau)\equiv O_{\bQ}(\tau)-\langle O_{\bQ}\rangle\). If \(O_{\bQ}\) commutes with the Hamiltonian, this further reduces to
\begin{equation}
\frac{\partial^2 \ln Z}{\partial \lambda_{\bQ}^2}
=
\beta^2\Bigl(\langle O_{\bQ}^2\rangle-\langle O_{\bQ}\rangle^2\Bigr).
\label{eq:susceptibility_variance}
\end{equation}
For the staggered N\'eel and VBS probes used here one should keep the general integrated-correlation formula, and in the SSE reweighting language the natural objects are the cumulants of the time-integrated source insertion introduced below.

For finite \((L,\beta)\),
\begin{equation}
Z(\lambda_{\bQ}) = \Tr\, e^{-\beta\left(H_0-\lambda_{\bQ}O_{\bQ}\right)}
\label{eq:Z_def}
\end{equation}
is analytic in \(\lambda_{\bQ}\in\mathbb C\) \cite{itzykson1983distribution,bena2005statistical}. We define the normalized generating function
\begin{equation}
G(\lambda_{\bQ}) \equiv \frac{Z(\lambda_{\bQ})}{Z(0)},
\qquad
G(0)=1,
\label{eq:G_def}
\end{equation}
whose zeros are the Lee--Yang zeros \cite{yang1952statistical,lee1952statistical,itzykson1983distribution,bena2005statistical}. The same object may be written exactly as
\begin{equation}
G(\lambda_{\bQ})
=
\Bigl\langle T_\tau\exp\!\Bigl[\lambda_{\bQ}\!\int_{0}^{\beta}\! d\tau\,O_{\bQ}(\tau)\Bigr]\Bigr\rangle_{0},
\label{eq:G_anchor}
\end{equation}
where \(T_\tau\) denotes imaginary-time ordering, \(O_{\bQ}(\tau)=e^{\tau H_0}O_{\bQ}e^{-\tau H_0}\) is now evolved with the anchor Hamiltonian \(H_0\), and \(\langle\cdots\rangle_0\) denotes the corresponding anchor ensemble average \cite{bena2005statistical,blythe2003lee}. In the SSE implementation with a source that is diagonal in the sampling basis, analytic continuation changes only the corresponding diagonal matrix elements, so each stored configuration \(\mathcal C\) acquires a ratio factor \(R(\lambda_{\bQ};\mathcal C)\), and
\begin{equation}
G(\lambda_{\bQ})=\avg{R(\lambda_{\bQ};\mathcal C)}_{0}.
\label{eq:G_ratio_estimator}
\end{equation}
This is the standard reweighting representation used in recent Lee--Yang and related QMC calculations \cite{demidio2023lee,ding2024reweight,wang2025bipartite,ding2025tracking,ding2025evaluating,wang2026addressing}.

Expanding \(\ln G\) about \(\lambda_{\bQ}=0\) generates the integrated connected correlation functions,
\begin{equation}
\ln G(\lambda_{\bQ})
=
\sum_{n=1}^{\infty}\frac{\lambda_{\bQ}^{\,n}}{n!}
\int_0^\beta d\tau_1\cdots \int_0^\beta d\tau_n\,
\langle T_\tau O_{\bQ}(\tau_1)\cdots O_{\bQ}(\tau_n)\rangle_{0,c},
\label{eq:cumulant_expansion}
\end{equation}
where the subscript \(c\) denotes the connected part. In the diagonal-source SSE/path-integral representation, these coefficients are equivalently the cumulants of a c-number configuration functional,
\begin{equation}
X[\mathcal C] \equiv \int_{0}^{\beta}\! d\tau\, O_{\bQ}[\mathcal C,\tau],
\label{eq:X_def}
\end{equation}
where \(O_{\bQ}[\mathcal C,\tau]\) is the configuration-space value of the source insertion at imaginary time \(\tau\).

For a purely imaginary source,
\begin{equation}
\lambda_{\bQ}=\ii g_{\bQ},
\qquad
g_{\bQ}\in\mathbb R,
\label{eq:imag_source_choice}
\end{equation}
the configuration-space representation becomes the characteristic function
\begin{equation}
G(\ii g_{\bQ})=\langle e^{\ii g_{\bQ} X[\mathcal C]}\rangle_{0}.
\label{eq:charfunc_def}
\end{equation}
It is convenient to introduce the time-averaged configuration observable
\begin{equation}
m_{\rm cfg}[\mathcal C]\equiv \frac{X[\mathcal C]}{\beta}.
\label{eq:m_cfg_def}
\end{equation}
The corresponding anchor distribution is
\begin{equation}
P_0(m)\equiv
\Bigl\langle \delta\!\bigl(m-m_{\rm cfg}[\mathcal C]\bigr)\Bigr\rangle_{0},
\label{eq:P0_def}
\end{equation}
which gives the Fourier representation
\begin{equation}
G(\ii g_{\bQ})=\int dm\, P_0(m)\, e^{\ii g_{\bQ}\beta m}.
\label{eq:charfunc_fourier}
\end{equation}
These formulas should be understood in the diagonal-source configuration representation; Eq.~\eqref{eq:G_anchor} remains the exact operator statement.

For the rest of this subsection, we write
\(
\lambda \equiv \lambda_{\bQ}
\),
\(
V \equiv L^d
\),
and
\(
m_{\bQ} \equiv O_{\bQ}/V
\),
where \(V\) is the spatial volume and \(m_{\bQ}\) is the order-parameter density conjugate to the source \(\lambda\). We also use
\(
G_{L,\beta}(\lambda)\equiv Z_{L,\beta}(\lambda)/Z_{L,\beta}(0)
\).

A basic advantage of the Lee--Yang formulation is that source derivatives of \(\ln Z\) are controlled directly by the nearby zeros \cite{itzykson1983distribution,blythe2003lee,bena2005statistical}. If \(\lambda_1,\dots,\lambda_{n_{\mathrm z}}\) denote the zeros of \(G_{L,\beta}\) in a chosen neighborhood of the real axis, then one may write
\begin{equation}
G_{L,\beta}(\lambda)
=
A^{\rm loc}_{L,\beta}(\lambda)
\prod_{j=1}^{n_{\mathrm z}}
\left(1-\frac{\lambda}{\lambda_j}\right),
\qquad
A^{\rm loc}_{L,\beta}(\lambda)\neq 0
\ \text{and analytic}.
\label{eq:G_local_factorization_clean}
\end{equation}
Taking derivatives gives
\begin{equation}
\frac{\partial^n \ln Z_{L,\beta}}{\partial \lambda^n}
=
\frac{\partial^n \ln A^{\rm loc}_{L,\beta}}{\partial \lambda^n}
-
(n-1)!\sum_{j=1}^{n_{\mathrm z}}
\frac{1}{(\lambda_j-\lambda)^n},
\qquad n\ge 1.
\label{eq:nth_derivative_zero_sum_clean}
\end{equation}
Equation~\eqref{eq:nth_derivative_zero_sum_clean} makes the logic explicit: derivatives of \(\ln Z\) contain a smooth analytic-background part and a singular part set by inverse powers of the distances to the nearby zeros. In this precise sense, real-axis response functions are most sensitive when a Lee--Yang zero lies nearby.

It is also now clear why the second derivative, rather than the first derivative, is the natural finite-size example at the symmetric point. In the symmetry-resolved channels used here, the source couples to an operator that is odd under a symmetry of the anchor ensemble, so \(G_{L,\beta}(\lambda)=G_{L,\beta}(-\lambda)\). Therefore,
\begin{equation}
\frac{\partial \ln Z_{L,\beta}}{\partial \lambda}\Big|_{\lambda=0}=0
\label{eq:first_derivative_zero_by_symmetry}
\end{equation}
for every finite system. The first derivative becomes discontinuous only after taking the thermodynamic limit in the coexistence regime; at finite \(L\) that discontinuity is rounded, and at the symmetry point it vanishes identically. The first nontrivial source derivative at \(\lambda=0\) is thus the second derivative, i.e. the integrated susceptibility. If the leading zeros are axis-pinned,
\(
\lambda_{1,2}=\pm \ii g_{\mathrm{LY}}^{(1)}
\),
then Eq.~\eqref{eq:nth_derivative_zero_sum_clean} yields
\begin{equation}
\frac{\partial^2 \ln Z_{L,\beta}}{\partial \lambda^2}\Big|_{\lambda=0}
=
\frac{2}{\bigl[g_{\mathrm{LY}}^{(1)}\bigr]^2}
+
O\!\left(\frac{1}{\bigl[g_{\mathrm{LY}}^{(2)}\bigr]^2}\right)
+
\frac{\partial^2 \ln A^{\rm loc}_{L,\beta}}{\partial \lambda^2}\Big|_{\lambda=0},
\label{eq:second_derivative_nearest_zero_clean}
\end{equation}
where \(g_{\mathrm{LY}}^{(2)}\) denotes the next low-lying imaginary-axis zero. Thus the leading zero sets the singular scale of the response.

The situation is different for equal-time observables such as the squared order-parameter density \(m_{\bQ}^2\) or a fixed-distance spin correlator. These quantities are perfectly well defined at a first-order transition, but at coexistence they are not observables of a single scale-invariant state. Rather, for periodic boundary conditions, the finite-volume partition function is asymptotically a sum of the competing phase contributions \cite{privman1983finite,binder1987finite,borgs1990rigorous,biskup2000general,biskup2004partition},
\begin{equation}
Z_{L,\beta}(\lambda)
\approx
q_a\,e^{-\beta V f_a(\lambda)}
+
q_b\,e^{-\beta V f_b(\lambda)},
\label{eq:two_phase_partition_clean}
\end{equation}
up to corrections that are exponentially small in \(L\). Here \(f_a\) and \(f_b\) are the metastable free-energy densities of the two competing phases, and \(q_a\), \(q_b\) are their degeneracy factors. Consequently, for any bounded equal-time observable \(X\),
\begin{equation}
\langle X\rangle_{L,\beta}
\approx
p_a\,X_a
+
p_b\,X_b,
\qquad
p_\alpha=
\frac{q_\alpha e^{-\beta V f_\alpha}}
{q_a e^{-\beta V f_a}+q_b e^{-\beta V f_b}},
\quad
\alpha\in\{a,b\},
\label{eq:observable_mixture_clean}
\end{equation}
where \(X_\alpha\) denotes the expectation value of \(X\) in phase \(\alpha\). In particular, \(m_{\bQ}^2\) and fixed-distance spin correlators are phase-weighted mixtures of their pure-phase values. They are therefore well defined, but at coexistence they do not approach a unique critical power law. Their finite-size behavior is controlled by the phase weights and by the physics within each phase, not by a single critical fixed point.

Sharp first-order finite-size laws arise instead for source derivatives of \(\ln Z\), because differentiating Eq.~\eqref{eq:two_phase_partition_clean} acts directly on the phase weights. Each derivative brings down a factor proportional to \(\beta V\). For example, the integrated susceptibility per unit volume,
\begin{equation}
\chi^{\rm int}_{L,\beta}
\equiv
\frac{1}{\beta V}
\frac{\partial^2 \ln Z_{L,\beta}}{\partial \lambda^2}\Big|_{\lambda=0},
\label{eq:chi_int_clean}
\end{equation}
has a phase-mixing contribution proportional to \(\beta V\), so its peak obeys the standard first-order scaling law
\begin{equation}
\chi^{\rm int}_{L,\beta,\max}\sim \beta V.
\label{eq:chi_peak_volume_clean}
\end{equation}
Equivalently, in the Lee--Yang language the leading zero approaches the real axis as
\begin{equation}
g_{\mathrm{LY}}^{(1)}(L,\beta)\sim \frac{1}{\beta V},
\label{eq:gly_inverse_volume_clean}
\end{equation}
and Eq.~\eqref{eq:second_derivative_nearest_zero_clean} then reproduces the same spacetime-volume scaling for the singular part of the susceptibility.

This is the sense in which the leading Lee--Yang zeros reveal weakly first-order or pseudocritical drift more cleanly than real-axis observables. At coexistence, \(m_{\bQ}^2\) and fixed-distance correlators are rounded mixtures of competing phases and need not follow any unique critical power law, whereas the low-lying zeros continue to track the singularity scale directly through their distance from the real axis \cite{gorbenko2018walking,ma2020theory,janke2001strength,kenna2013finite,deger2019determination,wada2025locating}.

\subsection{Symmetry constraints and imaginary-axis scans}
\label{subsec:sym_constraints_imag_axis}

For Hermitian \(H_0\) and \(O_{\bQ}\),
\begin{equation}
Z(\lambda_{\bQ}^*) = Z(\lambda_{\bQ})^*,
\label{eq:Z_reality_property}
\end{equation}
so \(G(\lambda_{\bQ}^*)=G(\lambda_{\bQ})^*\) and Lee--Yang zeros occur in complex-conjugate pairs
\(\{\lambda_{\bQ}^{(k)},(\lambda_{\bQ}^{(k)})^*\}\) \cite{blythe2003lee,bena2005statistical}. An additional discrete symmetry may flip the probed channel,
\begin{equation}
U H_0 U^{-1}=H_0,
\qquad
U O_{\bQ} U^{-1}=-O_{\bQ}.
\label{eq:U_odd_channel}
\end{equation}
The physical content of Eq.~\eqref{eq:U_odd_channel} is that the chosen order-parameter channel carries a \(\mathbb Z_2\)-odd character under an exact symmetry of the symmetric anchor Hamiltonian: the finite-size anchor ensemble is invariant under \(U\), while the channel operator changes sign. Any such \(U\) is sufficient for the arguments below, since it implies that the source deformation is mapped as \(H(\lambda_{\bQ})\mapsto H(-\lambda_{\bQ})\), and therefore \(Z(\lambda_{\bQ})=Z(-\lambda_{\bQ})\). Equivalently, the anchor ensemble does not distinguish between the two symmetry-related signs of the probed channel, so odd moments vanish and the corresponding fluctuation distribution is even. For the N\'eel channel one may use any symmetry of \(H_0\) that flips \(M^z\), such as a one-site translation when available or, more generally, a global \(\pi\) spin rotation. For translation-resolved VBS/PSS channels, the relevant \(U\) is a lattice symmetry that interchanges the two staggered subclasses \cite{tinkham1964group,sachdev2011quantum,demidio2023lee}. In these cases the anchor distribution of the time-averaged configuration observable is even,
\begin{equation}
P_0(m)=P_0(-m).
\label{eq:P0_evenness}
\end{equation}
Along the imaginary axis, \(\lambda_{\bQ}=\ii g_{\bQ}\), this gives
\begin{equation}
G(\ii g_{\bQ})
=
\langle e^{\ii g_{\bQ}X[\mathcal C]}\rangle_0
=
\langle \cos(g_{\bQ}X[\mathcal C])\rangle_0
\in \mathbb R
\label{eq:theory_G_real_imag_axis}
\end{equation}
in the exact symmetric ensemble. Equivalently, the sine part cancels between symmetry-related fluctuations with opposite signs of \(X[\mathcal C]\); in the present SSE estimators this corresponds to the two staggered \(\mathbb Z_2\)-related subclasses entering with opposite phases, e.g. \((\uparrow\downarrow)\) versus \((\downarrow\uparrow)\) vertices in the N\'eel channel and the two staggered bond/plaquette subclasses in the VBS/PSS channels. This symmetry-induced near-reality is what makes staggered, symmetry-resolved source scans display particularly clear Lee--Yang structure in one-dimensional imaginary-axis scans. By contrast, for deformations without such an operator-odd symmetry, for example generic uniform or Fisher-type complexifications of couplings, the evenness constraint is absent, \(G(\ii g)\) is generally complex along the scan direction, and the nearest zeros need not be tied to that axis \cite{yang1952statistical,lee1952statistical,blythe2003lee,bena2005statistical}. In practice, finite Monte Carlo statistics leave a small residual imaginary part even in the symmetry-resolved scans, since we do not explicitly symmetrize the estimator. We therefore identify low-lying Lee--Yang zeros from peaks of \(-\ln|G(\ii g_{\bQ})|\), equivalently minima of \(|G|\), rather than by imposing an exact real-root condition \cite{demidio2023lee,wada2025locating}.

%============================================================
\section{SSE reweighting estimators for symmetry-resolved generators}
\label{sec:ratio_and_factorization}
%============================================================

%------------------------------------------------------------
\subsection{General ratio-estimator framework}
\label{subsec:anchor_ratio_factorization}
%------------------------------------------------------------

We sample stochastic-series-expansion (SSE) configurations \(\mathcal C\) from a real, symmetry-preserving anchor Hamiltonian \(H_0\) and evaluate Lee--Yang generators for complex deformations by reweighting \cite{sandvik1991quantum,sandvik1999stochastic,syljuasen2002quantum,sandvik2010computational,demidio2023lee,ding2024reweight}. In this subsection, the symbol \(\lambda\) stands for whichever complex source coordinate is being scanned. For such a general source, let \(W(\mathcal C;\lambda)\) be the SSE weight of a stored configuration. The corresponding configuration-level ratio is
\begin{equation}
R(\lambda;\mathcal C)\equiv \frac{W(\mathcal C;\lambda)}{W(\mathcal C;0)}.
\label{eq:Rcfg_def_SM}
\end{equation}
Its anchor average gives the normalized generator,
\begin{equation}
G(\lambda)\equiv \expval{R(\lambda;\mathcal C)}_{0}
=
\frac{Z(\lambda)}{Z(0)}.
\label{eq:R_def}
\end{equation}
This is the standard ratio-estimator representation of the normalized partition function used in reweighting-based Monte Carlo analyses \cite{ferrenberg1988new,ferrenberg1989optimized,demidio2023lee,ding2024reweight,wang2025bipartite}.

In the main text we specialize to imaginary-axis scans, \(\lambda=\ii g\), and correspondingly write \(R(\ii g;\mathcal C)\) and \(G(\ii g)\). In the Supplemental Material we retain the general notation \(\lambda\) in the formal setup and then switch to the model-specific scan variables used in the data, namely
\begin{equation}
p \in \{h,\xi_x,\xi_y,\eta_x,\eta_y,\eta_{+},\eta_{-}\},
\label{eq:axes_set}
\end{equation}
with complex deformations \(\ii p\). Here \(h\) is the N\'eel source, \(\xi_x\) and \(\xi_y\) are bond-VBS sources, \(\eta_x\) and \(\eta_y\) are interaction-supported VBS sources, and \(\eta_{\pm}\) are checkerboard-PSS sources. When scanning a selected source or source pair, all other coordinates are set to zero. In the SSE implementation, deformations set to zero contribute an identically unit ratio factor at the configuration level \cite{demidio2023lee,ding2024reweight,wang2025bipartite}.

For the deformations studied here, analytic continuation modifies only a specified subset of local diagonal matrix elements along the operator string. The configuration ratio therefore factorizes into contributions from the affected operator classes, and for multi-parameter scans it further factorizes by symmetry sector within each class. This structure is explicit in Eqs.~\eqref{eq:Rh} and \eqref{eq:RJx_SM}--\eqref{eq:Rpss_minus_SM}, and is the basis for efficient accumulation on source grids \cite{sandvik1999stochastic,syljuasen2002quantum,sandvik2010computational,demidio2023lee,ding2024reweight}.

To control the dynamic range, we evaluate \(G(\lambda)\) in a max-shifted form. Writing
\begin{equation}
a_{\mathrm{cfg}}(\lambda)\equiv \ln\!\abs{R(\lambda;\mathcal C)},
\qquad
b_{\mathrm{cfg}}(\lambda)\equiv \Arg\!\big(R(\lambda;\mathcal C)\big),
\end{equation}
and defining \(a_{\max}(\lambda)=\max_{\mathrm{cfg}} a_{\mathrm{cfg}}(\lambda)\) over the stored ensemble, we compute
\begin{equation}
G(\lambda)
=
\frac{e^{a_{\max}(\lambda)}}{N_{\mathrm{cfg}}}\sum_{\mathrm{cfg}}
\exp\!\big(a_{\mathrm{cfg}}(\lambda)-a_{\max}(\lambda)\big)\,e^{\ii b_{\mathrm{cfg}}(\lambda)}.
\label{eq:R_stable}
\end{equation}
Here \(N_{\mathrm{cfg}}\) is the number of stored configurations and \(\Arg\) denotes the principal complex phase. Equation~\eqref{eq:R_stable} is used only as a numerically stable implementation of the anchor average in Eq.~\eqref{eq:R_def}. Physical conclusions are drawn from the finite-size scaling of the resulting \(G(\lambda)\) \cite{demidio2023lee,ding2024reweight,wada2025locating}.

Before giving the explicit ratio formulas, we note a sign-convention point. The formal sections above use \(H(\lambda)=H_0-\lambda O\), which is convenient for the general Lee--Yang derivations. In the actual SSE implementation, however, the imaginary source is inserted through the positive coefficients appearing in the operator-string weights of \(-H\), so the effective sign of the deformation depends on the microscopic probe. Here \(g\) stands schematically for the relevant real scan variable (\(h,\xi_\mu,\eta_\mu,\eta_\pm\)). With the conventions used here, the N\'eel and bond-supported VBS scans correspond effectively to \(H^{\rm SSE}=H_0+\ii g\,O\), whereas the interaction-supported VBS and checkerboard-PSS scans correspond effectively to \(H^{\rm SSE}=H_0-\ii g\,O\). This distinction only changes the orientation of the imaginary-axis scan, i.e. a channel-dependent relabeling \(g\to -g\), and therefore does not affect the Lee--Yang zero set, the profile of \(|G|\), or the associated finite-size scaling. The channel-specific ratio expressions below are the definitions used in the numerical analysis.

%------------------------------------------------------------
\subsection{N\'eel-channel estimator}
\label{subsec:neel_channel}
%------------------------------------------------------------

N\'eel order resides in the \(\bQ=(\pi,\pi)\) sector \cite{sandvik2007evidence,sandvik2010continuous,kaul2013bridging}. We use the \(z\)-component staggered magnetization
\begin{equation}
M^z=\sum_{\br}(-1)^{x+y}S^z_{\br},
\label{eq:Neel_order}
\end{equation}
and in the SSE implementation the corresponding imaginary deformation is
\begin{equation}
H_h^{\rm SSE}=H_0+\ii h\,M^z,
\qquad
h\in\mathbb R.
\label{eq:Neel_field}
\end{equation}
The symmetry-resolved \((\pi,\pi)\) generator is obtained from the scan \(G_h(\ii h)\) \cite{demidio2023lee,wada2025locating}.

On the bipartite lattice we orient each nearest-neighbor bond from \(i\in A\) to \(j\in B\). Let
\(
s=(\mu,\epsilon)\in\{x+,x-,y+,y-\}
\),
where \(\mu\in\{x,y\}\) is the bond orientation and \(\epsilon=\pm\) labels the two translation-related subclasses; the corresponding real anchor denominator is \(J_s\). For each stored configuration \(\mathcal C\), the counts
\(n^{(s)}_{\uparrow\downarrow}\) and \(n^{(s)}_{\downarrow\uparrow}\) record diagonal SSE vertices in sector \(s\) with ordered local bond states \((\uparrow_i\downarrow_j)\) and \((\downarrow_i\uparrow_j)\), respectively. Their diagonal matrix elements are shifted from \(J_s/2\) to \(J_s/2\mp \ii h/N_c\), so
\begin{equation}
\alpha_s=\frac{2}{N_c\,J_s},
\label{eq:alpha_s}
\end{equation}
with \(N_c=4\) on the square lattice, and the configuration ratio factorizes as
\begin{equation}
R_h(\ii h;\mathcal C)
=
\prod_{s\in\{x+,x-,y+,y-\}}
\bigl(1-\ii \alpha_s h\bigr)^{n^{(s)}_{\uparrow\downarrow}}
\bigl(1+\ii \alpha_s h\bigr)^{n^{(s)}_{\downarrow\uparrow}}.
\label{eq:Rh}
\end{equation}
The corresponding generator is
\begin{equation}
G_h(\ii h)=\avg{R_h(\ii h;\mathcal C)}_0 .
\end{equation}

%------------------------------------------------------------
\subsection{Bond-supported VBS estimators}
\label{subsec:vbs_bond_channels}
%------------------------------------------------------------

Columnar VBS order is naturally diagnosed in the bond-energy sectors at \((\pi,0)\) and \((0,\pi)\) \cite{sandvik2007evidence,sandvik2010continuous,nahum2015emergent,nahum2015deconfined,shao2016quantum}. Defining
\begin{equation}
B_x(\br)=\bS_{\br}\!\cdot\!\bS_{\br+\hat x},
\qquad
B_y(\br)=\bS_{\br}\!\cdot\!\bS_{\br+\hat y},
\end{equation}
the corresponding channel operators are
\(
D_x=\sum_{\br}(-1)^x B_x(\br)
\)
and
\(
D_y=\sum_{\br}(-1)^y B_y(\br)
\)
\cite{sandvik2007evidence,sandvik2010continuous,nahum2015deconfined,wang2017deconfined}.

In the SSE implementation we continue the real bond denominators as
\(
J_{x\epsilon}\to J_{x\epsilon}+\ii\,\epsilon\,\xi_x
\)
and
\(
J_{y\epsilon}\to J_{y\epsilon}+\ii\,\epsilon\,\xi_y
\),
where \(\epsilon=\pm\) records the sign of the staggering factor: \(\epsilon=(-1)^x\) for \(x\)-bonds and \(\epsilon=(-1)^y\) for \(y\)-bonds. Because \(P_{ij}=\tfrac14-\bS_i\!\cdot\!\bS_j\), this corresponds, up to an irrelevant source-dependent constant, to
\(
H_{Jx}^{\rm SSE}=H_0+\ii\xi_x D_x
\)
and
\(
H_{Jy}^{\rm SSE}=H_0+\ii\xi_y D_y
\).
If \(N_{Jx\epsilon}\) and \(N_{Jy\epsilon}\) denote the sector-resolved insertion counts, then
\begin{align}
R_{Jx}(\ii\xi_x;\mathcal C)
&=
\left(1+\ii\frac{\xi_x}{J_{x+}}\right)^{N_{Jx+}}
\left(1-\ii\frac{\xi_x}{J_{x-}}\right)^{N_{Jx-}},
\label{eq:RJx_SM}
\\[1mm]
R_{Jy}(\ii\xi_y;\mathcal C)
&=
\left(1+\ii\frac{\xi_y}{J_{y+}}\right)^{N_{Jy+}}
\left(1-\ii\frac{\xi_y}{J_{y-}}\right)^{N_{Jy-}}.
\label{eq:RJy_SM}
\end{align}
The corresponding generators are
\begin{equation}
G_{Jx}(\ii\xi_x)=\avg{R_{Jx}(\ii\xi_x;\mathcal C)}_0,
\qquad
G_{Jy}(\ii\xi_y)=\avg{R_{Jy}(\ii\xi_y;\mathcal C)}_0 .
\end{equation}

%------------------------------------------------------------
\subsection{Interaction-supported VBS estimators}
\label{subsec:vbs_plaquette_channels}
%------------------------------------------------------------

In J-Q models, the multispin interaction terms furnish alternative microscopic realizations of the same columnar VBS symmetry sectors as the bond-supported probes \cite{sandvik2007evidence,lou2009antiferromagnetic,shao2016quantum}. For J-Q$_2$ the local object is an elementary plaquette carrying two parallel singlet projectors, while for J-Q$_3$ it is a \(2\times 3\) or \(3\times 2\) rectangle carrying three adjacent parallel singlet projectors:
\begin{equation}
H_{Q_2}
=
-Q_{2x}\sum_{\square_x} P_{i,i+\hat x}P_{j,j+\hat x}
-
Q_{2y}\sum_{\square_y} P_{i,i+\hat y}P_{j,j+\hat y},
\label{eq:Q2_cartoon}
\end{equation}
\begin{equation}
H_{Q_3}
=
-Q_{3x}\sum_{r_x} \prod_{m=1}^{3} P_{i_m,i_m+\hat x}
-
Q_{3y}\sum_{r_y} \prod_{m=1}^{3} P_{j_m,j_m+\hat y}.
\label{eq:Q3_cartoon}
\end{equation}
Here the imaginary continuation acts directly on the multispin couplings, so the effective deformation keeps the formal sign convention. Introducing
\(
Q_{p\mu,\epsilon}\to Q_{p\mu}+\ii\,\epsilon\,\eta_\mu
\)
for \(p\in\{2,3\}\), with \(\epsilon=(-1)^x\) in the \(x\)-channel and \(\epsilon=(-1)^y\) in the \(y\)-channel, gives the interaction-supported VBS scans.

With sector-resolved insertion counts \(N_{Q2x\pm}\) and \(N_{Q2y\pm}\), the J-Q$_2$ ratio factors are
\begin{align}
R_{Q2x}(\ii\eta_x;\mathcal C)
&=
\left(1+\ii\frac{\eta_x}{Q_{2x}}\right)^{N_{Q2x+}}
\left(1-\ii\frac{\eta_x}{Q_{2x}}\right)^{N_{Q2x-}},
\label{eq:RQ2x_SM}
\\[1mm]
R_{Q2y}(\ii\eta_y;\mathcal C)
&=
\left(1+\ii\frac{\eta_y}{Q_{2y}}\right)^{N_{Q2y+}}
\left(1-\ii\frac{\eta_y}{Q_{2y}}\right)^{N_{Q2y-}}.
\label{eq:RQ2y_SM}
\end{align}
For J-Q$_3$, with \(N_{Q3x\pm}\) and \(N_{Q3y\pm}\) the corresponding rectangle-insertion counts,
\begin{align}
R_{Q3x}(\ii\eta_x;\mathcal C)
&=
\left(1+\ii\frac{\eta_x}{Q_{3x}}\right)^{N_{Q3x+}}
\left(1-\ii\frac{\eta_x}{Q_{3x}}\right)^{N_{Q3x-}},
\label{eq:RQ3x_SM}
\\[1mm]
R_{Q3y}(\ii\eta_y;\mathcal C)
&=
\left(1+\ii\frac{\eta_y}{Q_{3y}}\right)^{N_{Q3y+}}
\left(1-\ii\frac{\eta_y}{Q_{3y}}\right)^{N_{Q3y-}}.
\label{eq:RQ3y_SM}
\end{align}
For joint scans in either model,
\begin{equation}
R_{Qp}(\ii\eta_x,\ii\eta_y;\mathcal C)
=
R_{Qpx}(\ii\eta_x;\mathcal C)\,R_{Qpy}(\ii\eta_y;\mathcal C),
\qquad
p\in\{2,3\},
\label{eq:RQp_xy_SM}
\end{equation}
and
\begin{equation}
G_{Qp}(\ii\eta_x,\ii\eta_y)=\avg{R_{Qp}(\ii\eta_x,\ii\eta_y;\mathcal C)}_0,
\qquad
p\in\{2,3\}.
\end{equation}

%------------------------------------------------------------
\subsection{Checkerboard-resolved PSS estimator}
\label{subsec:pss_channel}
%------------------------------------------------------------

For the checkerboard J-Q model, plaquette observables are naturally refined by the checkerboard decomposition, since the microscopic multi-spin interaction acts only on one checkerboard sublattice of plaquettes \cite{zhao2019symmetry,sun2021emergent}. We label plaquettes by
\(\bR_\square=(x(\square),y(\square))\) and define
\begin{equation}
\epsilon_{\rm cb}(\square)\equiv (-1)^{x(\square)+y(\square)}=\pm 1,
\qquad
\epsilon_{\rm row}(\square)\equiv (-1)^{y(\square)}=\pm 1.
\label{eq:PSS_eps_defs_SM}
\end{equation}
For a local plaquette density \(\mathcal O(\square)\), the projected row-staggered operators are
\begin{equation}
O^{\rm PSS}_{s}\equiv \sum_{\square}\frac{1+s\,\epsilon_{\rm cb}(\square)}{2}\,
\epsilon_{\rm row}(\square)\,\mathcal O(\square),
\qquad s=\pm 1,
\label{eq:PSS_ops_SM}
\end{equation}
with \(O^{\rm PSS}_{+}\equiv O^{\rm PSS}_{e}\) and \(O^{\rm PSS}_{-}\equiv O^{\rm PSS}_{o}\). When both checkerboard families are active,
\begin{equation}
O^{\rm PSS}_{e}+O^{\rm PSS}_{o}=O^{\rm plaq}_{(0,\pi)},\qquad
O^{\rm PSS}_{e}-O^{\rm PSS}_{o}=O^{\rm plaq}_{(\pi,0)}.
\label{eq:PSS_to_clean_ops_SM}
\end{equation}
In CBJQ only one checkerboard family is active, so the SSE probe reduces to the corresponding projected operator.

The PSS deformation is
\begin{equation}
H_{\mathrm{pss}}^{\rm SSE}(\ii\eta_{+},\ii\eta_{-})
=
H_0-\ii\eta_{+}\,O^{\rm PSS}_{+}-\ii\eta_{-}\,O^{\rm PSS}_{-}.
\label{eq:PSS_source_coupling_SM}
\end{equation}
Let \(N_{Q\alpha,\rho}\) count insertions on checkerboard family \(\alpha=\pm\) with row parity \(\rho=\pm\). Then
\begin{equation}
R_{\mathrm{pss}}(\ii\eta_{+},\ii\eta_{-};\mathcal C)
=
R_{\mathrm{pss},+}(\ii\eta_{+};\mathcal C)\,R_{\mathrm{pss},-}(\ii\eta_{-};\mathcal C),
\label{eq:Rpss_factor_SM}
\end{equation}
with
\begin{align}
R_{\mathrm{pss},+}(\ii\eta_{+};\mathcal C)
&=
\left(1+\ii\frac{\eta_{+}}{Q_{+}}\right)^{N_{Q+,+}}
\left(1-\ii\frac{\eta_{+}}{Q_{+}}\right)^{N_{Q+,-}},
\label{eq:Rpss_plus_SM}
\\[1mm]
R_{\mathrm{pss},-}(\ii\eta_{-};\mathcal C)
&=
\left(1+\ii\frac{\eta_{-}}{Q_{-}}\right)^{N_{Q-,+}}
\left(1-\ii\frac{\eta_{-}}{Q_{-}}\right)^{N_{Q-,-}}.
\label{eq:Rpss_minus_SM}
\end{align}
The generator is
\begin{equation}
G_{\mathrm{pss}}(\ii\eta_{+},\ii\eta_{-})
=
\avg{R_{\mathrm{pss}}(\ii\eta_{+},\ii\eta_{-};\mathcal C)}_0 .
\end{equation}
For CBJQ only one checkerboard family carries nonzero \(Q\), so in practice
\begin{equation}
R_{\mathrm{pss}}(\ii\eta,0;\mathcal C)=R_{\mathrm{pss},+}(\ii\eta;\mathcal C),
\qquad
G_{\mathrm{pss}}(\ii\eta,0)=\avg{R_{\mathrm{pss},+}(\ii\eta;\mathcal C)}_0 .
\label{eq:Rpss_CBJQ_reduce_SM}
\end{equation}

%============================================================
\section{Numerical details and supplementary interaction-supported results}
\label{sec:sm_numerics_plaq}
%============================================================

%------------------------------------------------------------
% Figure: Interaction-supported VBS estimators for JQ3 and JQ2
%------------------------------------------------------------
\begin{figure}[t]
  \centering
  \includegraphics[width=0.35\linewidth]{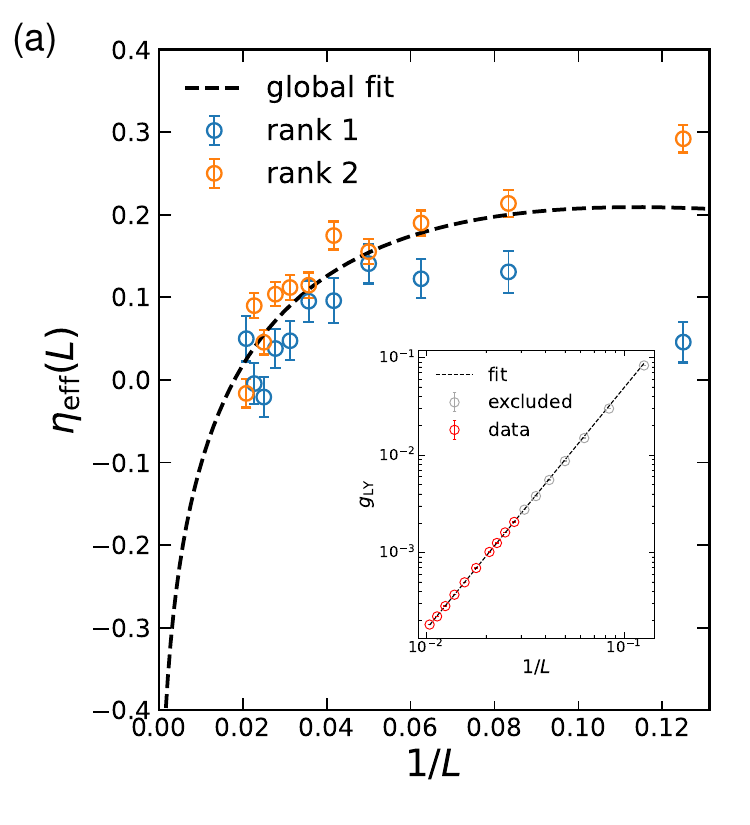}%
  \includegraphics[width=0.35\linewidth]{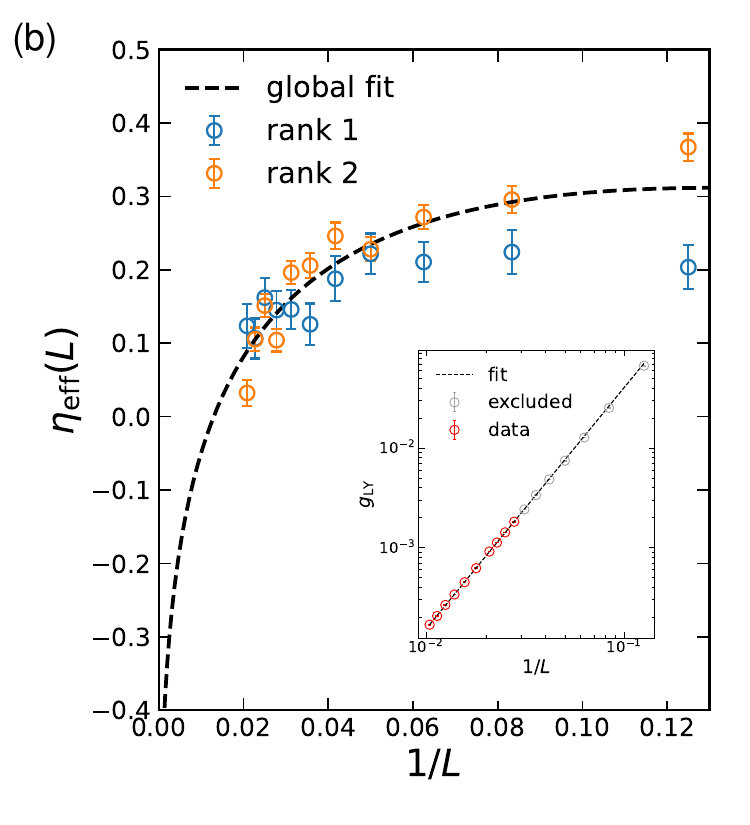}
  \caption{
    Leading Lee--Yang zeros \(g_{\mathrm{LY}}\) obtained from interaction-supported columnar VBS probes, using \(Q_3\)-rectangle-supported operators in the \(2{+}1\)D J-Q$_3$ model at \(J_c=J/Q_3=0.67045\) \cite{lou2009antiferromagnetic,zhao2022scaling,takahashi2024so5} in panel (a), and \(Q_2\)-plaquette-supported operators in the \(2{+}1\)D J-Q$_2$ model at \(J_c=J/Q_2=0.045\) \cite{sandvik2010continuous,suwa2016level,shao2016quantum,sandvik2011thermodynamics,takahashi2024so5} in panel (b). In each panel, the two-size estimator \(\eta_{\mathrm{eff}}(L)\) is constructed from \((L,2L)\) pairs using the first two zeros, and the dashed line shows the first-order reference form \(\eta+c_1L^{-\omega}+c_2L^{-2\omega}\) with fixed \(\eta=-1\), where \(c_1\) and \(c_2\) are fitting coefficients and \(\omega\) is an effective correction exponent. The interaction-supported results closely track the bond-supported VBS results shown in the main text, including the systematic downward drift of \(\eta_{\mathrm{eff}}(L)\). Insets: leading \(g_{\mathrm{LY}}\) versus \(1/L\), fitted to \(a\,L^{-y_h}\), where \(a\) is a nonuniversal amplitude and the fitted \(\eta\) is defined by \(y_h=(d+z+2-\eta)/2\), excluding \(L\le 32\).
  }
  \label{fig:jq23_plaq_maxima_eta_fit}
\end{figure}

%------------------------------------------------------------
\subsection{Numerical details}
\label{subsec:sm_numerical_details}
%------------------------------------------------------------

All simulations were performed within the stochastic series expansion (SSE) quantum Monte Carlo framework with loop updates, using the symmetry-preserving anchor Hamiltonians described in the main text \cite{sandvik1991quantum,sandvik1999stochastic,syljuasen2002quantum,sandvik2010computational,evertz2003loop}. We studied \(L\times L\) square lattices with periodic boundary conditions using a fixed-aspect-ratio scaling protocol \(\beta=L/2\), chosen to remain in the expected \(z=1\) scaling window for the square-lattice models considered here \cite{sandvik2007evidence,melko2008scaling,sandvik2010continuous,kaul2013bridging,shao2016quantum}. For each parameter point, we accumulated \(10^8\) Monte Carlo steps. System sizes extend up to \(L=128\) for the columnar dimerized Heisenberg model and up to \(L=96\) for the CBJQ, J-Q$_2$, and J-Q$_3$ models. Lee--Yang generators were evaluated by reweighting configurations sampled from the anchor ensemble to complex source fields through the ratio estimators introduced in the main text \cite{ding2024reweight,demidio2023lee,wada2025locating}. The range of system sizes and the sampling effort are comparable to those of previous large-scale SSE studies of square-lattice quantum magnets \cite{sandvik2007evidence,melko2008scaling,sandvik2010continuous,shao2016quantum,zhao2019symmetry} and are sufficient to resolve the slow finite-size drift discussed in the main text.

%------------------------------------------------------------
\subsection{Interaction-supported VBS probes in J-Q$_2$ and J-Q$_3$}
\label{subsec:sm_plaq_vbs_jq}
%------------------------------------------------------------

As an additional robustness check, we also computed Lee--Yang zeros using interaction-supported VBS probes in the J-Q$_2$ and J-Q$_3$ models. Specifically, we used the \(Q_2\)-plaquette-supported operators in J-Q$_2$ and the \(Q_3\)-rectangle-supported operators in J-Q$_3$, both in the same columnar symmetry sectors as the bond-supported probes \(D_x\) and \(D_y\) used in the main text \cite{sandvik2007evidence,sandvik2010continuous,nahum2015emergent,nahum2015deconfined,shao2016quantum}. The resulting leading-zero flow and the corresponding two-size estimator \(\eta_{\mathrm{eff}}(L)\) are shown in Fig.~\ref{fig:jq23_plaq_maxima_eta_fit}.

The interaction-supported results closely parallel the bond-supported results presented in the main text. In particular, both J-Q$_3$ and J-Q$_2$ display the same qualitative downward drift of \(\eta_{\mathrm{eff}}(L)\) with increasing \(L\), together with the same slow flow toward the first-order benchmark. This consistency is expected on symmetry grounds: bond-supported, plaquette-supported, and rectangle-supported columnar VBS operators belonging to the same lattice-symmetry sector can mix under coarse graining \cite{sachdev2011quantum,kaul2013bridging,nahum2015deconfined,wang2017deconfined}. The close numerical agreement between the different probes indicates that the observed Lee--Yang drift is a property of the columnar VBS channel itself, rather than a consequence of a particular microscopic estimator. For this reason, we use the bond-supported VBS probes in the main text and present the interaction-supported results here as a supplementary consistency check.

%============================================================
\section{Coupling dependence of the Lee--Yang flow in the J-Q$_2$ model}
\label{sec:sm_jq2_real_coupling}
%============================================================

\begin{figure}[t]
  \centering
  \includegraphics[width=\linewidth]{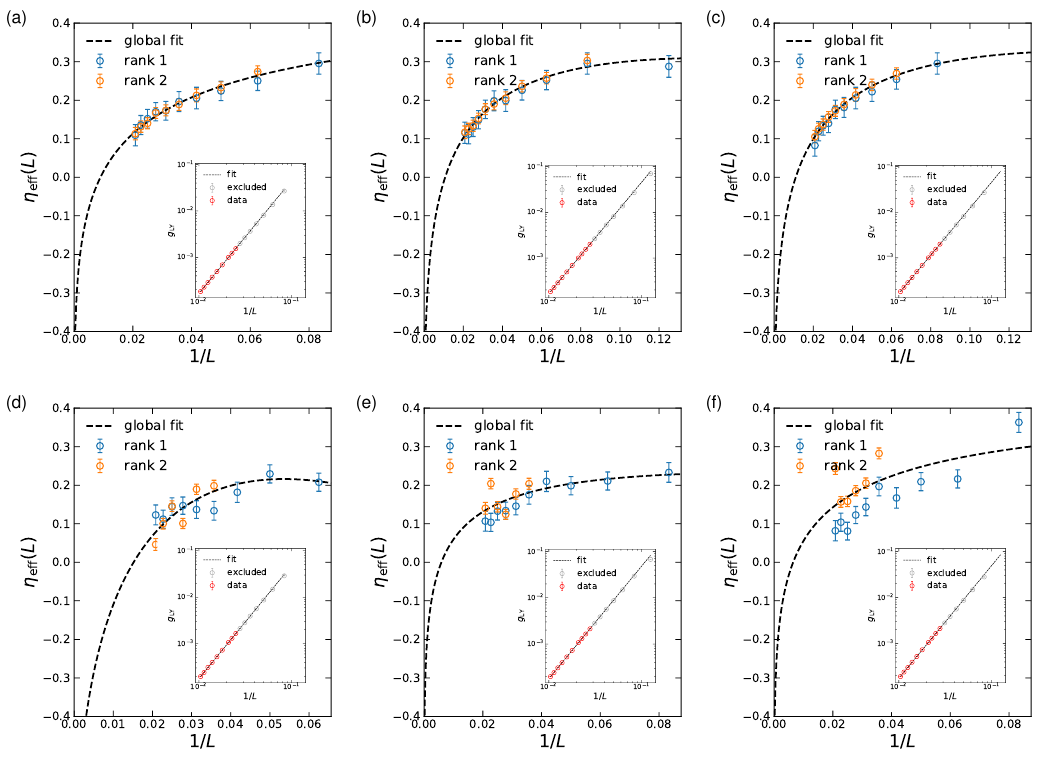}
  \caption{
    Leading Lee--Yang zeros \(g_{\mathrm{LY}}\) and two-size estimators \(\eta_{\mathrm{eff}}(L)\) for the \(2{+}1\)D J-Q$_2$ model. The columns correspond to \(J/Q_2=0.04500\), \(0.04505\), and \(0.04510\), respectively. Panels (a)--(c) show zeros obtained from the imaginary N\'eel source \(\ii h\) conjugate to \(M^z\), and panels (d)--(f) show zeros obtained from the imaginary columnar-VBS source \(\ii\xi_x\) conjugate to \(D_x\). The two-size estimator is constructed from \((L,2L)\) pairs for the first two zeros. Dashed curves are constrained fits to the first-order reference form \(\eta+c_1L^{-\omega}+c_2L^{-2\omega}\), with \(\eta=-1\) fixed. From left to right, the fitted effective correction exponents are \(\omega=0.20\pm0.11\), \(0.256\pm0.040\), and \(0.254\pm0.077\) in the N\'eel channel, and \(0.434\pm0.058\), \(0.189\pm0.092\), and \(0.189\pm0.107\) in the VBS channel. The constrained fit in panel (f) is less stable and is included as a reference for the finite-size trend. Insets show fits of the leading zero to \(g_{\mathrm{LY}}=aL^{-y_h}\), with \(y_h=(5-\eta)/2\). The corresponding effective exponents are \(\eta=0.138\pm0.021\), \(0.132\pm0.021\), and \(0.126\pm0.021\) in panels (a)--(c), and \(0.134\pm0.018\), \(0.132\pm0.018\), and \(0.102\pm0.023\) in panels (d)--(f).
  }
  \label{fig:jq2_maxima_eta_2x3}
\end{figure}

Figure~\ref{fig:jq2_maxima_eta_2x3} compares the N\'eel- and VBS-channel Lee--Yang-zero flow at \(J/Q_2=0.04500\), \(0.04505\), and \(0.04510\), three closely spaced couplings in the vicinity of the N\'eel--VBS transition \cite{takahashi2024so5}.

The leading-zero exponents agree within uncertainties across the three couplings and between the two order-parameter channels. At each coupling, the two-size estimators obtained from the first two zeros decrease systematically with increasing \(L\) in both the N\'eel and VBS channels. The VBS data at \(J/Q_2=0.04510\) exhibit the largest finite-size corrections but retain the same overall trend.

The agreement across the three couplings demonstrates that the observed two-channel flow is robust over this interval. For a fixed detuning into either ordered phase and at scales beyond the associated crossover length, spacetime-volume pinching occurs only for the source conjugate to that order, while the competing channel retains a finite Lee--Yang edge. Over the available sizes, all three couplings instead exhibit a common downward flow in both channels, with no resolved channel-selective crossover. The coupling robustness and channel coherence therefore reinforce the inferred slow crossover toward spacetime-volume scaling.

\clearpage

\bibliography{ref}